\documentclass[aps,prd,floatfix,superscriptaddress,nofootinbib,11pt]{revtex4-2}
\usepackage{amsmath,bm,amssymb,amsthm,mathrsfs,mathtools,multirow}
\usepackage[title]{appendix}
\usepackage{adjustbox}
\usepackage{rotating}
\usepackage{subfigure,color}
\usepackage{ragged2e}
\justifying\let\raggedright\justifying

\allowdisplaybreaks[4]

\newenvironment{rotatepage}%
    {\clearpage\pagebreak[4]\global\pdfpageattr\expandafter{\the\pdfpageattr/Rotate 90}}%
    {\clearpage\pagebreak[4]\global\pdfpageattr\expandafter{\the\pdfpageattr/Rotate 0}}%
\usepackage{tikz}
\usetikzlibrary{backgrounds}
\usetikzlibrary{shapes,matrix,trees}
\usetikzlibrary{arrows.meta}
\usetikzlibrary{positioning}			% For "above of=" commands
\usetikzlibrary{calc,through}			% For coordinates
\usetikzlibrary{decorations.pathreplacing}     % For curly braces
\usepackage{pgffor}                                        % For repeating patterns
\usetikzlibrary{decorations.pathmorphing}	% For Feynman Diagrams
\usetikzlibrary{decorations.markings}
\tikzset{
	vector/.style={decorate, decoration={snake}, draw},
	provector/.style={decorate, decoration={snake,amplitude=2.5pt}, draw},
	antivector/.style={decorate, decoration={snake,amplitude=-2.5pt}, draw},
	fermion/.style={draw=black, postaction={decorate},
		decoration={markings,mark=at position .6 with {\arrow[draw=black]{>}}}},
	fermionbar/.style={draw=black, postaction={decorate},
		decoration={markings,mark=at position .6 with {\arrow[draw=black]{<}}}},
	fermionnoarrow/.style={draw=black},
	gluon/.style={decorate, draw=black,
		decoration={coil,amplitude=4pt, segment length=5pt}},
	scalar/.style={dashed,draw=black, postaction={decorate},
		decoration={markings,mark=at position .6 with {\arrow[draw=black]{>}}}},
	scalarbar/.style={dashed,draw=black, postaction={decorate},
		decoration={markings,mark=at position .6 with {\arrow[draw=black]{<}}}},
	scalarnoarrow/.style={dashed,draw=black},
	inert/.style={dotted,draw=black, postaction={decorate},
		decoration={markings,mark=at position .5 with {\arrow[draw=black]{>}}}},
	electron/.style={draw=black, postaction={decorate},
		decoration={markings,mark=at position .5 with {\arrow[draw=black]{>}}}},
	bigvector/.style={decorate, decoration={snake,amplitude=4pt}, draw},
	photon/.style={decorate, draw=black,decoration={snake,amplitude=4pt, segment length=5pt} }
}

\definecolor{ccblue}{rgb}{0.0,0.4,0.8}
\usepackage[]{hyperref}
\hypersetup{  colorlinks=true,
	linkcolor=ccblue,
	urlcolor=ccblue,
	citecolor=ccblue}

\begin{document}
%%title%%
\title{The Decomposition of Neutron-Antineutron Oscillation Operators}
	
%%author%%
\author{Shao-Long Chen}
\email[E-mail: ]{chensl@mail.ccnu.edu.cn}
\affiliation{Key Laboratory of Quark and Lepton Physics (MoE) and Institute of Particle Physics, Central China Normal University, Wuhan 430079, China}
\affiliation{Center for High Energy Physics, Peking University, Beijing 100871, China}
\author{Yu-Qi Xiao}
\email[E-mail: ]{xiaoyq@mails.ccnu.edu.cn}
\affiliation{Key Laboratory of Quark and Lepton Physics (MoE) and Institute of Particle Physics, Central China Normal University, Wuhan 430079, China}

%%abstract%%
\begin{abstract}
We study the systematic decomposition of the dimension nine neutron-antineutron oscillation operators at tree and one-loop levels. We discuss the topologies' generation and the assignment of the chiral quarks. The completed lists of the decompositions are provided. We furthermore show an example that the neutron-antineutron oscillation occurs at one-loop level, with the tiny neutrino mass being generated via the scotogenic model and proton decay being evaded.
\end{abstract}
	
\maketitle
	
%%sec1%%    
\section{Introduction}\label{sec1}	
The baryon number violation is required to produce the observed matter-antimatter asymmetry in the Universe~\cite{Sakharov:1967dj}. The neutron-antineutron ($n-\bar{n}$)  oscillation is proposed to be a mechanism involved in generating baryon asymmetry as the process violates the baryon number by two units $|\Delta B|=2$~\cite{Kuzmin:1970nx}. The observation of the $n-\bar{n}$ oscillation shall provide implications for probing the origin of the matter-antimatter asymmetry.

Currently, the $n-\bar{n}$ oscillation is not experimentally observed. The limit on the $n-\bar{n}$ oscillation lifetime from dinucleon decays at Super-Kamiokande (SK) experiment is the most stringent $\tau_{n-\bar{n}}^{\text{SK}}\geq4.7\times10^{8}~{\text{s}}$~\cite{Super-Kamiokande:2020bov}, while ILL experiment obtains the limit from the free $n-\bar{n}$ oscillation as $\tau_{n-\bar{n}}^{\text{ILL}}\geq8.6\times10^{7}~{\text{s}}$~\cite{Baldo-Ceolin:1994hzw}. The sensitivity of future $n-\bar{n}$ oscillation experiments NNBAR program, which searches for free $n-\bar{n}$ oscillation, is around $\tau_{n-\bar{n}}^{\text{NNBAR}}\geq3\times10^{9}~{\text{s}}$~\cite{Addazi:2020nlz}. The expected sensitivity of the DUNE experiment reaches $\tau_{n-\bar{n}}^{\text{DUNE}}\geq5.53\times10^{8}~{\text{s}}$~\cite{DUNE:2020ypp} for an assumed exposure of 400 kton$\cdot$years, and the sensitivity will be further improved by the Hyper-Kamiokande experiment~\cite{Hyper-Kamiokande:2018ofw}.

In the effective field theory approach, the $n-\overline{n}$ oscillation can be described via dimension-nine operators $\mathcal{O}_{n-\overline{n}}\propto(udd)^{2}/\Lambda^{5}$, where $\Lambda$ is denoted as the new physics scale~\cite{Chang:1980ey,Kuo:1980ew,Rao:1982gt,Caswell:1982qs,Rao:1983sd,Buchoff:2015qwa,Oosterhof:2019dlo,Rinaldi:2019thf}. The simple way to generate the operators at tree level has been explored and realized in many different models~\cite{Mohapatra:1980qe,Dev:2015uca,Allahverdi:2017edd,Grojean:2018fus,Bell:2018mgg}, see reviews in~\cite{Mohapatra:2009wp,Babu:2013yww,Babu:2013jba,Phillips:2014fgb,Proceedings:2020nzz,Dev:2022jbf}. The basis for the operators was constructed as
\begin{equation}
	\begin{aligned}
		\mathcal{O}^{1}_{\chi_{1}\chi_{2}\chi_{3}}&=[\overline{u_{i}^{c}}P_{\chi_{1}}u_{j}][\overline{d_{k}^{c}}P_{\chi_{2}}d_{l}][\overline{d_{m}^{c}}P_{\chi_{3}}d_{n}]T^{SSS},\\
		\mathcal{O}^{2}_{\chi_{1}\chi_{2}\chi_{3}}&=[\overline{u_{i}^{c}}P_{\chi_{1}}d_{j}][\overline{u_{k}^{c}}P_{\chi_{2}}d_{l}][\overline{d_{m}^{c}}P_{\chi_{3}}d_{n}]T^{SSS},\\
		\mathcal{O}^{3}_{\chi_{1}\chi_{2}\chi_{3}}&=[\overline{u_{i}^{c}}P_{\chi_{1}}d_{j}][\overline{u_{k}^{c}}P_{\chi_{2}}d_{l}][\overline{d_{m}^{c}}P_{\chi_{3}}d_{n}]T^{AAS},
	\end{aligned}
\end{equation}
where $i,j,k,l,m,n$ are color indices, $P_{\chi_{i}}=P_{L,R}=\frac{1}{2}(1\mp\gamma_{5})$ are chiral projectors, and $T$ is the color tensor
\begin{equation}
	\begin{aligned}
		T^{SSS}&=\epsilon_{ikm}\epsilon_{jln}+\epsilon_{jkm}\epsilon_{iln}+\epsilon_{ilm}\epsilon_{jkn}+\epsilon_{jlm}\epsilon_{ikn}\,,\\
		T^{AAS}&=\epsilon_{ijm}\epsilon_{kln}+\epsilon_{ijn}\epsilon_{klm}\,.
	\end{aligned}
\end{equation}
The number of independent operators is 14, which can be derived using the Fierz relation $\mathcal{O}^{2}_{\chi\chi\chi^{\prime}}-\mathcal{O}^{1}_{\chi\chi\chi^{\prime}}=\mathcal{O}^{3}_{\chi\chi\chi^{\prime}}$. The independent effective operators $\mathcal{O}_{i}$ are defined as~\cite{Buchoff:2015qwa,Rinaldi:2019thf}
\begin{equation}
\begin{gathered}
    \mathcal{O}_{1}=-4\mathcal{O}^{3}_{RRR}\,,\quad   \mathcal{O}_{2}=-4\mathcal{O}^{3}_{LRR}\,,\quad\mathcal{O}_{3}=-4\mathcal{O}^{3}_{LLR}\,,\\
    \mathcal{O}_{4}=-\frac{4}{5}\mathcal{O}^{1}_{RRR}-\frac{16}{5}\mathcal{O}^{2}_{RRR}\,,\quad\mathcal{O}_{5}=\mathcal{O}^{1}_{RLL}\,,\\
    \mathcal{O}_{6}=-4\mathcal{O}^{2}_{RLL}\,,\quad\mathcal{O}_{7}=-\frac{4}{3}\mathcal{O}^{1}_{LLR}-\frac{8}{3}\mathcal{O}^{2}_{LLR}\,.\\
\end{gathered}
\end{equation}
The $n-\bar{n}$ oscillation can be described by the effective Lagrangian 
\begin{align}
\mathcal{L}_{\text{eff}}=\sum\limits_{i=1,2,3,5}(\mathcal{C}_{i}(\mu)\mathcal{O}_{i}(\mu)+\mathcal{C}_{i}^{P}(\mu)\mathcal{O}_{i}^{P}(\mu))\,,
\end{align}
and the transition rate is 
\begin{align}
\tau^{-1}_{n-\bar{n}}=\bigg|\sum\limits_{i=1,2,3,5}(\mathcal{C}_{i}(\mu)\mathcal{M}_{i}(\mu)+\mathcal{C}_{i}^{P}(\mu)\mathcal{M}_{i}^{P}(\mu))\bigg|\,,
\end{align}
in which $\mathcal{C}_{i}(\mu)$ are the Wilson coefficients at the new physics scale $\mu=\mu_{NP}$, $\mathcal{O}^{P}_{i}$ are the operators that parity transformed from $\mathcal{O}_{i}$, and $\mathcal{M}_{i}(\mu)$ is the transition matrix element of the operator of which value can be evolved from the lattice-QCD scale after considering RG running effects~\cite{Buchoff:2015qwa,Rinaldi:2019thf,Rinaldi:2018osy}.

In this work, we systematically discuss the decompositions of the $n-\bar{n}$ oscillation $d=9$ operators. Inspired by the general recipes and the discussions of the decomposition of $0\nu\beta\beta$ decay $d=9$ operators~\cite{Bonnet:2012kh,Chen:2021rcv,Antusch:2008tz},  we detail the decompositions of the $n-\bar{n}$ operators at tree level and one-loop level with the mediators considered as scalars or fermions.

We organize our paper as follows. In Sec.~\ref{sec2}, we discuss how to generate the topologies and diagrams and assign the chiral quarks to the external legs. We discuss which vertices of form $q-q-I$ are vanishing in Appendix~\ref{appendixA} and give the completed lists of decompositions of the neutron-antineutron $d=9$ operator at tree level and one-loop level in Appendix~\ref{appendixB}. In Sec.~\ref{example}, we provide an example of one-loop model and discuss the neutron-antineutron oscillation rate. Finally, we concluded our result in Sec.~\ref{sec4}.

\section{Topologies, Diagrams and Decompositions}\label{sec2}
In this section, we will detail the decompositions and give the lists. In the following discussions, we consider the internal particles to be scalars or fermions, as adding vectors may call for new gauge symmetry beyond the Standard Model.

\subsection{Generation of the topologies and diagrams}
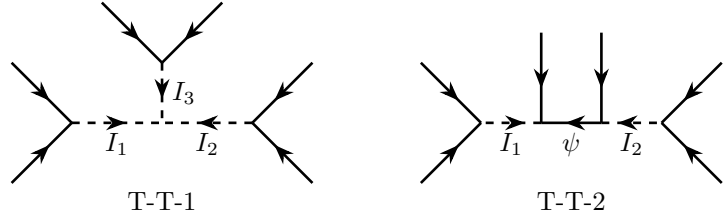
\begin{figure}[htb]
\centering
	\begin{tikzpicture}[line width=1pt, scale=0.8,>=Stealth]
	%%%%
	\path (0:0) coordinate (a0);
	\path (0:1.5) coordinate (a1);
	\path (90:1) coordinate (a2);
	\path(180:1.5) coordinate (a3);
	%%%%
	\draw [scalar] (a1)--(a0);
	\draw [scalar] (a2)--(a0);
	\draw [scalar] (a3)--(a0);
	\draw [fermionbar] (a2)--(1,2);
	\draw [fermionbar] (a2)--(-1,2);
	\draw [fermionbar] (a1)--(2.5,1);
	\draw [fermionbar] (a1)--(2.5,-1);
	\draw [fermionbar] (a3)--(-2.5,1);
	\draw [fermionbar] (a3)--(-2.5,-1);
	%%%%
	\node[below] at (180:0.75) {$I_{1}$};
	\node[below] at (0:0.75) {$I_{2}$};
	\node[right] at (0,0.5) {$I_{3}$};
	\node at (0,-1.3) {\rm{T-T-1}};
	\end{tikzpicture}\qquad\qquad
	\begin{tikzpicture}[line width=1pt, scale=0.8,>=Stealth]
	%%%% 
	\path (0:0) coordinate (a0);
	\path (0:1.5) coordinate (a1);
	\path(180:1.5) coordinate (a3);
	%%%%
	\draw [scalarbar] (0.5,0)--(1.5,0);
	\draw [scalarbar] (-0.5,0)--(-1.5,0);
	\draw [fermion] (0.5,0)--(-0.5,0);
	\draw [fermionbar] (a1)--(2.5,1);
	\draw [fermionbar] (a1)--(2.5,-1);
	\draw [fermionbar] (a3)--(-2.5,1);
	\draw [fermionbar] (a3)--(-2.5,-1);
	\draw [fermionbar] (0.5,0)--(0.5,1.5);
	\draw [fermionbar] (-0.5,0)--(-0.5,1.5);
	%%%%
	\node[below] at (180:1) {$I_{1}$};
	\node[below] at (0:1) {$I_{2}$};
	\node[below] at (0,0) {$\psi$};
	\node at (0,-1.3) {\rm{T-T-2}};
	\end{tikzpicture}
	\caption{The tree level diagrams of neutron-antineutron oscillation.}\label{tree-level}
\end{figure}

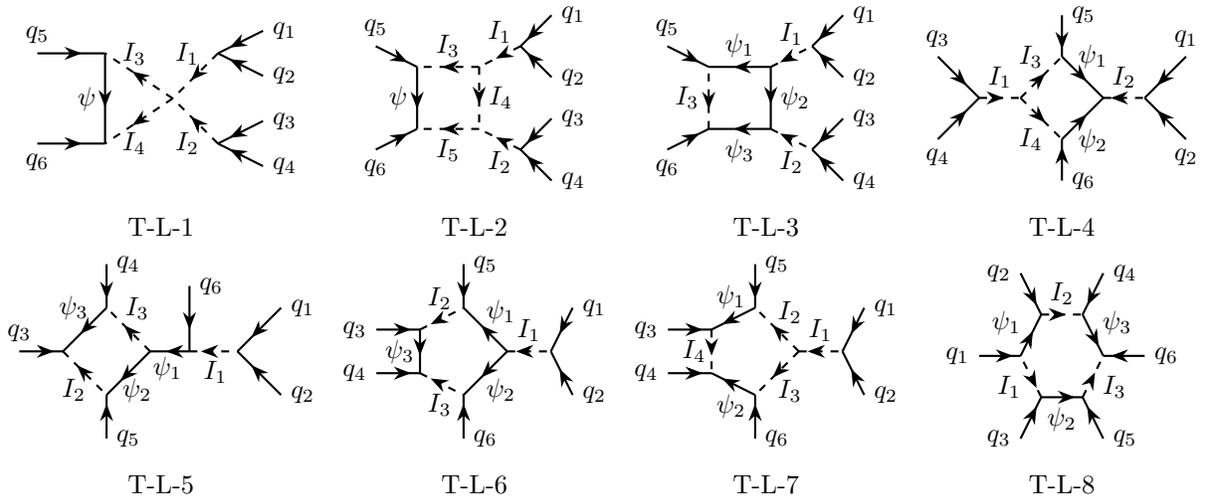
\begin{figure}[b]
\begin{tabular}{cccc}
	\begin{tikzpicture}[line width=0.8pt, scale=0.6,>=Stealth]
	%%%%
	\path (0,0) coordinate (a0);
	\path (1,1) coordinate (a1);
	\path(2,0.5) coordinate (a2);
	\path(2,1.5) coordinate (a3);
	\path (1,-1) coordinate (a4);
	\path(2,-1.5) coordinate (a5);
	\path (2,-0.5) coordinate (a6);
	\path(-1.5,1) coordinate (a7);
	\path(-3,1) coordinate (a8);
	\path(-1.5,-1) coordinate (a9);
	\path(-3,-1) coordinate (a10);
	\path(-3,-2.2) coordinate (a11);
	%%%%%%%
	\draw [fermionbar] (a1)--(a2);
	\draw [fermionbar] (a1)--(a3);
	\draw [fermionbar] (a4)--(a5);
	\draw [fermionbar] (a4)--(a6);
	\draw [fermionbar] (a7)--(a8);
	\draw [fermionbar] (a9)--(a10);
	\draw [fermion] (a7)--(a9);
	\draw [scalar] (a0)--(a7);
	\draw [scalar] (a0)--(a9);
	\draw [scalar] (a1)--(a0);
	\draw [scalar] (a4)--(a0);
	%%%%
	\node[right] at (a3) {$q_{1}$};
	\node[right] at (a2) {$q_{2}$};
	\node[right] at (a6) {$q_{3}$};
	\node[right] at (a5) {$q_{4}$};
	\node[above] at (a8) {$q_{5}$};
	\node[below] at (a10) {$q_{6}$};
	\node at (0.32,1) {$I_{1}$};
	\node at (0.32,-1) {$I_{2}$};
	\node at (-0.85,1) {$I_{3}$};
	\node at (-0.85,-1) {$I_{4}$};
	\node at (-1.9,0) {$\psi$};
	\end{tikzpicture}&
	\begin{tikzpicture}[line width=0.8pt, scale=0.55,>=Stealth]
	%%%%
	\path (0,0.75) coordinate (a0);
	\path (1,1.25) coordinate (a1);
	\path(1.75,2) coordinate (a2);
	\path(1.75,0.5) coordinate (a3);
	\path(0,-0.75) coordinate (a4);
	\path (1.75,-0.5) coordinate (a5);
	\path (1.75,-2) coordinate (a6);
	\path(-1.5,-0.75) coordinate (a7);
	\path(-1.5,0.75) coordinate (a8);
	\path(-2.5,1.25) coordinate (a9);
	\path(-2.5,-1.25) coordinate (a10);
	\path(1,-1.25) coordinate (a11);
	%%%%%%%
	\draw [fermionbar] (a1)--(a2);
	\draw [fermionbar] (a1)--(a3);
	\draw [fermionbar] (a11)--(a5);
	\draw [fermionbar] (a11)--(a6);
	\draw [fermionbar] (a8)--(a9);
	\draw [fermionbar] (a7)--(a10);
	\draw [fermionbar] (a7)--(a8);
	\draw [scalar] (a4)--(a7);
	\draw [scalar] (a0)--(a4);
	\draw [scalar] (a0)--(a8);
	\draw [scalar] (a1)--(a0);
	\draw [scalar] (a11)--(a4);
	%%%%
	\node[right] at (a2) {$q_{1}$};
	\node[right] at (a3) {$q_{2}$};
	\node[right] at (a5) {$q_{3}$};
	\node[right] at (a6) {$q_{4}$};
	\node[below] at (a10) {$q_{6}$};
	\node[above] at (a9) {$q_{5}$};
	\node[above] at (-0.75,0.75) {$I_{3}$};
	\node[below] at (-0.75,-0.75) {$I_{5}$};
	\node[above] at (0.5,1.1) {$I_{1}$};
	\node[below] at (0.5,-1.05) {$I_{2}$};
	\node[right] at (0,0) {$I_{4}$};
	\node[left] at (-1.5,0) {$\psi$};
	\end{tikzpicture}&
	\begin{tikzpicture}[line width=0.8pt, scale=0.55,>=Stealth]
	%%%%
	\path (0,0.75) coordinate (a0);
	\path (1,1.25) coordinate (a1);
	\path(1.75,2) coordinate (a2);
	\path(1.75,0.5) coordinate (a3);
	\path(0,-0.75) coordinate (a4);
	\path (1.75,-0.5) coordinate (a5);
	\path (1.75,-2) coordinate (a6);
	\path(-1.5,-0.75) coordinate (a7);
	\path(-1.5,0.75) coordinate (a8);
	\path(-2.5,1.25) coordinate (a9);
	\path(-2.5,-1.25) coordinate (a10);
	\path(1,-1.25) coordinate (a11);
	%%%%%%%
	\draw [fermionbar] (a1)--(a2);
	\draw [fermionbar] (a1)--(a3);
	\draw [fermionbar] (a11)--(a5);
	\draw [fermionbar] (a11)--(a6);
	\draw [fermionbar] (a8)--(a9);
	\draw [fermionbar] (a7)--(a10);
	\draw [fermion] (a0)--(a8);
	\draw [fermion] (a0)--(a4);
	\draw [fermion] (a4)--(a7);
	\draw [scalarbar] (a0)--(a1);
	\draw [scalarbar] (a4)--(a11);
	\draw [scalarbar] (a7)--(a8);
	%%%%
	\node[right] at (a2) {$q_{1}$};
	\node[right] at (a3) {$q_{2}$};
	\node[right] at (a5) {$q_{3}$};
	\node[right] at (a6) {$q_{4}$};
	\node[below] at (a10) {$q_{6}$};
	\node[above] at (a9) {$q_{5}$};
	\node[above] at (-0.75,0.75) {$\psi_{1}$};
	\node[below] at (-0.75,-0.75) {$\psi_{3}$};
	\node[above] at (0.5,1.05) {$I_{1}$};
	\node[below] at (0.5,-1.05) {$I_{2}$};
	\node[right] at (0,0) {$\psi_{2}$};
	\node[left] at (-1.5,0) {$I_{3}$};
	\end{tikzpicture}
	&
	\begin{tikzpicture}[line width=0.8pt, scale=0.55,>=Stealth]
	%%%%
	\path (0,0) coordinate (a0);
	\path (2,0) coordinate (a1);
	\path(3,1) coordinate (a2);
	\path(3,-1) coordinate (a3);
	\path (-2,0) coordinate (a4);
	\path(-3,1) coordinate (a5);
	\path (-3,-1) coordinate (a6);
	\path(0,1) coordinate (a7);
	\path(0,2) coordinate (a8);
	\path(0,-1) coordinate (a9);
	\path(0,-2) coordinate (a10);
	\path(-1,0) coordinate (a11);
	\path(1,0) coordinate (a12);
	%%%%%%%
	\draw [fermionbar] (a1)--(a2);
	\draw [fermionbar] (a1)--(a3);
	\draw [fermionbar] (a4)--(a5);
	\draw [fermionbar] (a4)--(a6);
	\draw [fermionbar] (a7)--(a8);
	\draw [fermionbar] (a9)--(a10);
	\draw [fermion] (a7)--(a12);
	\draw [fermion] (a9)--(a12);
	\draw [scalar] (a11)--(a7);
	\draw [scalar] (a11)--(a9);
	\draw [scalarbar] (a12)--(a1);
	\draw [scalar] (a4)--(a11);
	%%%%%%
	\node[above] at (a2) {$q_{1}$};
	\node[below] at (a3) {$q_{2}$};
	\node[above] at (a5) {$q_{3}$};
	\node[below] at (a6) {$q_{4}$};
	\node[right] at (a10) {$q_{6}$};
	\node[right] at (a8) {$q_{5}$};
	\node[above] at (-1.5,0) {$I_{1}$};
	\node[above] at (1.5,0) {$I_{2}$};
	\node[above] at (-0.75,0.5) {$I_{3}$};
	\node[below] at (-0.75,-0.5) {$I_{4}$};
	\node[above] at (0.75,0.5) {$\psi_{1}$};
	\node[below] at (0.75,-0.5) {$\psi_{2}$};	
	\end{tikzpicture}\\
	T-L-1&T-L-2&T-L-3&T-L-4\\
           \begin{tikzpicture}[line width=0.8pt, scale=0.58,>=Stealth]
	%%%%
	\path (0,0) coordinate (a0);
	\path (2,0) coordinate (a1);
	\path(3,1) coordinate (a2);
	\path(3,-1) coordinate (a3);
	\path (0.9,0) coordinate (a4);
	\path(0.9,1.5) coordinate (a5);
	\path (0,0) coordinate (a6);
	\path(-1,1) coordinate (a7);
	\path(-1,2) coordinate (a8);
	\path(-1,-1) coordinate (a9);
	\path(-1,-2) coordinate (a10);
	\path(-2,0) coordinate (a11);
	\path(-3,0) coordinate (a12);
	%%%%%%%
	\draw [fermionbar] (a1)--(a2);
	\draw [fermionbar] (a1)--(a3);
	\draw [fermionbar] (a4)--(a5);
	\draw [fermionbar] (a7)--(a8);
	\draw [fermionbar] (a9)--(a10);
	\draw [fermionbar] (a11)--(a12);
 	\draw [fermion] (a4)--(a6);
	\draw [fermion] (a7)--(a11);
	\draw [fermionbar] (a9)--(a6);
	\draw [scalar] (a6)--(a7);
	\draw [scalarbar] (a11)--(a9);
	\draw [scalarbar] (a4)--(a1);
	%%%%
	\node [right] at (a2) {$q_{1}$};
	\node [right] at (a3) {$q_{2}$};
	\node [above] at (a12) {$q_{3}$};
	\node [right] at (a8) {$q_{4}$};	
	\node [right] at (a10) {$q_{5}$};
	\node [right] at (a5) {$q_{6}$};		
	\node [below] at (0.4,0.1) {$\psi_{1}$};	
	\node [below] at (1.5,0) {$I_{1}$};		
	\node [below] at (-0.3,-0.4) {$\psi_{2}$};	
	\node [below] at (-1.75,-0.4) {$I_{2}$};	
	\node [above] at (-0.3,0.5) {$I_{3}$};	
	\node [above] at (-1.75,0.5) {$\psi_{3}$};	
	\end{tikzpicture}&
	\begin{tikzpicture}[line width=0.8pt, scale=0.58,>=Stealth]
	%%%%
	\draw [fermion] (1,0)--(0,1);
	\draw [fermion] (1,0)--(0,-1);
	\draw [fermionbar] (0,1)--(0,2);
	\draw [fermionbar] (0,-1)--(0,-2);
	\draw [fermion] (-1,0.5)--(-1,-0.5);
	\draw [fermionbar] (-1,0.5)--(-2,0.5);
	\draw [fermionbar] (-1,-0.5)--(-2,-0.5);
	\draw [fermion] (2.5,1)--(2,0);
	\draw [fermion] (2.5,-1)--(2,0);
	\draw [scalarbar] (-1,0.5)--(0,1);
	\draw [scalarbar] (-1,-0.5)--(0,-1);
	\draw [scalarbar] (1,0)--(2,0);
	%%%%
	\node [right] at (2.5,1) {$q_{1}$};
	\node [right] at (2.5,-1) {$q_{2}$};
	\node [left] at (-2,0.5) {$q_{3}$};
	\node [left] at (-2,-0.5) {$q_{4}$};
	\node [right] at (0,2) {$q_{5}$};
	\node [right] at (0,-2) {$q_{6}$};
	\node [above] at (-0.55,0.75) {$I_{2}$};
	\node [below] at (-0.55,-0.75) {$I_{3}$};
	\node [left] at (-0.9,0) {$\psi_{3}$};
	\node [right] at (0.3,0.9) {$\psi_{1}$};
	\node [right] at (0.3,-0.9) {$\psi_{2}$};
	\node [above] at (1.5,0) {$I_{1}$};
	\end{tikzpicture}&
	\begin{tikzpicture}[line width=0.8pt, scale=0.58,>=Stealth]
	%%%%
	\draw [fermionbar] (0,1)--(0,2);
	\draw [fermionbar] (0,-1)--(0,-2);
	\draw [fermionbar] (-1,0.5)--(-2,0.5);
	\draw [fermionbar] (-1,-0.5)--(-2,-0.5);
	\draw [fermion] (2.5,1)--(2,0);
	\draw [fermion] (2.5,-1)--(2,0);
	\draw [fermionbar] (-1,0.5)--(0,1);
	\draw [fermionbar] (-1,-0.5)--(0,-1);
	\draw [scalar] (1,0)--(0,1);
	\draw [scalar] (1,0)--(0,-1);
	\draw [scalar] (-1,0.5)--(-1,-0.5);
	\draw [scalarbar] (1,0)--(2,0);
	%%%%
	\node [right] at (2.5,1) {$q_{1}$};
	\node [right] at (2.5,-1) {$q_{2}$};
	\node [left] at (-2,0.5) {$q_{3}$};
	\node [left] at (-2,-0.5) {$q_{4}$};
	\node [right] at (0,2) {$q_{5}$};
	\node [right] at (0,-2) {$q_{6}$};
	\node [above] at (-0.55,0.75) {$\psi_{1}$};
	\node [below] at (-0.55,-0.75) {$\psi_{2}$};
	\node [left] at (-0.9,0) {$I_{4}$};
	\node [right] at (0.3,0.9) {$I_{2}$};
	\node [right] at (0.3,-0.9) {$I_{3}$};
	\node [above] at (1.5,0) {$I_{1}$};
	\end{tikzpicture}&
	\begin{tikzpicture}[line width=0.8pt, scale=0.55,>=Stealth]
	%%%%
	\draw [fermionbar] (1,0)--(2,0);
	\draw [fermionbar] (0.5,1)--(1,2);
	\draw [fermionbar] (-0.5,1)--(-1,2);
	\draw [fermionbar] (-1,0)--(-2,0);
	\draw [fermionbar] (-0.5,-1)--(-1,-2);
	\draw [fermionbar] (0.5,-1)--(1,-2);
	\draw [fermion] (-1,0)--(-0.5,1);
	\draw [fermionbar] (1,0)--(0.5,1);
	\draw [fermionbar] (0.5,-1)--(-0.5,-1);
	\draw [scalar] (-0.5,1)--(0.5,1);
	\draw [scalarbar] (-0.5,-1)--(-1,0);
	\draw [scalar] (0.5,-1)--(1,0);
	%%%%
	\node[above] at (0,1) {$I_{2}$};
	\node[left] at (-0.75,-0.75) {$I_{1}$};
	\node[right] at (0.75,-0.75) {$I_{3}$};
	\node[below] at (0,-1) {$\psi_{2}$}; 
	\node[right] at (0.75,0.8) {$\psi_{3}$};
	\node[left] at (-0.75,0.8) {$\psi_{1}$};
	\node[right] at (2,0) {$q_{6}$};
	\node[right] at (1,2) {$q_{4}$};
	\node[right] at (1,-2) {$q_{5}$};
	\node[left] at (-1,-2) {$q_{3}$};
	\node[left] at (-2,0) {$q_{1}$};
	\node[left] at (-1,2) {$q_{2}$};
	\end{tikzpicture}\\
	T-L-5&T-L-6&T-L-7&T-L-8
	\end{tabular}
	\caption{The genuine diagrams of neutron-antineutron oscillation operator at one-loop level}\label{T-L-G}
\end{figure}

\begin{figure}[tb]
\begin{tabular}{ccccccc}
	\begin{tikzpicture}[line width=0.8pt, scale=0.7,>=Stealth]
	%%%%
	\path (0,0) coordinate (a0);
	\path (0,1.75) coordinate (a1);
	\path(0.5,2.5) coordinate (a2);
	\path(-0.5,2.5) coordinate (a3);
	\path (1.25,0) coordinate (a4);
	\path(2,0.5) coordinate (a5);
	\path (2,-0.5) coordinate (a6);
	\path(-1.25,0) coordinate (a7);
	\path(-2,0.5) coordinate (a8);
	\path(-2,-0.5) coordinate (a9);
	\path(0,1) coordinate (a10);
	\path(-0.5,0) coordinate (a11);
	\path(0.5,0) coordinate (a12);
	%%%%%%%
	\draw [fermionbar] (a1)--(a2);
	\draw [fermionbar] (a1)--(a3);
	\draw [fermionbar] (a4)--(a5);
	\draw [fermionbar] (a4)--(a6);
	\draw [fermionbar] (a7)--(a8);
	\draw [fermionbar] (a7)--(a9);
	\draw [fermionbar] (a11)--(a10);
	\draw [fermionbar] (a11)--(a12);
	\draw [fermionbar] (a12)--(a10);
	\draw [scalarbar] (a10)--(a1);
	\draw [scalarbar] (a11)--(a7);
	\draw [scalarbar] (a12)--(a4);
	%%%%
	\node[left] at (90:1.3) {$I_{1}$};
	\node[below] at (180:1) {$I_{2}$};
	\node[below] at (0:1) {$I_{3}$};
	\node[left] at (120:0.5) {$\psi_{1}$};
	\node[right] at (60:0.5) {$\psi_{2}$};
	\node[below] at (0,0) {$\psi_{3}$};
	\node[left] at (-0.5,2.5) {$q_{1}$};
	\node[right] at (0.5,2.5) {$q_{2}$};
	\node[above] at (-2,0.5) {$q_{3}$};
	\node[below] at (-2,-0.5) {$q_{4}$};
	\node[above] at (2,0.5) {$q_{5}$};
	\node[below] at (2,-0.5) {$q_{6}$};
	\end{tikzpicture}&
	\begin{tikzpicture}[line width=0.8pt, scale=0.7,>=Stealth]
	%%%%
	\path (0,0) coordinate (a0);
	\path (0,1.75) coordinate (a1);
	\path(0.5,2.5) coordinate (a2);
	\path(-0.5,2.5) coordinate (a3);
	\path (1.25,0) coordinate (a4);
	\path(2,0.5) coordinate (a5);
	\path (2,-0.5) coordinate (a6);
	\path(-1.25,0) coordinate (a7);
	\path(-2,0.5) coordinate (a8);
	\path(-2,-0.5) coordinate (a9);
	\path(0,1) coordinate (a10);
	\path(-0.5,0) coordinate (a11);
	\path(0.5,0) coordinate (a12);
	%%%%%%%
	\draw [fermionbar] (a1)--(a2);
	\draw [fermionbar] (a1)--(a3);
	\draw [fermionbar] (a4)--(a5);
	\draw [fermionbar] (a4)--(a6);
	\draw [fermionbar] (a7)--(a8);
	\draw [fermionbar] (a7)--(a9);
	\draw [scalarbar] (a11)--(a10);
	\draw [scalarbar] (a11)--(a12);
	\draw [scalarbar] (a12)--(a10);
	\draw [scalarbar] (a10)--(a1);
	\draw [scalarbar] (a11)--(a7);
	\draw [scalarbar] (a12)--(a4);
	\node[left] at (-0.5,2.5) {$q_{1}$};
	\node[right] at (0.5,2.5) {$q_{2}$};
	\node[above] at (-2,0.5) {$q_{3}$};
	\node[below] at (-2,-0.5) {$q_{4}$};
	\node[above] at (2,0.5) {$q_{5}$};
	\node[below] at (2,-0.5) {$q_{6}$};
	\node[left] at (90:1.3) {$I_{1}$};
	\node[below] at (180:1) {$I_{2}$};
	\node[below] at (0:1) {$I_{3}$};
	\node[left] at (120:0.5) {$I_{4}$};
	\node[right] at (60:0.5) {$I_{5}$};
	\node[below] at (0,0) {$I_{6}$};
	%%%%
	\end{tikzpicture}&
	\begin{tikzpicture}[line width=0.8pt, scale=0.7,>=Stealth]
	%%%%
	\path (0,0) coordinate (a0);
	\path (1.5,0) coordinate (a1);
	\path(2.25,1) coordinate (a2);
	\path(2.25,-1) coordinate (a3);
	\path (-1.75,0) coordinate (a4);
	\path(-2.5,-1) coordinate (a5);
	\path (-2.5,1) coordinate (a6);
	\path(-0.5,1) coordinate (a7);
	\path(-0.5,2) coordinate (a8);
	\path(0.75,0) coordinate (a9);
	\path(0.75,1.25) coordinate (a10);
	\path(-1,0) coordinate (a11);
	\path(0,0) coordinate (a12);
	%%%%%%%
	\draw [fermionbar] (a1)--(a2);
	\draw [fermionbar] (a1)--(a3);
	\draw [fermionbar] (a4)--(a5);
	\draw [fermionbar] (a4)--(a6);
	\draw [fermionbar] (a7)--(a8);
	\draw [fermionbar] (a9)--(a10);
	\draw [fermion] (a11)--(a12);
	\draw [fermion] (a9)--(a12);
	\draw [fermion] (a11)--(a7);
	\draw [scalarbar] (a11)--(a4);
	\draw [scalarbar] (a12)--(a7);
	\draw [scalar] (a1)--(a9);
	%%%%
	\node[left] at (-0.5,1.8) {$q_{5}$};
	\node[right] at (0.8,1.2) {$q_{6}$};
	\node[above] at (-2.5,1) {$q_{1}$};
	\node[below] at (-2.5,-1) {$q_{2}$};
	\node[above] at (2.2,1) {$q_{3}$};
	\node[below] at (2.2,-1) {$q_{4}$};
	\node[below] at (-1.5,0) {$I_{1}$};
	\node[below] at (1.35,0) {$I_{2}$};
	\node[below] at (0.55,0) {$\psi_{1}$};
	\node[below] at (-0.4,0) {$\psi_{3}$};
	\node[left] at (-0.75,0.55) {$\psi_{2}$};
	\node at (0.18,0.5) {$I_{3}$};
	\end{tikzpicture}&
	\begin{tikzpicture}[line width=0.8pt, scale=0.7,>=Stealth]
	%%%%
	\path (0,0) coordinate (a0);
	\path (1.5,0) coordinate (a1);
	\path(2.25,1) coordinate (a2);
	\path(2.25,-1) coordinate (a3);
	\path (-1.75,0) coordinate (a4);
	\path(-2.5,-1) coordinate (a5);
	\path (-2.5,1) coordinate (a6);
	\path(-0.5,1) coordinate (a7);
	\path(-0.5,2) coordinate (a8);
	\path(0.75,0) coordinate (a9);
	\path(0.75,1.25) coordinate (a10);
	\path(-1,0) coordinate (a11);
	\path(0,0) coordinate (a12);
	%%%%%%%
	\draw [fermionbar] (a1)--(a2);
	\draw [fermionbar] (a1)--(a3);
	\draw [fermionbar] (a4)--(a5);
	\draw [fermionbar] (a4)--(a6);
	\draw [fermionbar] (a7)--(a8);
	\draw [fermionbar] (a9)--(a10);
	\draw [fermion] (a9)--(a12);
	\draw [fermionbar] (a12)--(a7);
	\draw [scalar] (a11)--(a12);
	\draw [scalar] (a11)--(a7);
	\draw [scalarbar] (a11)--(a4);
	\draw [scalar] (a1)--(a9);
	%%%%
	\node[left] at (-0.5,1.8) {$q_{5}$};
	\node[right] at (0.8,1.2) {$q_{6}$};
	\node[above] at (-2.5,1) {$q_{1}$};
	\node[below] at (-2.5,-1) {$q_{2}$};
	\node[above] at (2.2,1) {$q_{3}$};
	\node[below] at (2.2,-1) {$q_{4}$};
	\node[below] at (-1.5,0) {$I_{1}$};
	\node[below] at (1.35,0) {$I_{2}$};
	\node[below] at (0.55,0) {$\psi_{1}$};
	\node[below] at (-0.4,0) {$I_{4}$};
	\node[left] at (-0.75,0.55) {$I_{3}$};
	\node at (0.18,0.5) {$\psi_{2}$};	
	\end{tikzpicture}\\
	T-L-9&T-L-10&T-L-11&T-L-12\\
	\begin{tikzpicture}[line width=0.8pt, scale=0.65,>=Stealth]
	%%%%
	\path (0,0) coordinate (a0);
	\path (1,1) coordinate (a1);
	\path(2,1.5) coordinate (a2);
	\path(2,0.5) coordinate (a3);
	\path (2,-0.5) coordinate (a6);
	\path(2,-1.5) coordinate (a5);
	\path (1,-1) coordinate (a4);
	\path(-1,1) coordinate (a7);
	\path(-2,1) coordinate (a8);
	\path(-1,-1) coordinate (a9);
	\path(-2,-1) coordinate (a10);
	\path(0,0) coordinate (a11);
	\path(0.75,0) coordinate (a12);
	%%%%%%%
	\draw [fermionbar] (a1)--(a2);
	\draw [fermionbar] (a1)--(a3);
	\draw [fermionbar] (a4)--(a5);
	\draw [fermionbar] (a4)--(a6);
	\draw [fermionbar] (a7)--(a8);
	\draw [fermionbar] (a9)--(a10);
	\draw [fermion] (a11)--(a7);
	\draw [fermionbar] (a9)--(a11);
	\draw [scalarbar] (a11)--(a12);
	\draw [scalarbar] (a12)--(a1);
	\draw [scalarbar] (a12)--(a4);
	\draw [scalar] (a7)--(a9);
	%%%%
	\node[above] at (-2,1) {$q_{5}$};
	\node[below] at (-2,-1) {$q_{6}$};
	\node[right] at (2,1.5) {$q_{1}$};
	\node[right] at (2,0.5) {$q_{2}$};
	\node[right] at (2,-0.5) {$q_{3}$};
	\node[right] at (2,-1.5) {$q_{4}$};
	\node[left] at (-1,0) {$I_{4}$};
	\node[above] at (-0.4,0.5) {$\psi_{1}$};
	\node[below] at (-0.4,-0.5) {$\psi_{2}$};
	\node[below] at (0.3,0) {$I_{3}$};
	\node at (1.25,0.4) {$I_{1}$};
	\node at (1.25,-0.4) {$I_{2}$};
	\end{tikzpicture}&
	\begin{tikzpicture}[line width=0.8pt, scale=0.65,>=Stealth]
	%%%%
	\path (0,0) coordinate (a0);
	\path (1,1) coordinate (a1);
	\path(2,1.5) coordinate (a2);
	\path(2,0.5) coordinate (a3);
	\path (2,-0.5) coordinate (a6);
	\path(2,-1.5) coordinate (a5);
	\path (1,-1) coordinate (a4);
	\path(-1,1) coordinate (a7);
	\path(-2,1) coordinate (a8);
	\path(-1,-1) coordinate (a9);
	\path(-2,-1) coordinate (a10);
	\path(0,0) coordinate (a11);
	\path(0.75,0) coordinate (a12);
	%%%%%%%
	\draw [fermionbar] (a1)--(a2);
	\draw [fermionbar] (a1)--(a3);
	\draw [fermionbar] (a4)--(a5);
	\draw [fermionbar] (a4)--(a6);
	\draw [fermionbar] (a7)--(a8);
	\draw [fermionbar] (a9)--(a10);
	\draw [fermionbar] (a9)--(a7);
	\draw [scalarbar] (a11)--(a12);
	\draw [scalarbar] (a12)--(a1);
	\draw [scalarbar] (a9)--(a11);
	\draw [scalarbar] (a12)--(a4);
	\draw [scalarbar] (a7)--(a11);
	%%%%
	\node[above] at (-2,1) {$q_{5}$};
	\node[below] at (-2,-1) {$q_{6}$};
	\node[right] at (2,1.5) {$q_{1}$};
	\node[right] at (2,0.5) {$q_{2}$};
	\node[right] at (2,-0.5) {$q_{3}$};
	\node[right] at (2,-1.5) {$q_{4}$};
	\node[left] at (-1,0) {$\psi$};
	\node[above] at (-0.4,0.5) {$I_{4}$};
	\node[below] at (-0.4,-0.5) {$I_{5}$};
	\node[below] at (0.3,0) {$I_{3}$};
	\node at (1.25,0.4) {$I_{1}$};
	\node at (1.25,-0.4) {$I_{2}$};
	\end{tikzpicture}&
	\begin{tikzpicture}[line width=0.8pt, scale=0.65,>=Stealth]
	%%%%
	\path (0,0) coordinate (a0);
	\path (1.5,0) coordinate (a1);
	\path(2.25,1) coordinate (a2);
	\path(2.25,-1) coordinate (a3);
	\path (-0.25,0) coordinate (a4);
	\path(-0.25,1.5) coordinate (a5);
	\path (0.5,1.5) coordinate (a6);
	\path(-2,1) coordinate (a7);
	\path(-3,1) coordinate (a8);
	\path(-2,-1) coordinate (a9);
	\path(-3,-1) coordinate (a10);
	\path(0.5,0) coordinate (a11);
	\path(-1,0) coordinate (a12);
	\path(-1,-1.5) coordinate (a13);
	%%%%%%%
	\draw [fermionbar] (a1)--(a2);
	\draw [fermionbar] (a1)--(a3);
	\draw [fermionbar] (a4)--(a5);
	\draw [fermionbar] (a11)--(a6);
	\draw [fermionbar] (a7)--(a8);
	\draw [fermionbar] (a9)--(a10);
	\draw [fermion] (a12)--(a7);
	\draw [fermionbar] (a9)--(a12);
	\draw [fermionbar] (a4)--(a11);
	\draw [scalarbar] (a9)--(a7);
	\draw [scalarbar] (a12)--(a4);
	\draw [scalar] (a1)--(a11);
	%%%%
	\node[left] at (-0.25,1.5) {$q_{5}$};
	\node[right] at (0.5,1.5) {$q_{6}$};
	\node[above] at (2.2,1) {$q_{1}$};
	\node[below] at (2.2,-1) {$q_{2}$};
	\node[above] at (-3,1) {$q_{3}$};
	\node[below] at (-3,-1) {$q_{4}$};
	\node[left] at (-2,0) {$I_{3}$};
	\node at (-1.2,0.8) {$\psi_{2}$};
	\node at (-1.2,-0.8) {$\psi_{3}$};
	\node[below] at (0.2,0) {$\psi_{1}$};
	\node[below] at (1,0) {$I_{1}$};
	\node[below] at (-0.6,0) {$I_{2}$};
	\end{tikzpicture}&
	\begin{tikzpicture}[line width=0.8pt, scale=0.65,>=Stealth]
	%%%%
	\path (0,0) coordinate (a0);
	\path (1.5,0) coordinate (a1);
	\path(2.25,1) coordinate (a2);
	\path(2.25,-1) coordinate (a3);
	\path (-0.25,0) coordinate (a4);
	\path(-0.25,1.5) coordinate (a5);
	\path (0.5,1.5) coordinate (a6);
	\path(-2,1) coordinate (a7);
	\path(-3,1) coordinate (a8);
	\path(-2,-1) coordinate (a9);
	\path(-3,-1) coordinate (a10);
	\path(0.5,0) coordinate (a11);
	\path(-1,0) coordinate (a12);
	\path(-1,-1.5) coordinate (a13);
	%%%%%%%
	\draw [fermionbar] (a1)--(a2);
	\draw [fermionbar] (a1)--(a3);
	\draw [fermionbar] (a4)--(a5);
	\draw [fermionbar] (a11)--(a6);
	\draw [fermionbar] (a7)--(a8);
	\draw [fermionbar] (a9)--(a10);
	\draw [fermionbar] (a9)--(a7);
	\draw [fermionbar] (a4)--(a11);
	\draw [scalarbar] (a9)--(a12);
	\draw [scalarbar] (a7)--(a12);
	\draw [scalarbar] (a12)--(a4);
	\draw [scalar] (a1)--(a11);
	%%%%
	\node[left] at (-0.25,1.5) {$q_{5}$};
	\node[right] at (0.5,1.5) {$q_{6}$};
	\node[above] at (2.2,1) {$q_{1}$};
	\node[below] at (2.2,-1) {$q_{2}$};
	\node[above] at (-3,1) {$q_{3}$};
	\node[below] at (-3,-1) {$q_{4}$};
	\node[left] at (-2,0) {$\psi_{2}$};
	\node at (-1.2,0.8) {$I_{3}$};
	\node at (-1.2,-0.8) {$I_{4}$};
	\node[below] at (0.2,0) {$\psi_{1}$};
	\node[below] at (1,0) {$I_{1}$};
	\node[below] at (-0.6,0) {$I_{2}$};	
	\end{tikzpicture}\\
	T-L-13&T-L-14&T-L-15&T-L-16\\
\end{tabular}
\caption{The non-genuine diagrams of neutron-antineutron oscillation operator at one-loop level}\label{T-L-NG}
\end{figure}
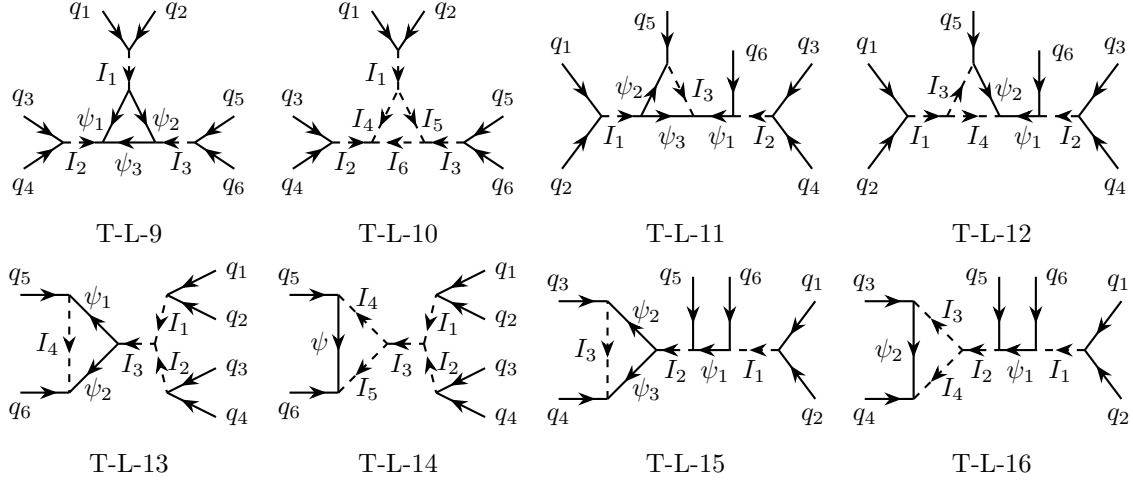

To list all possible decompositions, we need to generate the topologies and diagrams first. There are two topologies at tree level, and the internal particles could be three scalars, or two scalars and one fermion, as shown in Fig.~\ref{tree-level}. We denote the internal scalars as $I_{i}~(i=1,2,3)$ and the internal fermion as $\psi$.
The topologies have many more varieties at one-loop level. Excluding topologies with self-energy and tadpole, the number of vertices in the loop can be 3, 4, 5, or 6. The internal scalars and fermions are denoted as $I_{i}~(i=1,2,...)$ and $\psi_{i}~(i=1,2,...)$. The topologies are genuine or non-genuine, where the internal loop in non-genuine topologies can be compressed into a renormalizable vertex~\cite{Cepedello:2018rfh}. It is necessary to note that the arrows of the internal fermions are arbitrary, and the fermion mass insertions are implicit. Whether the fermion masses insert depends on the chiral operators that the external chiral quarks give to the vertices. We divide the genuine topologies into three classes, class 1 (2+2+1+1): T-L-1, T-L-2, T-L-3, T-L-4; class 2 (2+1+1+1+1): T-L-5, T-L-6, T-L-7; class 3 (1+1+1+1+1+1): T-L-8. The class 1 contains the topologies with three vertices and four vertices in the loop, which have two independent external legs that do not couple directly with other external legs, so we denote class 1 as 2+2+1+1. The class 2 and the class 3 have four and six independent external legs, respectively. We show the corresponding diagrams in Fig.~\ref{T-L-G}.
The classification of the topologies is convenient for us to assign the external legs with the chiral quarks and list possible quantum numbers of $SU(2)$ and $U(1)_{Y}$ hypercharge. The more independent legs, the more possible decompositions. Here we still show the non-genuine topologies and list the decomposition in the following subsection. Similarly, we divide the non-genuine topologies into class $1^{\prime}$~(2+2+2): T-L-9, T-L-10, class $2^{\prime}$~(1+1+2+2): T-L-11, T-L-12, T-L-13, T-L-14, and class $3^{\prime}$~(1+1+1+1+2): T-L-15, T-L-16, as shown in Fig.~\ref{T-L-NG}.

\subsection{Assignment of chiral quarks and decompositions of the operators}\label{IIB}
The next step is to assign the chiral quarks to the external legs. The total hypercharge of the external legs is zero, leading to three possible cases if one considers the chiral quarks
\begin{align}
{\text{(i)}}~4d_{R}\,,~2u_{R}\,, \qquad
{\text{(ii)}}~2d_{R}\,,~2d_{L}\,,~2u_{L}\,,\qquad
{\text{(iii)}}~3d_{R}\,,~1d_{L}\,,~1u_{R}\,,~1u_{L}\,.
\end{align}
The corresponding processes are
\begin{align}
{\text{(i)}}~&d_{R}d_{R}u_{R}\leftrightarrow \overline{d_{R}}\,\overline{d_{R}}\overline{u_{R}}\,,\\
{\text{(ii)}}~&d_{R}d_{L}u_{L}\leftrightarrow \overline{d_{R}}\,\overline{d_{L}}\overline{u_{L}}\,,\quad d_{L}d_{L}u_{L}\leftrightarrow \overline{d_{R}}\,\overline{d_{R}}\overline{u_{L}}\,,\\
{\text{(iii)}}~&d_{L}d_{R}u_{L}\leftrightarrow \overline{d_{R}}\,\overline{d_{R}}\overline{u_{R}}\,,\quad d_{R}d_{L}u_{R}\leftrightarrow \overline{d_{R}}\,\overline{d_{R}}\overline{u_{L}}\,.
\end{align}
Based on the acknowledgment that the quarks should be both right-handed or both left-handed for the scalar quark current $(\overline{q^{c}}q^{(\prime)})$, one can find out the combination of the external legs in different classes.
For example, case (iii) can be combined as
\begin{align}
{\text{T-T-1~and~class~}}1^{\prime}:~&(d_{R}d_{R})(d_{R}u_{R})(d_{L}u_{L})\,,\\
{\text{T-T-2,~class~1,~and~class~}}2^{\prime}:~&(d_{R}d_{R})(d_{R}u_{R})(d_{L})(u_{L})\,,\quad(d_{R}d_{R})(d_{R})(u_{R})(d_{L}u_{L})\,,\notag\\
                                                                                      &(d_{R})(d_{R})(d_{R}u_{R})(d_{L}u_{L})\,,\\
{\text{class~2~and~class~}}3^{\prime}:~&(d_{R})(d_{R})(d_{R})(u_{R})(d_{L}u_{L})\,,\quad(d_{R})(d_{R})(d_{R}u_{R})(d_{L})(u_{L})\,,\notag\\
                                                                        &(d_{R}d_{R})(d_{R})(u_{R})(d_{L})(u_{L})\,,\\
{\text{class}}~3:~&(d_{R})(d_{R})(d_{R})(u_{R})(d_{L})(u_{L})\,.
\end{align}
We label the quark current as $(qq^{(\prime)})$ and the independent external quark as $(q)$. After the assignment, the decompositions can be listed with the interactions required as $SU(3)_{C}\times SU(2)_{L}\times U(1)_{Y}$ gauge invariant. For the $SU(3)_{C}$ symmetry, we obey the condition that $3\otimes3=\overline{3}\oplus 6$, $3\otimes\overline{3}=1\oplus 8$, $3\otimes6\supset8$, $3\otimes\overline{6}\supset\overline{3}$, $3\otimes 8\supset3\oplus\overline{6}$, $6\otimes6\supset\overline{6}$, $6\otimes\overline{6}\supset1\oplus8$, $6\otimes8\supset\overline{3}\oplus6$, $8\otimes8\supset1\oplus8\oplus8$. Here we do not consider the $SU(3)$ representation higher than 8. The $SU(2)_{L}$ gauge symmetry gives the rules that $2\otimes2=1\oplus3$, $2\otimes3\supset2$, $3\otimes3\supset1\oplus3$. The hypercharge $Y$ is conserved at every vertex.  

The hypercharge of the internal particles in the diagrams at tree level can be easily determined after the assignment of external legs. However, one should be careful when constructing the UV model as some vertices are vanishing. In Appendix~\ref{appendixA}, we list the vertices and determine whether they can contribute to neutron-antineutron oscillation. The decomposition at one-loop level is relatively more complicated than tree level. The internal lines can not be determined directly for the $U(1)$ gauge symmetry until a parameter $\alpha$ is introduced. One can just set the hypercharge of one internal particle as $\alpha$, and the other hypercharge can be fixed when one gives an exact value to $\alpha$. We detail the decompositions in Appendix~\ref{appendixB}.

\section{An example model of neutron-antineutron oscillation at one-loop level}\label{example}
\begin{table}[b]
\begin{tabular}{||c|c||c|c||}
\hline
SM particles&Quantum Number&New Particles&Quantum Number\\
\hline
$\Phi=(\phi^{+},\phi^{0})^{T}$&$(1,2,1/2,+1)$&$S_{1}$&$(3,1,-1/3,-1)$\\
$Q_{L}=(U_{L},D_{L})^{T}$&$(3,2,1/6,+1)$&$S_{2}=(S_{2}^{+2/3},S_{2}^{-1/3})^{T}$&$(3,2,1/6,-1)$\\
$L_{L}=(\nu_{L},E_{L})^{T}$&$(1,2,-1/2,+1)$&$\eta=(\eta^{+},\eta^{0})^{T}$&$(1,2,1/2,-1)$\\
$U_{R}$&$(3,1,+2/3,+1)$&$N_{i}$&$(1,1,0,-1)$\\
$D_{R}$&$(3,1,-1/3,+1)$&$\psi=(\psi^{+2/3},\psi^{-1/3})^{T}$&$(3,2,1/6,-1)$\\
$E_{R}$&$(1,1,-1,+1)$&&\\
\hline
\end{tabular}
\caption{The corresponding quantum numbers of the SM particles and new particles under the symmetry $SU(3)_{C}\times SU(2)_{L}\times U(1)_{Y}\times \mathbb{Z}_{2}$.}\label{model_QNs}
\end{table}

In this section, we will give an example of realizing the tiny neutrino mass and $n-\bar{n}$ oscillation simultaneously via a one-loop model. We focus on the Scotogenic-type neutrino mass model, which has been widely discussed, see for instance~\cite{Ma:2006km,Farzan:2009ji,Liao:2009fm,Cai:2011qr,Ma:2012ez,Hehn:2012kz,Ferreira:2016sbb,Nomura:2019lnr,Cacciapaglia:2020psm,Beniwal:2020hjc,Rosenlyst:2021tdr,Dcruz:2022dao,Arora:2022hza}. In addition, we consider the one-loop neutron-antineutron diagram T-L-8. The original Scotogenic model~\cite{Ma:2006km} introduces right-handed neutrinos $N_{i},(i=1,2,3)$ and an $SU(2)_{L}$ doublet scalar $\eta$ with an extra $\mathbb{Z}_{2}$ symmetry. The internal particle of T-L-8 $\psi_{3}$ is considered to be $N_{i}$, while $I_{1}$ is to be $\eta$. The color numbers of the internal particles can be easily found. For $SU(2)_{L}$ quantum numbers, one can choose the numbers of the other internal particles after fixing $I_{1}$ at 2 and $\psi_{3}$ at 1 from Table \ref{T-L-8_SU32} and Table \ref{T-L-8_SU2}. Here we select the row in blue of the third column in Table \ref{T-L-8_SU2}, and the external quarks $q_{1,2,3,6}$ can be determined as four singlets and $q_{4,5}$ as two doublets which correspond to the case (iii) we have shown in Sec. \ref{IIB}. Then one can find out the rows in Table \ref{T-L-8_U1-1} and \ref{T-L-8_U1-2}, which are in accordance with the chirality of the external quarks, and the hypercharges of the particles can be determined.

In this toy model, part of the additional particles is introduced as: three right-handed neutrinos $N_{i},(i=1,2,3)$, an $SU(2)_{L}$ doublet fermion $\psi$, a singlet scalar $S_{1}$ and two doublet scalars $S_{2},\eta$, with the quantum numbers of these particles under $SU(3)_{C}\times SU(2)_{L}\times U(1)_{Y}\times \mathbb{Z}_{2}$ summarized in Table~\ref{model_QNs}. 
Here we introduce $\mathbb{Z}_{2}$ symmetry to forbid the tree-level contribution to neutrino mass. We set all the Standard Model (SM) particles as $+1$ and new particles as $-1$ under the $\mathbb{Z}_{2}$ symmetry.
The Yukawa interactions of these particles in the fermion weak eigenstates can be written as
\begin{align}
-\mathcal{L}_{Y}\supset~&y_{N}^{ij}\overline{(L_{L}^{i})^{c}}i\sigma_{2}\eta N_{j}^{c}+y_{\psi}^{i}\overline{(L_{L}^{i\alpha})^{c}}i\sigma_{2}\psi^{\alpha}S_{1}^{*\bar{\alpha}}+y_{1}^{i}\overline{\psi^{\alpha}}i\sigma_{2}\eta^{*}U_{R}^{i\alpha}\notag\\
&+y_{2}^{ij}\overline{(D_{R}^{i\alpha})^{c}}N_{j}S_{1}^{*\bar{\alpha}}+y_{3}^{i}\overline{(Q_{L}^{i\alpha})^{c}}i\sigma_{2}\psi^{\beta}S_{1}^{\gamma}\epsilon^{\alpha\beta\gamma}+y_{4}^{ij}\overline{(Q_{L}^{i\alpha})^{c}}S_{2}^{*\bar{\alpha}}N_{j}^{c}\notag\\
&+y_{5}^{i}\overline{(\psi^{\alpha})^{c}}i\sigma_{2}S_{2}^{\beta}D_{R}^{i\gamma}\epsilon^{\alpha\beta\gamma}+y_{6}^{i}\overline{\psi^{\alpha}}\eta D_{R}^{i\alpha}+{\text{h.c.}}\,.
\end{align}
We suppose the right-handed neutrino $N_{1,2}$ couple with leptons, $N_{3}$ couples with quarks, and $y_{\psi}=0$ to avoid proton decay. Hence the coefficient $y_{N}^{ij}$ is a $3\times2$ matrix, and $y_{2}^{ij},y_{4}^{ij}$ are $3\times1$ matrices. We show the corresponding Feynman diagrams of proton decay in Fig.~\ref{proton_decay}, where the red vertex is zero in our assumption. 
\begin{figure}[t]
    \begin{tikzpicture}[line width=1pt, scale=0.85,>=Stealth]
	%%%%
	\path (0,0) coordinate (a0);
	\path (2,1) coordinate (a1);
	\path (-2,-1) coordinate (a4);
	\path(-1,1) coordinate (a7);
	\path(-2,1) coordinate (a8);
	\path(1,-1) coordinate (a9);
	\path(2,-1) coordinate (a10);
	\path(-1,-1) coordinate (a11);
	\path(1,1) coordinate (a12);
	%%%%%%%
	\draw [fermionbar] (a7)--(a8);
	\draw [fermionbar] (a9)--(a10);
	\draw [scalarbar] (a7)--(a12);
	\draw [fermionbar] (a9)--(a12);
	\draw [fermion] (a11)--(a7);
	\draw [scalar] (a11)--(a9);
	\draw [fermionbar] (a12)--(a1);
	\draw [fermionbar] (a11)--(a4);
        \draw [fermion] (-2,1.8)--(2,1.8);
	%%%%
	\node[right] at (a10) {$e_{L}$};
	\node[left] at (a8) {$u_{L}$};
	\node[left] at (-2,-1) {$u_{R}$};
	\node[right] at (2,1) {$d_{R}$};
	\node at (-0.65,0) {$\psi$};
	\node at (0,-0.65) {$\eta$};
	\node at (0,0.65) {$S_{1}$};
	\node at (0.6,0) {$N_{3}$};	
        \node at (a9) {\textcolor{red}{$\bullet$}};
        \node[left] at (-2,1.8) {$d$};
        \node[right] at (2,1.8) {$d$};
        \end{tikzpicture}\qquad
        \begin{tikzpicture}[line width=1pt, scale=0.85,>=Stealth]
	%%%%
	\path (0,0) coordinate (a0);
	\path (2,1) coordinate (a1);
	\path (-2,-1) coordinate (a4);
	\path(-1,1) coordinate (a7);
	\path(-2,1) coordinate (a8);
	\path(1,-1) coordinate (a9);
	\path(2,-1) coordinate (a10);
	\path(-1,-1) coordinate (a11);
	\path(1,1) coordinate (a12);
	%%%%%%%
	\draw [fermionbar] (a7)--(a8);
	\draw [fermionbar] (a9)--(a10);
	\draw [scalar] (a7)--(a12);
	\draw [fermionbar] (a9)--(a12);
	\draw [fermionbar] (a11)--(a7);
	\draw [scalarbar] (a11)--(a9);
	\draw [fermionbar] (a12)--(a1);
	\draw [fermionbar] (a11)--(a4);
        \draw [fermion] (-2,1.8)--(2,1.8);
	%%%%
	\node[right] at (a10) {$e_{L}$};
	\node[left] at (a8) {$u_{R}$};
	\node[left] at (-2,-1) {$u_{L}$};
	\node[right] at (2,1) {$d_{R}$};
	\node at (-0.65,0) {$\psi$};
	\node at (0,-0.6) {$S_{1}$};
	\node at (0,0.65) {$\eta$};
	\node at (0.65,0) {$\psi$};	
        \node at (a9) {\textcolor{red}{$\bullet$}};
        \node[left] at (-2,1.8) {$d$};
        \node[right] at (2,1.8) {$d$};
\end{tikzpicture}
\caption{The Feynman diagrams of proton decay. Our model assumes the red vertex to be zero to forbid proton decay.}\label{proton_decay}
\end{figure}
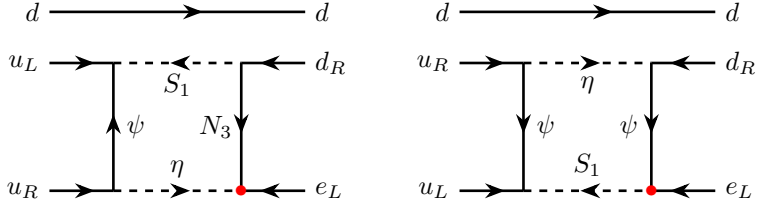
\begin{figure}[t]
\begin{tikzpicture}[line width=1pt, scale=0.6,>=Stealth]
	%%%%
	\draw [fermion] (3.5,0)--(2,0);
	\draw [fermion] (2,0)--(0,0);
	\draw [fermion] (-2,0)--(0,0);
	\draw [fermion] (-3.5,0)--(-2,0);
	\draw [scalar] (0,2) arc(90:0:2);
	\draw [scalar] (0,2) arc(90:180:2);
	\draw [scalarbar] (0,2)--(-1.5,3.5);
	\draw [scalarbar] (0,2)--(1.5,3.5);
	%%%%
	\node[below] at (3.5,0) {$\nu_{L}$};
	\node[below] at (-3.5,0) {$\nu_{L}$};
	\node at (0,0) {$\times$};
	\node[below] at (0,0) {$N_{1,2}$};
	\node at (-2,1.5) {$\eta_{0}$};
	\node at (2,1.5) {$\eta_{0}$};
	\node[left] at (-1.5,3.5) {$\Phi_{0}$};
	\node[right] at (1.5,3.5) {$\Phi_{0}$};
\end{tikzpicture}\quad
\begin{tikzpicture}[line width=1pt, scale=0.8,>=Stealth]
	%%%%
	\draw [fermionbar] (1,0)--(2,0);
	\draw [fermionbar] (0.5,1)--(1,2);
	\draw [fermionbar] (-0.5,1)--(-1,2);
	\draw [fermionbar] (-1,0)--(-2,0);
	\draw [fermionbar] (-0.5,-1)--(-1,-2);
	\draw [fermionbar] (0.5,-1)--(1,-2);
	\draw [fermion] (-1,0)--(-0.5,1);
	\draw [fermionbar] (1,0)--(0.5,1);
	\draw [fermionbar] (0.5,-1)--(-0.5,-1);
	\draw [scalar] (-0.5,1)--(0.5,1);
	\draw [scalarbar] (-0.5,-1)--(-1,0);
	\draw [scalar] (0.5,-1)--(1,0);
	%%%%
	\node[above] at (0,1) {$S_{2}^{-2/3}$};
	\node[left] at (-0.75,-0.75) {$\eta^{+}$};
	\node[right] at (0.75,-0.75) {$S_{1}^{+1/3}$};
	\node[below] at (0,-1) {$\psi^{+2/3}$}; 
	\node[right] at (0.8,0.6) {$N_{3}$};
	\node[left] at (-0.7,0.6) {$\psi^{-1/3}$};
	\node[right] at (2,0) {$d_{R}$};
	\node[right] at (1,2) {$u_{L}$};
	\node[right] at (1,-2) {$d_{L}$};
	\node[left] at (-1,-2) {$d_{R}$};
	\node[left] at (-2,0) {$u_{R}$};
	\node[left] at (-1,2) {$d_{R}$};
\end{tikzpicture}\quad
\begin{tikzpicture}[line width=1pt, scale=0.8,>=Stealth]
	%%%%
	\draw [fermionbar] (1,0)--(2,0);
	\draw [fermionbar] (0.5,1)--(1,2);
	\draw [fermionbar] (-0.5,1)--(-1,2);
	\draw [fermionbar] (-1,0)--(-2,0);
	\draw [fermionbar] (-0.5,-1)--(-1,-2);
	\draw [fermionbar] (0.5,-1)--(1,-2);
	\draw [fermion] (-1,0)--(-0.5,1);
	\draw [fermionbar] (1,0)--(0.5,1);
	\draw [fermionbar] (0.5,-1)--(-0.5,-1);
	\draw [scalar] (-0.5,1)--(0.5,1);
	\draw [scalarbar] (-0.5,-1)--(-1,0);
	\draw [scalar] (0.5,-1)--(1,0);
	%%%%
	\node[above] at (0,1) {$S_{2}^{+1/3}$};
	\node[left] at (-0.75,-0.75) {$\eta^{0}$};
	\node[right] at (0.75,-0.75) {$S_{1}^{+1/3}$};
	\node[below] at (0,-1) {$\psi^{-1/3}$}; 
	\node[right] at (0.8,0.6) {$N_{3}$};
	\node[left] at (-0.7,0.6) {$\psi^{+2/3}$};
	\node[right] at (2,0) {$d_{R}$};
	\node[right] at (1,2) {$d_{L}$};
	\node[right] at (1,-2) {$u_{L}$};
	\node[left] at (-1,-2) {$d_{R}$};
	\node[left] at (-2,0) {$u_{R}$};
	\node[left] at (-1,2) {$d_{R}$};
\end{tikzpicture}
\caption{The Feynman diagrams of neutrino mass (left) and neutron-antineutron oscillation (middle and right) in the example model.}\label{mnu_nnbar}
\end{figure}
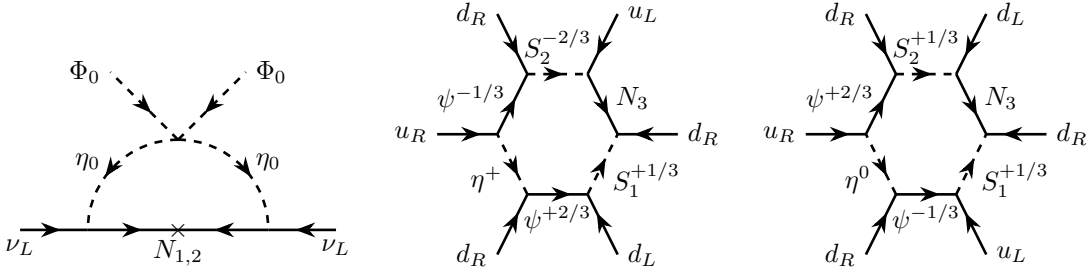

\paragraph{Neutrino mass and flavor anomalies}
The tiny neutrino mass can be induced via the scotogenic model in this toy model, as shown in Fig.~\ref{mnu_nnbar}. The neutral component of the scalar doublet $\eta$ is $\eta^{0}=\eta^{0}_{R}+i\eta^{0}_{I}$, and we suppose that $m_{0}^{2}=(m^{2}_{\eta^{0}_{R}}+m^{2}_{\eta^{0}_{I}})/2\simeq m_{N}^{2}$, where $m_{N}$ denotes the mass of right-handed neutrinos $N_{1,2}$. The neutrino mass can be estimated as
\begin{align}
m_{\nu}^{ij}=\dfrac{\lambda_{5}v^{2}}{16\pi^{2}}\sum\limits_{k=1,2}\dfrac{y_{N}^{ik}y_{N}^{jk}}{m_{N}}\,,
\end{align}
where $v=174~{\text{GeV}}$, and $\lambda_{5}$ comes from the scalar potential term $\frac{\lambda_{5}}{2}(\Phi^{\dagger}\eta)^{2}$. If one sets $m_{N}=2~{\text{TeV}}$, $y_{N}\sim0.05$, and $\lambda_{5}\sim10^{-7}$, the neutrino mass can be obtained as $m_{\nu}\sim0.05~{\text{eV}}$. This model can also contribute to $b\to s\ell\ell$ process via penguin diagrams that relate to the couplings $y_{2}^{23},y_{2}^{33}$ which do not contribute to $n-\bar{n}$ oscillation.

\begin{figure}[tb]
\includegraphics[height=8cm]{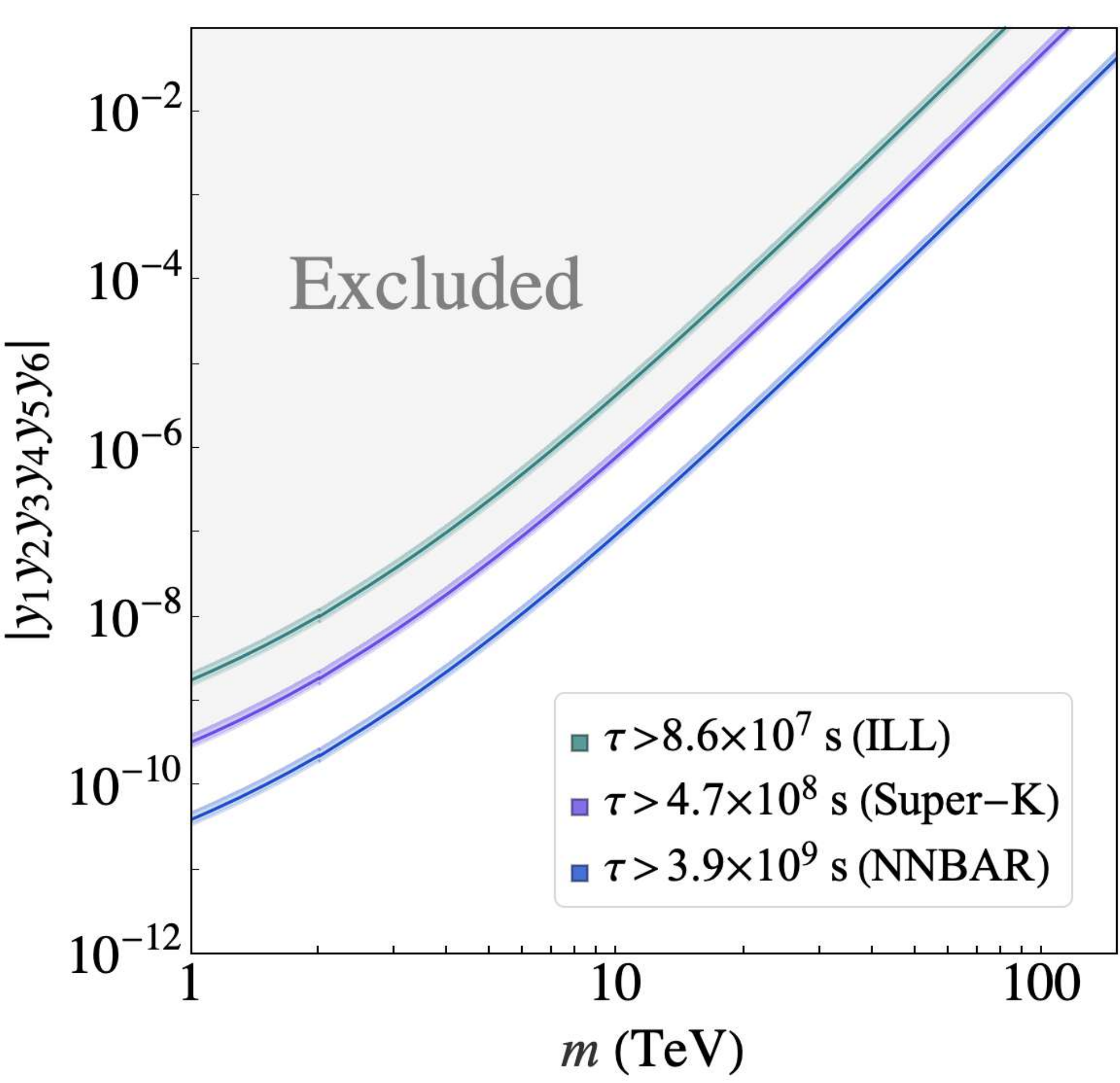}
\caption{The correlation between the coefficient $|y_{1}^{1}y_{2}^{13}y_{3}^{1}y_{4}^{\prime13}y_{5}^{1}y_{6}^{1}|\equiv|y_{1}y_{2}y_{3}y_{4}y_{5}y_{6}|$ and the mass $m\equiv m_{\psi}=m_{S_{i}}$. The narrow bands show the errors from the nuclear matrix elements. The Super-K experiment excludes the parameter region above the lines (the region in gray). We have fixed that $m_{N}=2~{\text{TeV}}$.}\label{nnbar_plot}
\end{figure}
\paragraph{Neutron-antineutron oscillation}
In this toy model, the $n-\bar{n}$ oscillation can happen at one-loop level. The corresponding effective operator is $\mathcal{O}_{2}=-4\mathcal{O}^{3}_{RLR}$ and the coefficient $\mathcal{C}_{2}=-\frac{1}{4}\mathcal{C}_{RLR}^{3}$ is
\begin{align}
\mathcal{C}_{2}&=\dfrac{m_{\psi}}{64\pi^{2}m_{N_{3}}^{6}}\bigg[\mathcal{Y}^{+}\times F_{6}(a,b,c,d,f,g^{+})-\mathcal{Y}^{0}\times \bigg(F_{6}(a,b,c,d,f,g^{0}_{R})-F_{6}(a,b,c,d,f,g^{0}_{I})\bigg)\bigg]\,,
\end{align}
where $\mathcal{Y}^{+}=y_{1}^{1}y_{2}^{13}y_{3}^{1}y_{4}^{\prime13}y_{5}^{1}y_{6}^{1}$ and $\mathcal{Y}^{0}=y_{1}^{1}y_{2}^{13}y_{3}^{\prime1}y_{4}^{13}y_{5}^{1}y_{6}^{1}$ with $y_{3}^{\prime1}=(V^{*}y_{3})^{1},y_{4}^{\prime13}=(V^{*}y_{4})^{13}$, and $V$ is the Cabibbo-Kobayashi-Maskawa (CKM) matrix. The loop integral is
\begin{align}
F_{6}(a,b,c,d,f,g)=\dfrac{1}{4(i\pi^{2})}\int d^{4}k\dfrac{k^{2}}{(k^{2}-a)(k^{2}-b)(k^{2}-c)(k^{2}-d)(k^{2}-f)(k^{2}-g)}\,,
%=&\dfrac{1}{4}\bigg[\frac{a^2 \ln{a}}{(a-b) (a-c) (a-d) (a-f) (a-g)}+\frac{b^2 \ln{b}}{(b-a) (b-c) (b-d) (b-f) (b-g)}\notag\\
%&+\frac{c^2 \ln{c}}{(c-a) (c-b) (c-d) (c-f) (c-g)}+\frac{d^2 \ln{d}}{(d-a) (d-b) (d-c) (d-f) (d-g)}\notag\\
%&+\frac{f^2 \ln{f}}{(f-a) (f-b) (f-c) (f-d) (f-g)}+\frac{g^2 \ln{g}}{(g-a) (g-b) (g-c) (g-d) (g-f)}\bigg]\,,
\end{align}
and the parameters $a,b,c,d,f,g$ are defined as
\begin{align}
a\equiv\dfrac{m^{2}_{N_{3}}}{m^{2}_{N_{3}}}\,,\;
b=c\equiv\dfrac{m^{2}_{\psi}}{m^{2}_{N_{3}}}\,,\;
d=f\equiv\dfrac{m^{2}_{S_{i}}}{m^{2}_{N_{3}}}\,,\;
g^{+}\equiv\dfrac{m^{2}_{\eta^{+}}}{m^{2}_{N_{3}}}\,,\;
g^{0}_{R}\equiv\dfrac{m^{2}_{\eta^{0}_{R}}}{m^{2}_{N_{3}}}\,,\;
g^{0}_{I}\equiv\dfrac{m^{2}_{\eta^{0}_{I}}}{m^{2}_{N_{3}}}\,.
\end{align}
If we suppose that $m_{N}=2~{\text{TeV}},~m_{\eta^{+}}=m_{\eta^{0}_{R}}=2~{\text{TeV}},~m_{\eta^{0}_{I}}=m_{\eta^{0}_{R}}-2\lambda_{5}v^{2}(\lambda_{5}=10^{-7}),~m_{\psi}=m_{S_{i}}$, i.e. $a=1,~b=c=d=f$,~$g^{+}=g^{0}_{R}=1$, and $g^{0}_{I}\simeq1$, the loop integral can be simplified as
\begin{align}
F_{6}(1,b,b,b,b,1)=\frac{-b^3-9 b^2+9 b+6 (b+1) b \ln{b}+1}{12 (b-1)^5 b}\,.
\end{align}
The transition rate for the neutron-antineutron oscillation is 
\begin{align}
\tau^{-1}_{n-\bar{n}}=1.52\times10^{9}\left|\dfrac{\mathcal{M}_{2}(\mu)}{{\text{GeV}}^{6}}\dfrac{\mathcal{C}_{2}(\mu)}{{\text{TeV}}^{-5s}}\right|\text{s}^{-1}\,,
\end{align}
where $\mathcal{M}_{2}(\mu)$ is the corresponding nuclear matrix element at scale $\mu$. The evolutions of the nuclear matrix elements from the lattice-QCD scale to the relevant heavy scale have been considered. The correlation between the coefficient $|y_{1}^{1}y_{2}^{13}y_{3}^{1}y_{4}^{\prime13}y_{5}^{1}y_{6}^{1}|$ and the mass $m\equiv m_{\psi}=m_{S_{i}}$ can be derived from the current experimental limits, as shown in Fig.~\ref{nnbar_plot}. The parameter region in gray is excluded by the Super-K experiment~\cite{Super-Kamiokande:2020bov}.

%%%Conclusion%%%
\section{Conclusion}\label{sec4}
In this paper, we discuss the decomposition of the neutron-antineutron dimension nine oscillation operators at tree level and one-loop level. We follow the usual steps that generate the topologies, assign the external legs, and determine the quantum numbers of the internal particles. We divide the topologies into different classes to serve the assignment of the external legs with chiral quarks. We reserve the non-genuine topologies and detail the decompositions of those non-genuine diagrams, as they could be significant if the couplings related to tree level diagrams are suppressed.

Furthermore, we consider an example toy model that can generate the $n-\bar{n}$ oscillation at one-loop level. The example requires $\mathbb{Z}_{2}$ symmetry to realize the tiny neutrino mass via the scotogenic model, and it can also contribute to the flavor anomalies process. We present the correlation between the couplings and the mass of new physics particles by using the current experimental limit on the $n-\bar{n}$ oscillation. The future $n-\bar{n}$ experiments can further explore the parameter region of the model.

%%%Acknowledgements%%%
\noindent{\bf Acknowledgements.}
This work is supported in part by the National Natural Science Foundation of China (12175082, 11775093) and the Fundamental Research Funds for the Central Universities (CCNULJ004).

%%%Appendix%%%
\newpage
\begin{appendices}
\section{The vertices $q-q-I$}\label{appendixA}
In this section, we will determine which vertices $q-q-I$ (quark-quark-internal scalar) are vanishing. These vertices commonly appear in the diagrams that we discuss. The possible scalars and the corresponding vertices are listed in Table~\ref{DQs}, consistent with~\cite{Davies:1990sc}. 
\begin{table}[hbt]
	\renewcommand\arraystretch{0.9}
	\centering
	\begin{tabular}{|c|c|c|c||c|c|c|c|}
	\hline\hline
	\multirow{2}*{Scalar}&\multirow{2}*{Vertices}&Weak&Contribute&\multirow{2}*{Scalar}&\multirow{2}*{Vertices}&Weak&Contribute\\
	~&~& eigenstate&to $n-\bar{n}$ &~&~& eigenstate& $n-\bar{n}$ \\
	\hline &
	\multirow{3}*{\begin{tikzpicture}[line width=0.8pt, scale=0.6,>=Stealth]
	\path (0:0) coordinate (a0);
	\path (0:1) coordinate (a1);
	\path (120:1) coordinate (a2);
	\path (240:1) coordinate (a3);
	\draw [fermion](a3)--(a0);
	\draw [fermion](a2)--(a0);
	\draw [scalar](a0)--(a1);
	\node[right] at (0:1) {$I$};
	\node[left] at (120:1) {$u_{L}$};
	\node[left] at (240:1) {$d_{L}$};
	\end{tikzpicture}}
	&&\multirow{3}*{\begin{tabular}{c}Yes\end{tabular}}&&
	\multirow{3}*{\begin{tikzpicture}[line width=0.8pt, scale=0.6,>=Stealth]
	\path (0:0) coordinate (a0);
	\path (0:1) coordinate (a1);
	\path (120:1) coordinate (a2);
	\path (240:1) coordinate (a3);
	\draw [fermion](a3)--(a0);
	\draw [fermion](a2)--(a0);
	\draw [scalar](a0)--(a1);
	\node[right] at (0:1) {$I$};
	\node[left] at (120:1) {$u_{L}$};
	\node[left] at (240:1) {$d_{L}$};
	\end{tikzpicture}}
	&\multirow{3}*{\begin{tabular}{c}antisymmetric\\(vanish)\end{tabular}}
	&\multirow{3}*{\begin{tabular}{c}Yes\end{tabular}}\\
	&~&symmetric&&&~&&\\
	$I\sim$&~&&&$I\sim$&~&&\\
	\cline{2-4}\cline{6-8}
	$(\overline{3},1,\frac{1}{3})$
	&\multirow{3}*{\begin{tikzpicture}[line width=0.8pt, scale=0.6,>=Stealth]
	\path (0:0) coordinate (a0);
	\path (0:1) coordinate (a1);
	\path (120:1) coordinate (a2);
	\path (240:1) coordinate (a3);
	\draw [fermion](a3)--(a0);
	\draw [fermion](a2)--(a0);
	\draw [scalar](a0)--(a1);
	\node[right] at (0:1) {$I$};
	\node[left] at (120:1) {$u_{R}$};
	\node[left] at (240:1) {$d_{R}$};
	\end{tikzpicture}}
	&&\multirow{3}*{\begin{tabular}{c}Yes\end{tabular}}&
	$(6,1,\frac{1}{3})$
	&\multirow{3}*{\begin{tikzpicture}[line width=0.8pt, scale=0.6,>=Stealth]
	\path (0:0) coordinate (a0);
	\path (0:1) coordinate (a1);
	\path (120:1) coordinate (a2);
	\path (240:1) coordinate (a3);
	\draw [fermion](a3)--(a0);
	\draw [fermion](a2)--(a0);
	\draw [scalar](a0)--(a1);
	\node[right] at (0:1) {$I$};
	\node[left] at (120:1) {$u_{R}$};
	\node[left] at (240:1) {$d_{R}$};
	\end{tikzpicture}}
	&&\multirow{3}*{\begin{tabular}{c}Yes\end{tabular}}\\
          &~&arbitrary&&&~&arbitrary&\\
          &~&&&&~&&\\
          \hline    
          %%%%%%%%%%%%%%%%%%%%%%%%%%%%%%%%%%
          \multirow{9}*{\begin{tabular}{c}$I\sim$\\$(\overline{3},3,\frac{1}{3})$\end{tabular}}&
	\multirow{3}*{\begin{tikzpicture}[line width=0.8pt, scale=0.6,>=Stealth]
	\path (0:0) coordinate (a0);
	\path (0:1) coordinate (a1);
	\path (120:1) coordinate (a2);
	\path (240:1) coordinate (a3);
	\draw [fermion](a3)--(a0);
	\draw [fermion](a2)--(a0);
	\draw [scalar](a0)--(a1);
	\node[right] at (0:1) {$I$};
	\node[left] at (120:1) {$u_{L}$};
	\node[left] at (240:1) {$d_{L}$};
	\end{tikzpicture}}
	&\multirow{3}*{\begin{tabular}{c}antisymmetric\\(vanish)\end{tabular}}
	&\multirow{3}*{\begin{tabular}{c}Yes\end{tabular}}
	&\multirow{9}*{\begin{tabular}{c}$I\sim$\\$(6,3,\frac{1}{3})$\end{tabular}}&
	\multirow{3}*{\begin{tikzpicture}[line width=0.8pt, scale=0.6,>=Stealth]
	\path (0:0) coordinate (a0);
	\path (0:1) coordinate (a1);
	\path (120:1) coordinate (a2);
	\path (240:1) coordinate (a3);
	\draw [fermion](a3)--(a0);
	\draw [fermion](a2)--(a0);
	\draw [scalar](a0)--(a1);
	\node[right] at (0:1) {$I$};
	\node[left] at (120:1) {$u_{L}$};
	\node[left] at (240:1) {$d_{L}$};
	\end{tikzpicture}}
	&
	&\multirow{3}*{\begin{tabular}{c}Yes\end{tabular}}\\
	&~&&&&~&symmetric&\\
	&~&&&&~&&\\
	\cline{2-4}\cline{6-8}
	&\multirow{3}*{\begin{tikzpicture}[line width=0.8pt, scale=0.6,>=Stealth]
	\path (0:0) coordinate (a0);
	\path (0:1) coordinate (a1);
	\path (120:1) coordinate (a2);
	\path (240:1) coordinate (a3);
	\draw [fermion](a3)--(a0);
	\draw [fermion](a2)--(a0);
	\draw [scalar](a0)--(a1);
	\node[right] at (0:1) {$I$};
	\node[left] at (120:1) {$u_{L}$};
	\node[left] at (240:1) {$u_{L}$};
	\end{tikzpicture}}
	&\multirow{3}*{\begin{tabular}{c}antisymmetric\\(vanish)\end{tabular}}
	&\multirow{3}*{\begin{tabular}{c}Yes\end{tabular}}&
	&\multirow{3}*{\begin{tikzpicture}[line width=0.8pt, scale=0.6,>=Stealth]
	\path (0:0) coordinate (a0);
	\path (0:1) coordinate (a1);
	\path (120:1) coordinate (a2);
	\path (240:1) coordinate (a3);
	\draw [fermion](a3)--(a0);
	\draw [fermion](a2)--(a0);
	\draw [scalar](a0)--(a1);
	\node[right] at (0:1) {$I$};
	\node[left] at (120:1) {$u_{L}$};
	\node[left] at (240:1) {$u_{L}$};
	\end{tikzpicture}}
	&&\multirow{3}*{\begin{tabular}{c}Yes\end{tabular}}\\
          &~&&&&~&symmetric&\\
          &~&&&&~&&\\
          \cline{2-4}\cline{6-8}
          	&\multirow{3}*{\begin{tikzpicture}[line width=0.8pt, scale=0.57,>=Stealth]
	\path (0:0) coordinate (a0);
	\path (0:1) coordinate (a1);
	\path (120:1) coordinate (a2);
	\path (240:1) coordinate (a3);
	\draw [fermion](a3)--(a0);
	\draw [fermion](a2)--(a0);
	\draw [scalar](a0)--(a1);
	\node[right] at (0:1) {$I$};
	\node[left] at (120:1) {$d_{L}$};
	\node[left] at (240:1) {$d_{L}$};
	\end{tikzpicture}}
	&\multirow{3}*{\begin{tabular}{c}antisymmetric\\(vanish)\end{tabular}}
	&&&\multirow{3}*{\begin{tikzpicture}[line width=0.8pt, scale=0.57,>=Stealth]
	\path (0:0) coordinate (a0);
	\path (0:1) coordinate (a1);
	\path (120:1) coordinate (a2);
	\path (240:1) coordinate (a3);
	\draw [fermion](a3)--(a0);
	\draw [fermion](a2)--(a0);
	\draw [scalar](a0)--(a1);
	\node[right] at (0:1) {$I$};
	\node[left] at (120:1) {$d_{L}$};
	\node[left] at (240:1) {$d_{L}$};
	\end{tikzpicture}}
	&&\multirow{3}*{\begin{tabular}{c}Yes\end{tabular}}\\
          &~&&{No}&&~&symmetric&\\
          &~&&&&~&&\\
          \hline    
	\multirow{3}*{\begin{tabular}{c}$I\sim$\\$(\overline{3},1,\frac{4}{3})$\end{tabular}}&
	\multirow{3}*{\begin{tikzpicture}[line width=0.8pt, scale=0.6,>=Stealth]
	\path (0:0) coordinate (a0);
	\path (0:1) coordinate (a1);
	\path (120:1) coordinate (a2);
	\path (240:1) coordinate (a3);
	\draw [fermion](a3)--(a0);
	\draw [fermion](a2)--(a0);
	\draw [scalar](a0)--(a1);
	\node[right] at (0:1) {$I$};
	\node[left] at (120:1) {$u_{R}$};
	\node[left] at (240:1) {$u_{R}$};
	\end{tikzpicture}}
	&\multirow{3}*{\begin{tabular}{c}antisymmetric\\(vanish)\end{tabular}}
	&\multirow{3}*{\begin{tabular}{c}{No}\end{tabular}}&
	\multirow{3}*{\begin{tabular}{c}$I\sim$\\$(6,1,\frac{4}{3})$\end{tabular}}&
	\multirow{3}*{\begin{tikzpicture}[line width=0.8pt, scale=0.6,>=Stealth]
	\path (0:0) coordinate (a0);
	\path (0:1) coordinate (a1);
	\path (120:1) coordinate (a2);
	\path (240:1) coordinate (a3);
	\draw [fermion](a3)--(a0);
	\draw [fermion](a2)--(a0);
	\draw [scalar](a0)--(a1);
	\node[right] at (0:1) {$I$};
	\node[left] at (120:1) {$u_{R}$};
	\node[left] at (240:1) {$u_{R}$};
	\end{tikzpicture}}
	&\multirow{3}*{\begin{tabular}{c}symmetric\end{tabular}}
	&\multirow{3}*{\begin{tabular}{c}Yes\end{tabular}}\\
	&&&&&&&\\
	&&&&&&&\\
	\hline
	\multirow{3}*{\begin{tabular}{c}$I\sim$\\$(\overline{3},1,-\frac{2}{3})$\end{tabular}}&
	\multirow{3}*{\begin{tikzpicture}[line width=0.8pt, scale=0.58,>=Stealth]
	\path (0:0) coordinate (a0);
	\path (0:1) coordinate (a1);
	\path (120:1) coordinate (a2);
	\path (240:1) coordinate (a3);
	\draw [fermion](a3)--(a0);
	\draw [fermion](a2)--(a0);
	\draw [scalar](a0)--(a1);
	\node[right] at (0:1) {$I$};
	\node[left] at (120:1) {$d_{R}$};
	\node[left] at (240:1) {$d_{R}$};
	\end{tikzpicture}}
	&\multirow{3}*{\begin{tabular}{c}antisymmetric\\(vanish)\end{tabular}}
	&\multirow{3}*{\begin{tabular}{c}{No}\end{tabular}}&
	\multirow{3}*{\begin{tabular}{c}$I\sim$\\$(6,1,-\frac{2}{3})$\end{tabular}}&
	\multirow{3}*{\begin{tikzpicture}[line width=0.8pt, scale=0.58,>=Stealth]
	\path (0:0) coordinate (a0);
	\path (0:1) coordinate (a1);
	\path (120:1) coordinate (a2);
	\path (240:1) coordinate (a3);
	\draw [fermion](a3)--(a0);
	\draw [fermion](a2)--(a0);
	\draw [scalar](a0)--(a1);
	\node[right] at (0:1) {$I$};
	\node[left] at (120:1) {$d_{R}$};
	\node[left] at (240:1) {$d_{R}$};
	\end{tikzpicture}}
	&\multirow{3}*{\begin{tabular}{c}symmetric\end{tabular}}
	&\multirow{3}*{\begin{tabular}{c}Yes\end{tabular}}\\
	&&&&&&&\\
	&&&&&&&\\\hline 
	%%%%%%%%%%%%%%%%%%%%%
         	\end{tabular}
\caption{The quantum numbers of the possible scalars under the symmetry $SU(3)_{C}\times SU(2)_{L}\times U(1)_{Y}$ and the corresponding vertices which can contribute to neutron-antineutron oscillation.}\label{DQs}
\end{table}
We transform the weak eigenstates to the mass eigenstates with the transition $U_{L}^{j}\rightarrow(V^{\dagger})_{jk}U_{L}^{k},~D_{L}^{j}\rightarrow D_{L}^{j},~E_{L}^{j}\rightarrow E_{L}$ and $\nu_{L}^{j}\rightarrow U_{jk}\nu_{L}^{k}$, where $V$ is the Cabibbo-Kobayashi-Maskawa (CKM) matrix, and $U$ is the Pontecorvo-Maki-Nakagawa-Sakata (PMNS) matrix. If the vertex does not vanish in the mass eigenstate, it can contribute to neutron-antineutron oscillation.

\paragraph*{An example:}\quad The Lagrangian involved with $I\sim(\overline{3},1,\frac{4}{3})$ in weak eigenbasis is
\begin{align}
-\mathcal{L}\supset x_{1}^{ij}\overline{(U_{R}^{i\alpha})^{c}}U_{R}^{j\beta}I^{*\gamma}\epsilon^{\alpha\beta\gamma}+{\rm{h.c.}}\,,
\end{align}
where
\begin{align}
x_{1}^{ij}\overline{(U_{R}^{i\alpha})^{c}}U_{R}^{j\beta}I^{*\gamma}\epsilon^{\alpha\beta\gamma}
&=x_{1}^{ij}\overline{(U_{R}^{j\beta})^{c}}U_{R}^{i\alpha}I^{*\gamma}\epsilon^{\alpha\beta\gamma}\notag\\
&=-x_{1}^{ij}\overline{(U_{R}^{j\alpha})^{c}}U_{R}^{i\beta}I^{*\gamma}\epsilon^{\alpha\beta\gamma}\notag\\
&=-x_{1}^{ji}\overline{(U_{R}^{i\alpha})^{c}}U_{R}^{j\beta}I^{*\gamma}\epsilon^{\alpha\beta\gamma}\,,
\end{align}
i.e. $x_{1}^{ij}=-x_{1}^{ji}$ which leads to $x_{1}^{11}=0$. The vertex also vanishes in mass eigenbasis.

\section{Decomposition of neutron-antineutron oscillation operators }\label{appendixB}
In this section, we give the complete lists of the decomposition of $n-\bar{n}$ operators at tree level and one-loop level. 
\begin{table}[h]\scriptsize
\adjustbox{valign=t}{
\renewcommand\arraystretch{1.552}
\centering
\begin{tabular}{|c|c|c|c|}
\hline
Operator&$S_{1}$&$S_{2}$&$S_{3}$\\
\hline
$(d_{R}d_{R})(d_{R}d_{R})(u_{R}u_{R})$&$(6,1,-\frac{2}{3})$&$(6,1,-\frac{2}{3})$&$(6,1,\frac{4}{3})$\\
\hline 
\multirow{2}*{$(d_{R}u_{R})(d_{R}u_{R})(d_{R}d_{R})$}&$(\overline{3},1,\frac{1}{3})$&$(\overline{3},1,\frac{1}{3})$&$(6,1,-\frac{2}{3})$\\
&$(6,1,\frac{1}{3})$&$(6,1,\frac{1}{3})$&$(6,1,-\frac{2}{3})$\\
\hline
$(d_{R}d_{R})(d_{L}d_{L})(u_{L}u_{L})$&$(6,1,-\frac{2}{3})$&$(6,3,\frac{1}{3})$&$(6,3,\frac{1}{3})$\\
\hline
\multirow{4}*{$(d_{L}u_{L})(d_{L}u_{L})(d_{R}d_{R})$}&$(\overline{3},1,\frac{1}{3})$&$(\overline{3},1,\frac{1}{3})$&$(6,1,-\frac{2}{3})$\\
&$(6,1,\frac{1}{3})$&$(6,1,\frac{1}{3})$&$(6,1,-\frac{2}{3})$\\
&$(\overline{3},3,\frac{1}{3})$&$(\overline{3},3,\frac{1}{3})$&$(6,1,-\frac{2}{3})$\\
&$(6,3,\frac{1}{3})$&$(6,3,\frac{1}{3})$&$(6,1,-\frac{2}{3})$\\
\hline
\multirow{2}*{$(d_{L}u_{L})(d_{R}u_{R})(d_{R}d_{R})$}&$(\overline{3},1,\frac{1}{3})$&$(\overline{3},1,\frac{1}{3})$&$(6,1,-\frac{2}{3})$\\
&$(6,1,\frac{1}{3})$&$(6,1,\frac{1}{3})$&$(6,1,-\frac{2}{3})$\\
\hline
\hline
Operator&$S_{1}$&$S_{2}$&$\psi$\\
\hline
$(d_{R}d_{R})(d_{R}d_{R})(u_{R})(u_{R})$&$(6,1,-\frac{2}{3})$&$(6,1,-\frac{2}{3})$&$(8,1,0)$\\
\hline
$(d_{R}d_{R})(u_{R}u_{R})(d_{R})(d_{R})$&$(6,1,-\frac{2}{3})$&$(6,1,\frac{4}{3})$&$(8,1,1)$\\
\hline
\multirow{4}*{$(d_{R}d_{R})(d_{R}u_{R})(d_{R})(u_{R})$}&$(6,1,-\frac{2}{3})$&$(\overline{3},1,\frac{1}{3})$&$(8,1,1)$\\
&$(6,1,-\frac{2}{3})$&$(6,1,\frac{1}{3})$&$(8,1,1)$\\
&$(6,1,-\frac{2}{3})$&$(\overline{3},1,\frac{1}{3})$&$(8,1,0)$\\
&$(6,1,-\frac{2}{3})$&$(6,1,\frac{1}{3})$&$(8,1,0)$\\
\hline
\multirow{4}*{$(d_{R}u_{R})(d_{R}u_{R})(d_{R})(d_{R})$}&$(\overline{3},1,\frac{1}{3})$&$(\overline{3},1,\frac{1}{3})$&$(1,1,0)$\\
&$(\overline{3},1,\frac{1}{3})$&$(\overline{3},1,\frac{1}{3})$&$(8,1,0)$\\
&$(\overline{3},1,\frac{1}{3})$&$(6,1,\frac{1}{3})$&$(8,1,0)$\\
&$(6,1,\frac{1}{3})$&$(6,1,\frac{1}{3})$&$(8,1,0)$\\
\hline
$(d_{R}d_{R})(d_{L}d_{L})(u_{L})(u_{L})$&$(6,1,-\frac{2}{3})$&$(6,3,\frac{1}{3})$&$(8,2,\frac{1}{2})$\\
\hline
\multirow{2}*{$(d_{R}d_{R})(u_{L}u_{L})(d_{L})(d_{L})$}&$(6,1,-\frac{2}{3})$&$(6,3,\frac{1}{3})$&$(8,2,\frac{1}{2})$\\
&$(6,1,-\frac{2}{3})$&$(\overline{3},3,\frac{1}{3})$&$(8,2,\frac{1}{2})$\\
\hline
\end{tabular}}\quad
\adjustbox{valign=t}{
\renewcommand\arraystretch{1.51}
\centering
\begin{tabular}{|c|c|c|c|}
\hline
Operator&$S_{1}$&$S_{2}$&$\psi$\\ 
\hline
\multirow{2}*{$(d_{L}d_{L})(u_{L}u_{L})(d_{R})(d_{R})$}&$(6,3,\frac{1}{3})$&$(6,3,\frac{1}{3})$&$(8,3,0)$\\
&$(6,3,\frac{1}{3})$&$(\overline{3},3,\frac{1}{3})$&$(8,3,0)$\\
\hline
\multirow{8}*{$(d_{L}u_{L})(d_{L}u_{L})(d_{R})(d_{R})$}&$(\overline{3},1,\frac{1}{3})$&$(\overline{3},1,\frac{1}{3})$&$(1,1,0)$\\
&$(\overline{3},1,\frac{1}{3})$&$(\overline{3},1,\frac{1}{3})$&$(8,1,0)$\\
&$(\overline{3},1,\frac{1}{3})$&$(6,1,\frac{1}{3})$&$(8,1,0)$\\
&$(6,1,\frac{1}{3})$&$(6,1,\frac{1}{3})$&$(8,1,0)$\\
&$(\overline{3},3,\frac{1}{3})$&$(\overline{3},3,\frac{1}{3})$&$(1,3,0)$\\
&$(\overline{3},3,\frac{1}{3})$&$(\overline{3},3,\frac{1}{3})$&$(8,3,0)$\\
&$(\overline{3},3,\frac{1}{3})$&$(6,3,\frac{1}{3})$&$(8,3,0)$\\
&$(6,3,\frac{1}{3})$&$(6,3,\frac{1}{3})$&$(8,3,0)$\\
\hline
\multirow{2}*{$(d_{L}u_{L})(d_{R}d_{R})(d_{L})(u_{L})$}&$(\overline{3},1,\frac{1}{3})$&$(6,1,-\frac{2}{3})$&$(8,2,-\frac{1}{2})$\\
&$(\overline{3},3,\frac{1}{3})$&$(6,1,-\frac{2}{3})$&$(8,2,-\frac{1}{2})$\\
&$(6,1,\frac{1}{3})$&$(6,1,-\frac{2}{3})$&$(8,2,-\frac{1}{2})$\\
&$(6,3,\frac{1}{3})$&$(6,1,-\frac{2}{3})$&$(8,2,-\frac{1}{2})$\\
\hline
\multirow{5}*{$(d_{L}u_{L})(d_{R}u_{R})(d_{R})(d_{R})$}&$(\overline{3},1,\frac{1}{3})$&$(\overline{3},1,\frac{1}{3})$&$(1,1,0)$\\
&$(\overline{3},1,\frac{1}{3})$&$(\overline{3},1,\frac{1}{3})$&$(8,1,0)$\\
&$(\overline{3},1,\frac{1}{3})$&$(6,1,\frac{1}{3})$&$(8,1,0)$\\
&$(6,1,\frac{1}{3})$&$(\overline{3},1,\frac{1}{3})$&$(8,1,0)$\\
&$(6,1,\frac{1}{3})$&$(6,1,\frac{1}{3})$&$(8,1,0)$\\
\hline
\multirow{4}*{$(d_{L}u_{L})(d_{R}d_{R})(d_{R})(u_{R})$}&$(\overline{3},1,\frac{1}{3})$&$(6,1,-\frac{2}{3})$&$(8,1,0)$\\
&$(\overline{3},1,\frac{1}{3})$&$(6,1,-\frac{2}{3})$&$(8,1,-1)$\\
&$(6,1,\frac{1}{3})$&$(6,1,-\frac{2}{3})$&$(8,1,0)$\\
&$(6,1,\frac{1}{3})$&$(6,1,-\frac{2}{3})$&$(8,1,-1)$\\
\hline
\multirow{2}*{$(d_{L}u_{L})(d_{R}d_{R})(d_{R})(u_{R})$}&$(6,1,-\frac{2}{3})$&$(\overline{3},1,\frac{1}{3})$&$(8,2,\frac{1}{2})$\\
&$(6,1,-\frac{2}{3})$&$(6,1,\frac{1}{3})$&$(8,2,\frac{1}{2})$\\
\hline
\end{tabular}}
\caption{The decomposition of $n-\overline{n}$ operators at tree level.}
\label{T-T-1/2-D}
\end{table}

%%%%%%%%%%%%%%%%%%%%%%%%%%%%%%%%%%%%%%%%%%%%%%%%%%%%%%%%%%%%%%%%%%%%%%%%%%%%%%%%%%%%%%%%%%%%%%%%%%%%%%%%%%%%%%%%%%%%%%
\begin{rotatepage}
\begin{sidewaystable}[p]\scriptsize
\adjustbox{valign=t}{
\renewcommand\arraystretch{1.177}
\begin{tabular}{||c|c|c|c|c|}
\hline
\multicolumn{5}{||c|}{$SU(3)_{C}$}\\
\hline
$I_{1}$&$I_{2}$&$I_{3}$&$I_{4}$&$\psi$\\
\hline
$\overline{3}$&$\overline{3}$&3&8&$\overline{3}$\\
\cline{3-5}
&&3&8&6\\
\cline{3-5}
&&3&1&$\overline{3}$\\
\cline{3-5}
&&$\overline{3}$&$\overline{3}$&1\\
\cline{3-5}
&&$\overline{3}$&$\overline{3}$&8\\
\cline{3-5}
&&$\overline{3}$&6&8\\
\cline{3-5}
&&6&$\overline{3}$&8\\
\cline{3-5}
&&6&6&8\\
\cline{3-5}
&&$\overline{6}$&8&$\overline{3}$\\
\cline{3-5}
&&$\overline{6}$&1&$\overline{3}$\\
\cline{3-5}
&&8&3&3\\
\cline{3-5}
&&8&3&$\overline{6}$\\
\cline{3-5}
&&1&3&3\\
\cline{3-5}
&&8&$\overline{6}$&3\\
\cline{3-5}
&&1&$\overline{6}$&3\\
\hline
6&6&3&8&$\overline{3}$\\
\cline{3-5}
&&3&8&6\\
\cline{3-5}
&&$\overline{3}$&6&8\\
\cline{3-5}
&&6&$\overline{3}$&8\\
\cline{3-5}
&&6&6&8\\
\cline{3-5}
&&$\overline{6}$&8&$\overline{3}$\\
\cline{3-5}
&&$\overline{6}$&1&$\overline{3}$\\
\cline{3-5}
&&8&3&3\\
\cline{3-5}
&&8&3&$\overline{6}$\\
\cline{3-5}
&&8&$\overline{6}$&3\\
\cline{3-5}
&&1&$\overline{6}$&3\\
\hline
\end{tabular}}\!\!
\adjustbox{valign=t}{
\renewcommand\arraystretch{1.177}
\begin{tabular}{|c|c|c|c|c||}
\hline
\multicolumn{5}{|c||}{$SU(3)_{C}$}\\
\hline
$I_{1}$&$I_{2}$&$I_{3}$&$I_{4}$&$\psi$\\
\hline
$\overline{3}$&6&3&8&$\overline{3}$\\
\cline{3-5}
&&3&1&$\overline{3}$\\
\cline{3-5}
&&3&8&6\\
\cline{3-5}
&&$\overline{3}$&$\overline{3}$&1\\
\cline{3-5}
&&$\overline{3}$&$\overline{3}$&8\\
\cline{3-5}
&&$\overline{3}$&6&8\\
\cline{3-5}
&&6&$\overline{3}$&8\\
\cline{3-5}
&&6&6&8\\
\cline{3-5}
&&$\overline{6}$&8&$\overline{3}$\\
\cline{3-5}
&&8&3&3\\
\cline{3-5}
&&1&3&3\\
\cline{3-5}
&&8&3&$\overline{6}$\\
\cline{3-5}
&&8&$\overline{6}$&3\\
\hline
6&$\overline{3}$&3&8&$\overline{3}$\\
\cline{3-5}
&&3&1&$\overline{3}$\\
\cline{3-5}
&&3&8&6\\
\cline{3-5}
&&$\overline{3}$&$\overline{3}$&1\\
\cline{3-5}
&&$\overline{3}$&$\overline{3}$&8\\
\cline{3-5}
&&$\overline{3}$&6&8\\
\cline{3-5}
&&6&$\overline{3}$&8\\
\cline{3-5}
&&6&6&8\\
\cline{3-5}
&&$\overline{6}$&8&$\overline{3}$\\
\cline{3-5}
&&8&3&3\\
\cline{3-5}
&&1&3&3\\
\cline{3-5}
&&8&3&$\overline{6}$\\
\cline{3-5}
&&8&$\overline{6}$&3\\
\hline
\end{tabular}}\quad
\adjustbox{valign=t}{
\renewcommand\arraystretch{0.78}
\begin{tabular}{||c|c|c|c|c|c|c|c|c|c|c||}
\hline
\multicolumn{11}{||c||}{$SU(2)_{L}$}\\
\hline
$q_{1}$&$q_{2}$&$q_{3}$&$q_{4}$&$q_{5}$&$q_{6}$&$I_{1}$&$I_{2}$&$I_{3}$&$I_{4}$&$\psi$\\
\hline
1&1&1&1&1&1&1&1&1&1&1\\
\cline{7-11}
&&&&&&1&1&2&2&2\\
\cline{7-11}
&&&&&&1&1&3&3&3\\
\hline
1&1&1&1&2&2&1&1&1&1&2\\
\cline{7-11}
&&&&&&1&1&2&2&1\\
\cline{7-11}
&&&&&&1&1&2&2&3\\
\cline{7-11}
&&&&&&1&1&3&3&2\\
\hline
1&1&2&2&1&1&1&1&1&1&1\\
\cline{7-11}
&&&&&&1&1&2&2&2\\
\cline{7-11}
&&&&&&1&1&3&3&3\\
\cline{7-11}
&&&&&&1&3&2&2&2\\
\cline{7-11}
&&&&&&1&3&3&3&3\\
\hline
1&1&2&2&2&2&1&1&1&1&2\\
\cline{7-11}
&&&&&&1&1&2&2&1\\
\cline{7-11}
&&&&&&1&1&2&2&3\\
\cline{7-11}
&&&&&&1&1&3&3&2\\
\cline{7-11}
&&&&&&1&3&1&3&2\\
\cline{7-11}
&&&&&&1&3&3&1&2\\
\cline{7-11}
&&&&&&1&3&2&2&1\\
\cline{7-11}
&&&&&&1&3&2&2&3\\
\cline{7-11}
&&&&&&1&3&3&3&2\\
\hline
2&2&2&2&1&1&1&1&1&1&1\\
\cline{7-11}
&&&&&&1&1&2&2&2\\
\cline{7-11}
&&&&&&1&1&3&3&3\\
\cline{7-11}
&&&&&&1&3&2&2&2\\
\cline{7-11}
&&&&&&1&3&3&3&3\\
\cline{7-11}
&&&&&&3&3&1&1&1\\
\cline{7-11}
&&&&&&3&3&2&2&2\\
\cline{7-11}
&&&&&&3&3&3&3&3\\
\cline{7-11}
&&&&&&3&3&1&3&3\\
\cline{7-11}
&&&&&&3&3&3&1&3\\
\hline
\end{tabular}}\quad
\adjustbox{valign=t}{
\renewcommand\arraystretch{1.825}
\begin{tabular}{||c|c|c|c|c|c|c|c|c|c|c||}
\hline
\multicolumn{6}{||c|}{Particles}&\multicolumn{5}{c||}{$U(1)_{Y}$}\\
\hline
$q_{1}$&$q_{2}$&$q_{3}$&$q_{4}$&$q_{5}$&$q_{6}$&$I_{1}$&$I_{2}$&$I_{3}$&$I_{4}$&$\psi$\\
\hline
$d_{R}$&$d_{R}$&$d_{R}$&$d_{R}$&$u_{R}$&$u_{R}$&$-\frac{2}{3}$&$-\frac{2}{3}$&$\alpha$&$-\alpha-\frac{4}{3}$&$\alpha+\frac{2}{3}$\\
\hline
$d_{R}$&$d_{R}$&$u_{R}$&$u_{R}$&$d_{R}$&$d_{R}$&$-\frac{2}{3}$&$+\frac{4}{3}$&$\alpha$&$-\alpha+\frac{2}{3}$&$\alpha-\frac{1}{3}$\\
\hline
$d_{R}$&$d_{R}$&$d_{R}$&$u_{R}$&$d_{R}$&$u_{R}$&\multirow{2}*{$-\frac{2}{3}$}&\multirow{2}*{$+\frac{1}{3}$}&\multirow{2}*{$\alpha$}&\multirow{2}*{$-\alpha-\frac{1}{3}$}&\multirow{2}*{$\alpha-\frac{1}{3}$}\\
\cline{1-6}
$d_{R}$&$d_{R}$&$d_{L}$&$u_{L}$&$d_{R}$&$u_{R}$&&&&&\\
\hline
$d_{R}$&$d_{R}$&$d_{R}$&$u_{R}$&$u_{R}$&$d_{R}$&\multirow{2}*{$-\frac{2}{3}$}&\multirow{2}*{$+\frac{1}{3}$}&\multirow{2}*{$\alpha$}&\multirow{2}*{$-\alpha-\frac{1}{3}$}&\multirow{2}*{$\alpha+\frac{2}{3}$}\\
\cline{1-6}
$d_{R}$&$d_{R}$&$d_{L}$&$u_{L}$&$u_{R}$&$d_{R}$&&&&&\\
\hline
$d_{R}$&$u_{R}$&$d_{R}$&$u_{R}$&$d_{R}$&$d_{R}$&\multirow{4}*{$+\frac{1}{3}$}&\multirow{4}*{$+\frac{1}{3}$}&\multirow{4}*{$\alpha$}&\multirow{4}*{$-\alpha+\frac{2}{3}$}&\multirow{4}*{$\alpha-\frac{1}{3}$}\\
\cline{1-6}
$d_{R}$&$u_{R}$&$d_{L}$&$u_{L}$&$d_{R}$&$d_{R}$&&&&&\\
\cline{1-6}
$d_{L}$&$d_{L}$&$u_{L}$&$u_{L}$&$d_{R}$&$d_{R}$&&&&&\\
\cline{1-6}
$d_{L}$&$u_{L}$&$d_{L}$&$u_{L}$&$d_{R}$&$d_{R}$&&&&&\\
\hline
$d_{R}$&$d_{R}$&$d_{R}$&$u_{R}$&$d_{L}$&$u_{L}$&\multirow{6}*{$-\frac{2}{3}$}&\multirow{6}*{$+\frac{1}{3}$}&\multirow{6}*{$\alpha$}&\multirow{6}*{$-\alpha-\frac{1}{3}$}&\multirow{6}*{$\alpha+\frac{1}{6}$}\\
\cline{1-6}
$d_{R}$&$d_{R}$&$d_{R}$&$u_{R}$&$u_{L}$&$d_{L}$&&&&&\\
\cline{1-6}
$d_{R}$&$d_{R}$&$d_{L}$&$d_{L}$&$u_{L}$&$u_{L}$&&&&&\\
\cline{1-6}
$d_{R}$&$d_{R}$&$u_{L}$&$u_{L}$&$d_{L}$&$d_{L}$&&&&&\\
\cline{1-6}
$d_{R}$&$d_{R}$&$u_{L}$&$d_{L}$&$u_{L}$&$d_{L}$&&&&&\\
\cline{1-6}
$d_{R}$&$d_{R}$&$u_{L}$&$d_{L}$&$d_{L}$&$u_{L}$&&&&&\\
\hline
\end{tabular}}
\caption{The possible quantum numbers of the fields in Topology-L-1. Once one determines the external quarks as, for example, $q_{1},q_{2},q_{3},q_{4}=d_{R}$, $q_{5},q_{6}=u_{R}$,  the quantum numbers of the internal particles under $SU(2)_{L}\times U(1)_{Y}$ symmetry can be figured out from the middle and right tables, i.e., $I_{1}\sim(1,-2/3),~I_{2}\sim(1,-2/3),~I_{3}\sim(1,\alpha),~I_{4}\sim(1,-\alpha-4/3),~\psi\sim(1,\alpha+2/3)$; or $I_{1}\sim(1,-2/3),~I_{2}\sim(1,-2/3),~I_{3}\sim(2,\alpha),~I_{4}\sim(2,-\alpha-4/3),~\psi\sim(2,\alpha+2/3)$; otherwise, $I_{1}\sim(1,-2/3),~I_{2}\sim(1,-2/3),~I_{3}\sim(3,\alpha),~I_{4}\sim(3,-\alpha-4/3),~\psi\sim(3,\alpha+2/3)$. According to Table~\ref{DQs}, the scalar $I\sim(6,1,-2/3)$ can contribute to the $n-\bar{n}$ oscillation in mass eigenstates, while $I\sim(\overline{3},1,-2/3)$ can not. Hence, the color number of the internal particles $I_{1,2}$ is 6, and the color numbers' decompositions of the other internal particles that are satisfied with the invariance of the $SU(3)_{C}$ gauge symmetry can be found in the ``6,6'' row of the table on the left.}\label{T-L-1-D}
\end{sidewaystable}

\begin{sidewaystable}[p]\scriptsize
\adjustbox{valign=t}{
\renewcommand\arraystretch{1.108}
\begin{tabular}{||c|c|c|c|c|c|} 
\hline
\multicolumn{6}{||c|}{$SU(3)_{C}$} \\ 
\hline
$I_{1}$ & $I_{2}$ & $\psi_{1}$ & $\psi_{2}$ & $\psi_{3}$ & $I_{3}$ \\ 
\hline
$I_{1}$ & $I_{2}$ & $I_{3}$ & $I_{4}$ & $I_{5}$ & $\psi$\\
\hline
 $\overline{3}$ & $\overline{3}$ & 3 & 3 & 1 & $\overline{3}$ \\ 
\cline{3-6}
&& 3 & 3 & 8 & $\overline{3}$ \\ 
\cline{3-6}
&& 3 & 3 & 8 & 6 \\ 
\cline{3-6}
&& 3 & $\overline{6}$ & 1 & $\overline{3}$ \\ 
\cline{3-6}
&& 3 & $\overline{6}$ & 8 & $\overline{3}$ \\
\hline
\end{tabular}}\!\!
\adjustbox{valign=t}{
\renewcommand\arraystretch{1.204}
\begin{tabular}{|c|c|c|c|c|c||} 
\hline
\multicolumn{6}{|c||}{$SU(3)_{C}$} \\ 
\hline
$I_{1}$ & $I_{2}$ & $\psi_{1}$ & $\psi_{2}$ & $\psi_{3}$ & $I_{3}$ \\ 
\hline
$I_{1}$ & $I_{2}$ & $I_{3}$ & $I_{4}$ & $I_{5}$ & $\psi$\\
\hline
$\overline{3}$ & 6 & 3 & 3 & 8 & $\overline{3}$ \\ 
\cline{3-6}
&& 3 & 3 & 8 & 6 \\ 
\cline{3-6}
&& 3 & $\overline{6}$ & 8 & $\overline{3}$ \\ 
\cline{3-6}
&& $\overline{3}$ & 8 & 3 & 3 \\ 
\cline{3-6}
&& $\overline{3}$ & 1 & 3 & 3 \\ 
\cline{3-6}
&& $\overline{3}$ & 8 & 3 & $\overline{6}$ \\ 
\cline{3-6}
&& 6 & 8 & $\overline{6}$ & 3 \\ 
\cline{3-6}
&& $\overline{6}$ & 3 & 1 & $\overline{3}$ \\ 
\cline{3-6}
&& $\overline{6}$ & 3 & 8 & $\overline{3}$ \\ 
\cline{3-6}
&& $\overline{6}$ & 3 & 8 & 6 \\
\cline{3-6}
&& 8 & $\overline{3}$ & $\overline{3}$ & 1 \\ 
\cline{3-6}
&& 8 & $\overline{3}$ & $\overline{3}$ & 8 \\ 
\cline{3-6}
&& 8 & $\overline{3}$ & 6 & 8 \\ 
\cline{3-6}
&& 1 & $\overline{3}$ & 6 & 8 \\ 
\cline{3-6}
&& 8 & 6 & $\overline{3}$ & 8 \\ 
\cline{3-6}
&& 8 & 6 & 6 & 8 \\
\hline
6 & $\overline{3}$ & 3 & 3 & 1 & $\overline{3}$ \\ 
\cline{3-6}
&& 3 & 3 & 8 & $\overline{3}$ \\ 
\cline{3-6}
&& 3 & 3 & 8 & 6 \\ 
\cline{3-6}
&& $\overline{3}$ & 8 & 3 & 3 \\ 
\cline{3-6}
&& $\overline{3}$ & 8 & 3 & $\overline{6}$ \\
\cline{3-6}
&& $\overline{3}$ & 8 & $\overline{6}$ & 3 \\  
\cline{3-6}
&& 6 & 8 & 3 & 3\\ 
\cline{3-6}
&& 6 & 1 & 3 & 3\\ 
\cline{3-6}
&& 6 & 8 & 3 & $\overline{6}$ \\ 
\cline{3-6}
&& $\overline{6}$ & $\overline{6}$ & 8 & $\overline{3}$ \\ 
\cline{3-6}
&& 8 & $\overline{3}$ & $\overline{3}$ & 1 \\ 
\cline{3-6}
&& 8 & $\overline{3}$ & $\overline{3}$ & 8 \\ 
\cline{3-6}
&& 8 & $\overline{3}$ & 6 & 8 \\ 
\cline{3-6}
&& 8 & 6 & $\overline{3}$ & 8 \\ 
\cline{3-6}
&& 1 & 6 & $\overline{3}$ & 8 \\ 
\cline{3-6}
&& 8 & 6 & 6 & 8 \\
\hline
\end{tabular}}\quad
\adjustbox{valign=t}{
\renewcommand\arraystretch{0.3}
\begin{tabular}{||c|c|c|c|c|c|c|c|c|c|c|c||} 
\hline
\multicolumn{12}{||c||}{$SU(2)_{L}$} \\ 
\hline
$q_{1}$&$q_{2}$&$q_{3}$&$q_{4}$&$q_{5}$&$q_{6}$&$I_{1}$&$I_{2}$&$I_{3}$&$I_{4}$&$I_{5}$&$\psi$\\
\hline
$q_{1}$&$q_{2}$&$q_{3}$&$q_{4}$&$q_{5}$&$q_{6}$&$I_{1}$&$I_{2}$&$\psi_{1}$&$\psi_{2}$&$\psi_{3}$&$I_{3}$\\
\hline
1&1&1&1&1&1&1&1&1&1&1&1\\
\cline{7-12}
&&&&&&1&1&2&2&2&2\\
\cline{7-12}
&&&&&&1&1&3&3&3&3\\
\hline
1&1&1&1&2&2&1&1&1&1&1&2\\
\cline{7-12}
&&&&&&1&1&2&2&2&1\\
\cline{7-12}
&&&&&&1&1&2&2&2&3\\
\cline{7-12}
&&&&&&1&1&3&3&3&2\\
\hline
1&1&2&2&1&1&1&1&1&1&1&1\\
\cline{7-12}
&&&&&&1&1&2&2&2&2\\
\cline{7-12}
&&&&&&1&1&3&3&3&3\\
\cline{7-12}
&&&&&&1&3&2&2&2&2\\
\cline{7-12}
&&&&&&1&3&3&3&3&3\\
\hline
2&2&1&1&1&1&1&1&1&1&1&1\\
\cline{7-12}
&&&&&&1&1&2&2&2&2\\
\cline{7-12}
&&&&&&1&1&3&3&3&3\\
\cline{7-12}
&&&&&&3&1&2&2&2&2\\
\cline{7-12}
&&&&&&3&1&3&3&3&3\\
\hline
1&1&2&2&2&2&1&1&1&1&1&2\\
\cline{7-12}
&&&&&&1&1&2&2&2&1\\
\cline{7-12}
&&&&&&1&1&2&2&2&3\\
\cline{7-12}
&&&&&&1&1&3&3&3&2\\
\cline{7-12}
&&&&&&1&3&1&1&3&2\\
\cline{7-12}
&&&&&&1&3&2&2&2&3\\
\cline{7-12}
&&&&&&1&3&2&2&2&1\\
\cline{7-12}
&&&&&&1&3&3&3&1&2\\
\cline{7-12}
&&&&&&1&3&3&3&3&2\\
\hline
2&2&2&2&1&1&1&1&1&1&1&1\\
\cline{7-12}
&&&&&&1&1&2&2&2&2\\
\cline{7-12}
&&&&&&1&1&3&3&3&3\\
\cline{7-12}
&&&&&&1&3&2&2&2&2\\
\cline{7-12}
&&&&&&1&3&3&3&3&3\\
\cline{7-12}
&&&&&&3&1&2&2&2&2\\
\cline{7-12}
&&&&&&3&1&3&3&3&3\\
\cline{7-12}
&&&&&&3&3&2&2&2&2\\
\cline{7-12}
&&&&&&3&3&3&3&3&3\\
\cline{7-12}
&&&&&&3&3&1&3&3&3\\
\cline{7-12}
&&&&&&3&3&3&1&1&1\\
\hline
2&2&1&1&2&2&1&1&1&1&1&2\\
\cline{7-12}
&&&&&&1&1&2&2&2&1\\
\cline{7-12}
&&&&&&1&1&2&2&2&3\\
\cline{7-12}
&&&&&&1&1&3&3&3&2\\
\cline{7-12}
&&&&&&3&1&2&2&2&1\\
\cline{7-12}
&&&&&&3&1&1&3&1&2\\
\cline{7-12}
&&&&&&3&1&3&1&3&2\\
\cline{7-12}
&&&&&&3&1&3&3&3&2\\
\cline{7-12}
&&&&&&3&1&2&2&2&3\\
\hline
\end{tabular}}\quad
\adjustbox{valign=t}{
\renewcommand\arraystretch{1.308}
\begin{tabular}{||c|c|c|c|c|c|c|c|c|c|c|c||}
\hline
\multicolumn{6}{||c|}{Particles}&\multicolumn{6}{c||}{$U(1)_{Y}$}\\
\hline
$q_{1}$&$q_{2}$&$q_{3}$&$q_{4}$&$q_{5}$&$q_{6}$&$I_{1}$&$I_{2}$&$I_{3}$&$I_{4}$&$I_{5}$&$\psi$\\
\hline
$q_{1}$&$q_{2}$&$q_{3}$&$q_{4}$&$q_{5}$&$q_{6}$&$I_{1}$&$I_{2}$&$\psi_{1}$&$\psi_{2}$&$\psi_{3}$&$I_{3}$\\
\hline
$d_{R}$&$d_{R}$&$d_{R}$&$d_{R}$&$u_{R}$&$u_{R}$&$-\frac{2}{3}$&$-\frac{2}{3}$&$\alpha$&$-\alpha-\frac{2}{3}$&$\alpha-\frac{2}{3}$&$-\alpha$\\
\hline
$d_{R}$&$d_{R}$&$u_{R}$&$u_{R}$&$d_{R}$&$d_{R}$&$-\frac{2}{3}$&$+\frac{4}{3}$&$\alpha$&$-\alpha-\frac{2}{3}$&$\alpha+\frac{4}{3}$&$-\alpha-1$\\
\hline
$u_{R}$&$u_{R}$&$d_{R}$&$d_{R}$&$d_{R}$&$d_{R}$&$+\frac{4}{3}$&$-\frac{2}{3}$&$\alpha$&$-\alpha+\frac{4}{3}$&$\alpha-\frac{2}{3}$&$-\alpha+1$\\
\hline  
$d_{R}$&$d_{R}$&$d_{R}$&$u_{R}$&$d_{R}$&$u_{R}$&\multirow{2}*{$-\frac{2}{3}$}&\multirow{2}*{$+\frac{1}{3}$}&\multirow{2}*{$\alpha$}&\multirow{2}*{$-\alpha-\frac{2}{3}$}&\multirow{2}*{$\alpha+\frac{1}{3}$}&\multirow{2}*{$-\alpha-1$}\\
\cline{1-6}
$d_{R}$&$d_{R}$&$d_{L}$&$u_{L}$&$d_{R}$&$u_{R}$&&&&&&\\
\hline
\end{tabular}}
\caption{The possible numbers of the fields in Topology-L-2 and Topology-L-3.}
\end{sidewaystable}

%%%%%%%%%%%%%%%%%%%%%%%%%%%%%%%%%%%%%%%%%%%%%%%%%%%%%%%%%%%%%%%%%%%%%%%%%%%%%%%%%%%%%%%%%%%

\begin{sidewaystable}[p]\scriptsize
\adjustbox{valign=t}{
\renewcommand\arraystretch{1.332}
\begin{tabular}{||c|c|c|c|c|c|}
\hline
\multicolumn{6}{||c|}{$SU(3)_{C}$}\\
\hline
$I_{1}$&$I_{2}$&$I_{3}$&$I_{4}$&$\psi_{1}$&$\psi_{2}$\\
\hline
$\overline{3}$&$\overline{3}$&3&3&$\overline{3}$&$\overline{3}$\\
\cline{3-6}
&&3&3&$\overline{3}$&6\\
\cline{3-6}
&&3&3&6&$\overline{3}$\\
\cline{3-6}
&&3&$\overline{6}$&$\overline{3}$&$\overline{3}$\\
\cline{3-6}
&&3&$\overline{6}$&6&$\overline{3}$\\
\cline{3-6}
&&$\overline{3}$&1&1&3\\
\cline{3-6}
&&$\overline{3}$&1&8&3\\
\cline{3-6}
&&$\overline{3}$&8&8&3\\
\cline{3-6}
&&$\overline{3}$&8&8&$\overline{6}$\\
\cline{3-6}
&&6&8&8&3\\
\cline{3-6}
&&6&8&8&$\overline{6}$\\
\cline{3-6}
&&$\overline{6}$&3&$\overline{3}$&$\overline{3}$\\
\cline{3-6}
&&$\overline{6}$&3&$\overline{3}$&6\\
\cline{3-6}
&&1&$\overline{3}$&3&1\\
\cline{3-6}
&&1&$\overline{3}$&3&8\\
\cline{3-6}
&&8&$\overline{3}$&3&8\\
\cline{3-6}
&&8&$\overline{3}$&$\overline{6}$&8\\
\cline{3-6}
&&8&6&3&8\\
\cline{3-6}
&&8&6&$\overline{6}$&8\\
\hline
6&6&3&3&$\overline{3}$&$\overline{3}$\\
\cline{3-6}
&&3&3&6&6\\
\cline{3-6}
&&$\overline{3}$&8&8&3\\
\cline{3-6}
&&$\overline{3}$&8&8&$\overline{6}$\\
\cline{3-6}
&&6&1&8&3\\
\cline{3-6}
&&6&8&8&3\\
\cline{3-6}
&&6&8&8&$\overline{6}$\\
\cline{3-6}
&&$\overline{6}$&$\overline{6}$&$\overline{3}$&$\overline{3}$\\
\cline{3-6}
&&8&$\overline{3}$&3&8\\
\cline{3-6}
&&8&$\overline{3}$&$\overline{6}$&8\\
\cline{3-6}
&&1&6&3&8\\
\hline
\end{tabular}}\!\!
\adjustbox{valign=t}{
\renewcommand\arraystretch{1.331}
\begin{tabular}{|c|c|c|c|c|c||}
\hline
\multicolumn{6}{|c||}{$SU(3)_{C}$}\\
\hline
$I_{1}$&$I_{2}$&$I_{3}$&$I_{4}$&$\psi_{1}$&$\psi_{2}$\\
\hline
6&6&8&6&3&8\\ 
\cline{3-6}
&&8&6&$\overline{6}$&8\\
\hline
$\overline{3}$&6&3&3&$\overline{3}$&$\overline{3}$\\
\cline{3-6}
&&3&3&6&6\\
\cline{3-6}
&&3&$\overline{6}$&$\overline{3}$&$\overline{3}$\\
\cline{3-6}
&&$\overline{3}$&1&8&3\\
\cline{3-6}
&&$\overline{3}$&8&8&3\\
\cline{3-6}
&&$\overline{3}$&8&8&$\overline{6}$\\
\cline{3-6}
&&6&8&8&3\\
\cline{3-6}
&&6&8&8&$\overline{6}$\\
\cline{3-6}
&&$\overline{6}$&3&$\overline{3}$&$\overline{3}$\\
\cline{3-6}
&&1&$\overline{3}$&3&8\\
\cline{3-6}
&&8&$\overline{3}$&3&8\\
\cline{3-6}
&&8&$\overline{3}$&$\overline{6}$&8\\
\cline{3-6}
&&8&6&3&8\\
\cline{3-6}
&&8&6&$\overline{6}$&8\\
\hline
6&$\overline{3}$&3&3&$\overline{3}$&$\overline{3}$\\
\cline{3-6}
&&3&3&$\overline{3}$&6\\
\cline{3-6}
&&3&3&6&$\overline{3}$\\
\cline{3-6}
&&$\overline{3}$&8&8&3\\
\cline{3-6}
&&$\overline{3}$&8&8&$\overline{6}$\\
\cline{3-6}
&&6&1&8&3\\
\cline{3-6}
&&6&8&8&3\\
\cline{3-6}
&&6&8&8&$\overline{6}$\\
\cline{3-6}
&&$\overline{6}$&$\overline{6}$&$\overline{3}$&$\overline{3}$\\
\cline{3-6}
&&8&$\overline{3}$&3&8\\
\cline{3-6}
&&8&$\overline{3}$&$\overline{6}$&8\\
\cline{3-6}
&&1&6&3&8\\
\cline{3-6}
&&8&6&3&8\\
\cline{3-6}
&&8&6&$\overline{6}$&8\\
\hline
\end{tabular}}\quad
\adjustbox{valign=t}{
\renewcommand\arraystretch{0.4}
\begin{tabular}{||c|c|c|c|c|c|c|c|c|c|c|c||}
\hline
\multicolumn{12}{||c||}{$SU(2)_{L}$}\\
\hline
$q_{1}$&$q_{2}$&$q_{3}$&$q_{4}$&$q_{5}$&$q_{6}$&$I_{1}$&$I_{2}$&$I_{3}$&$I_{4}$&$\psi_{1}$&$\psi_{2}$\\
\hline
1&1&1&1&1&1&1&1&1&1&1&1\\
\cline{7-12}
&&&&&&1&1&2&2&2&2\\
\cline{7-12}
&&&&&&1&1&3&3&3&3\\
\hline
1&1&1&1&2&2&1&1&1&1&2&2\\
\cline{7-12}
&&&&&&1&1&2&2&1&1\\
\cline{7-12}
&&&&&&1&1&2&2&3&3\\
\cline{7-12}
&&&&&&1&1&3&3&2&2\\
\hline
1&1&2&2&1&1&1&1&1&1&1&1\\
\cline{7-12}
&&&&&&1&1&2&2&2&2\\
\cline{7-12}
&&&&&&1&1&3&3&3&3\\
\cline{7-12}
&&&&&&1&3&2&2&2&2\\
\cline{7-12}
&&&&&&1&3&3&3&3&3\\
\hline
2&2&1&1&1&1&1&1&1&1&1&1\\
\cline{7-12}
&&&&&&1&1&2&2&2&2\\
\cline{7-12}
&&&&&&1&1&3&3&3&3\\
\cline{7-12}
&&&&&&3&1&2&2&2&2\\
\cline{7-12}
&&&&&&3&1&3&3&3&3\\
\hline
1&1&2&2&2&2&1&1&1&1&2&2\\
\cline{7-12}
&&&&&&1&1&2&2&1&1\\
\cline{7-12}
&&&&&&1&1&2&2&3&3\\
\cline{7-12}
&&&&&&1&1&3&3&2&2\\
\cline{7-12}
&&&&&&1&3&1&1&2&2\\
\cline{7-12}
&&&&&&1&3&2&2&1&3\\
\cline{7-12}
&&&&&&1&3&2&2&3&1\\
\cline{7-12}
&&&&&&1&3&2&2&3&3\\
\cline{7-12}
&&&&&&1&3&3&3&2&2\\
\hline
2&2&2&2&1&1&1&1&1&1&1&1\\
\cline{7-12}
&&&&&&1&1&2&2&2&2\\
\cline{7-12}
&&&&&&1&1&3&3&3&3\\
\cline{7-12}
&&&&&&1&3&2&2&2&2\\
\cline{7-12}
&&&&&&1&3&3&3&3&3\\
\cline{7-12}
&&&&&&3&1&2&2&2&2\\
\cline{7-12}
&&&&&&3&1&3&3&3&3\\
\cline{7-12}
&&&&&&3&3&2&2&2&2\\
\cline{7-12}
&&&&&&3&3&3&3&3&3\\
\cline{7-12}
&&&&&&3&3&1&3&1&3\\
\cline{7-12}
&&&&&&3&3&3&1&3&1\\
\hline
2&2&1&1&2&2&1&1&1&1&2&2\\
\cline{7-12}
&&&&&&1&1&2&2&1&1\\
\cline{7-12}
&&&&&&1&1&2&2&3&3\\
\cline{7-12}
&&&&&&1&1&3&3&2&2\\
\cline{7-12}
&&&&&&3&1&2&2&1&1\\
\cline{7-12}
&&&&&&3&1&1&3&2&2\\
\cline{7-12}
&&&&&&3&1&3&1&2&2\\
\cline{7-12}
&&&&&&3&1&3&3&2&2\\
\cline{7-12}
&&&&&&3&1&2&2&3&3\\
\hline
\end{tabular}}\quad
\adjustbox{valign=t}{
\renewcommand\arraystretch{1.363}
\begin{tabular}{||c|c|c|c|c|c|c|c|c|c|c|c||}
\hline
\multicolumn{6}{||c|}{Particles}&\multicolumn{6}{c||}{$U(1)_{Y}$}\\
\hline
$q_{1}$&$q_{2}$&$q_{3}$&$q_{4}$&$q_{5}$&$q_{6}$&$I_{1}$&$I_{2}$&$I_{3}$&$I_{4}$&$\psi_{1}$&$\psi_{2}$\\
\hline
$d_{R}$&$d_{R}$&$d_{R}$&$d_{R}$&$u_{R}$&$u_{R}$&$-\frac{2}{3}$&$-\frac{2}{3}$&$\alpha$&$-\alpha-\frac{2}{3}$&$\alpha+\frac{2}{3}$&$-\alpha$\\
\hline
$d_{R}$&$d_{R}$&$u_{R}$&$u_{R}$&$d_{R}$&$d_{R}$&$-\frac{2}{3}$&$+\frac{4}{3}$&$\alpha$&$-\alpha-\frac{2}{3}$&$\alpha-\frac{1}{3}$&$-\alpha-1$\\
\hline
$u_{R}$&$u_{R}$&$d_{R}$&$d_{R}$&$d_{R}$&$d_{R}$&$+\frac{4}{3}$&$-\frac{2}{3}$&$\alpha$&$-\alpha+\frac{4}{3}$&$\alpha-\frac{1}{3}$&$-\alpha+1$\\
\hline  
$d_{R}$&$d_{R}$&$d_{R}$&$u_{R}$&$d_{R}$&$u_{R}$&\multirow{2}*{$-\frac{2}{3}$}&\multirow{2}*{$+\frac{1}{3}$}&\multirow{2}*{$\alpha$}&\multirow{2}*{$-\alpha-\frac{2}{3}$}&\multirow{2}*{$\alpha-\frac{1}{3}$}&\multirow{2}*{$-\alpha$}\\
\cline{1-6}
$d_{R}$&$d_{R}$&$d_{L}$&$u_{L}$&$d_{R}$&$u_{R}$&&&&&&\\
\hline
$d_{R}$&$d_{R}$&$d_{R}$&$u_{R}$&$u_{R}$&$d_{R}$&\multirow{2}*{$-\frac{2}{3}$}&\multirow{2}*{$+\frac{1}{3}$}&\multirow{2}*{$\alpha$}&\multirow{2}*{$-\alpha-\frac{2}{3}$}&\multirow{2}*{$\alpha+\frac{2}{3}$}&\multirow{2}*{$-\alpha-1$}\\
\hline
\end{tabular}}
\caption{The possible quantum numbers of the fields in Topology-L-4.}
\end{sidewaystable}

%
%%%%%%%%%%%%%%%%%%%%%%%%%%%%%%%%%%%%%%%%%%%%%%%%%%%%%%%%%%%%%%%%%%%%%%%%%%%%%%%%%%%%%%%%%%%%%%%%%%%%%%%%

\begin{sidewaystable}[p]\scriptsize
\adjustbox{valign=t}{
\renewcommand\arraystretch{1.622}
\begin{tabular}{||c|c|c|c|c|c|}
\hline
\multicolumn{6}{||c|}{$SU(3)_{C}$}\\
\hline
$I_{1}$&$I_{2}$&$I_{3}$&$\psi_{1}$&$\psi_{2}$&$\psi_{3}$\\
\hline
$\overline{3}$&1&3&1&1&3\\
\hline
$\overline{3}$&1&3&8&8&3\\
\hline
$\overline{3}$&1&$\overline{6}$&8&8&3\\
\hline
$\overline{3}$&3&1&1&$\overline{3}$&$\overline{3}$\\
\hline
$\overline{3}$&3&8&1&$\overline{3}$&$\overline{3}$\\
\hline
$\overline{3}$&3&8&1&$\overline{3}$&6\\
\hline
$\overline{3}$&8&3&1&8&3\\
\hline
$\overline{3}$&8&$\overline{6}$&1&8&3\\
\hline
$\overline{3}$&8&3&1&8&$\overline{6}$\\
\hline
$\overline{3}$&6&$\overline{3}$&1&$\overline{6}$&8\\
\hline
$\overline{3}$&$\overline{3}$&$\overline{3}$&1&3&1\\
\hline
$\overline{3}$&$\overline{3}$&$\overline{3}$&1&3&8\\
\hline
$\overline{3}$&$\overline{3}$&6&1&3&8\\
\hline
$\overline{3}$&$\overline{6}$&8&1&6&$\overline{3}$\\
\hline
$\overline{3}$&3&1&8&$\overline{3}$&$\overline{3}$\\
\hline
$\overline{3}$&3&8&8&$\overline{3}$&$\overline{3}$\\
\hline
$\overline{3}$&3&8&8&$\overline{3}$&6\\
\hline
$\overline{3}$&6&$\overline{3}$&8&$\overline{6}$&8\\
\hline
$\overline{3}$&$\overline{6}$&8&8&6&$\overline{3}$\\
\hline
$\overline{3}$&8&3&8&8&3\\
\hline
$\overline{3}$&8&$\overline{6}$&8&8&3\\
\hline
$\overline{3}$&8&3&8&8&$\overline{6}$\\
\hline
$\overline{3}$&3&8&8&6&$\overline{3}$\\
\hline
$\overline{3}$&3&8&8&6&6\\
\hline
\end{tabular}}\!\!
\adjustbox{valign=t}{
\renewcommand\arraystretch{1.622}
\begin{tabular}{|c|c|c|c|c|c||}
\hline
\multicolumn{6}{|c||}{$SU(3)_{C}$}\\
\hline
$I_{1}$&$I_{2}$&$I_{3}$&$\psi_{1}$&$\psi_{2}$&$\psi_{3}$\\
\hline
$\overline{3}$&6&$\overline{3}$&8&3&8\\
\hline
$\overline{3}$&6&6&8&3&8\\
\hline
$\overline{3}$&$\overline{3}$&$\overline{3}$&8&$\overline{6}$&1\\
\hline
$\overline{3}$&$\overline{3}$&$\overline{3}$&8&$\overline{6}$&8\\
\hline
$\overline{3}$&$\overline{6}$&1&8&$\overline{3}$&$\overline{3}$\\
\hline
$\overline{3}$&$\overline{6}$&8&8&$\overline{3}$&$\overline{3}$\\
\hline
6&1&3&8&8&3\\
\hline
6&1&$\overline{3}$&8&8&3\\
\hline
6&3&1&8&$\overline{3}$&$\overline{3}$\\
\hline
6&3&8&8&$\overline{3}$&$\overline{3}$\\
\hline
6&3&8&8&$\overline{3}$&6\\
\hline
6&6&$\overline{3}$&8&$\overline{6}$&8\\
\hline
6&$\overline{6}$&8&8&6&$\overline{3}$\\
\hline
6&8&3&8&8&3\\
\hline
6&8&$\overline{6}$&8&8&3\\
\hline
6&8&3&8&8&$\overline{6}$\\
\hline
6&3&8&8&6&$\overline{3}$\\
\hline
6&3&8&8&6&6\\
\hline
6&6&$\overline{3}$&8&3&8\\
\hline
6&6&6&8&3&8\\
\hline
6&$\overline{3}$&$\overline{3}$&8&$\overline{6}$&1\\
\hline
6&$\overline{3}$&$\overline{3}$&8&$\overline{6}$&8\\
\hline
6&$\overline{6}$&1&8&$\overline{3}$&$\overline{3}$\\
\hline
6&$\overline{6}$&8&8&$\overline{3}$&$\overline{3}$\\
\hline
\end{tabular}}\quad
\adjustbox{valign=t}{
\renewcommand\arraystretch{0.295}
\begin{tabular}{||c|c|c|c|c|c|c|c|c|c|c|c|}
\hline
\multicolumn{12}{||c|}{$SU(2)_{L}$}\\
\hline
$q_{1}$&$q_{2}$&$q_{3}$&$q_{4}$&$q_{5}$&$q_{6}$&$I_{1}$&$I_{2}$&$I_{3}$&$\psi_{1}$&$\psi_{2}$&$\psi_{3}$\\
\hline
1&1&1&1&1&1&1&1&1&1&1&1\\
\cline{7-12}
&&&&&&1&2&2&1&2&2\\
\cline{7-12}
&&&&&&1&3&3&1&3&3\\
\hline
1&1&1&1&2&2&1&1&1&2&2&1\\
\cline{7-12}
&&&&&&1&2&2&2&1&2\\
\cline{7-12}
&&&&&&1&2&2&2&3&2\\
\cline{7-12}
&&&&&&1&3&3&2&2&3\\
\hline
2&2&1&1&1&1&1&1&1&1&1&1\\
\cline{7-12}
&&&&&&1&2&2&1&2&2\\
\cline{7-12}
&&&&&&1&3&3&1&3&3\\
\cline{7-12}
&&&&&&3&2&2&3&2&2\\ 
\cline{7-12}
&&&&&&3&3&3&3&3&3\\
\hline
1&1&2&2&1&1&1&1&1&1&1&2\\
\cline{7-12}
&&&&&&1&2&2&1&2&1\\
\cline{7-12}
&&&&&&1&2&2&1&2&3\\
\cline{7-12}
&&&&&&1&3&3&1&3&2\\
\hline
1&1&1&2&1&2&1&1&2&2&2&2\\
\cline{7-12}
&&&&&&1&2&1&2&1&1\\
\cline{7-12}
&&&&&&1&2&3&2&3&3\\
\cline{7-12}
&&&&&&1&3&2&2&2&2\\
\hline
1&1&2&1&1&2&1&1&2&2&2&1\\
\cline{7-12}
&&&&&&1&2&1&2&1&2\\
\cline{7-12}
&&&&&&1&2&3&2&3&2\\
\cline{7-12}
&&&&&&1&3&2&2&2&3\\
\hline
1&1&2&1&2&1&1&1&2&1&1&1\\
\cline{7-12}
&&&&&&1&2&1&1&2&2\\
\cline{7-12}
&&&&&&1&2&3&1&2&2\\
\cline{7-12}
&&&&&&1&3&2&1&3&3\\
\hline
1&1&1&2&2&1&1&1&2&1&1&2\\
\cline{7-12}
&&&&&&1&2&1&1&2&1\\
\cline{7-12}
&&&&&&1&2&3&1&2&3\\
\cline{7-12}
&&&&&&1&3&2&1&3&2\\
\hline
1&1&2&2&2&2&1&1&1&2&2&2\\
\cline{7-12}
&&&&&&1&1&3&2&2&2\\
\cline{7-12}
&&&&&&1&2&2&2&1&1\\
\cline{7-12}
&&&&&&1&2&2&2&1&3\\
\cline{7-12}
&&&&&&1&2&2&2&3&1\\
\cline{7-12}
&&&&&&1&2&2&2&3&3\\
\cline{7-12}
&&&&&&1&3&1&2&2&2\\
\cline{7-12}
&&&&&&1&3&3&2&2&2\\
\hline
2&2&1&1&2&2&1&1&2&2&2&2\\
\cline{7-12}
&&&&&&1&2&2&2&1&2\\
\cline{7-12}
&&&&&&1&2&2&2&3&2\\
\cline{7-12}
&&&&&&1&3&2&2&2&2\\
\cline{7-12}
&&&&&&3&1&2&2&2&2\\
\cline{7-12}
&&&&&&3&2&2&2&1&2\\
\cline{7-12}
&&&&&&3&2&2&2&3&2\\
\cline{7-12}
&&&&&&3&3&2&2&2&2\\
\hline
\end{tabular}}\!
\adjustbox{valign=t}{
\renewcommand\arraystretch{0.63}
\begin{tabular}{|c|c|c|c|c|c|c|c|c|c|c|c||}
\hline
\multicolumn{12}{|c||}{$SU(2)_{L}$}\\
\hline
$q_{1}$&$q_{2}$&$q_{3}$&$q_{4}$&$q_{5}$&$q_{6}$&$I_{1}$&$I_{2}$&$I_{3}$&$\psi_{1}$&$\psi_{2}$&$\psi_{3}$\\
\hline
2&2&2&2&1&1&1&1&1&1&1&2\\
\cline{7-12}
&&&&&&1&2&2&1&2&1\\
\cline{7-12}
&&&&&&1&2&2&1&2&3\\
\cline{7-12}
&&&&&&1&3&3&1&3&2\\
\cline{7-12}
&&&&&&3&2&2&3&2&1\\
\cline{7-12}
&&&&&&3&2&2&3&2&3\\
\cline{7-12}
&&&&&&3&3&3&3&3&2\\
\cline{7-12}
&&&&&&3&1&3&3&3&2\\
\cline{7-12}
&&&&&&3&3&1&3&1&2\\ 
\hline
2&2&1&2&1&2&1&1&2&2&2&2\\
\cline{7-12}
&&&&&&1&2&1&2&1&1\\
\cline{7-12}
&&&&&&1&2&3&2&3&3\\
\cline{7-12}
&&&&&&1&3&2&2&2&2\\
\cline{7-12}
&&&&&&3&1&2&2&2&2\\
\cline{7-12}
&&&&&&3&2&1&2&1&1\\
\cline{7-12}
&&&&&&3&2&3&2&3&3\\
\cline{7-12}
&&&&&&3&3&2&2&2&2\\
\hline    
2&2&2&1&1&2&1&1&2&1&2&1\\
\cline{7-12}
&&&&&&1&2&1&2&1&2\\
\cline{7-12}
&&&&&&1&2&3&2&3&2\\
\cline{7-12}
&&&&&&1&3&2&2&2&3\\
\cline{7-12}
&&&&&&3&1&2&2&2&1\\
\cline{7-12}
&&&&&&3&2&1&2&1&2\\
\cline{7-12}
&&&&&&3&2&3&2&3&2\\
\cline{7-12}
&&&&&&3&3&2&2&2&3\\
\hline
2&2&2&1&2&1&1&1&2&1&1&1\\
\cline{7-12}
&&&&&&1&2&1&1&2&2\\
\cline{7-12}
&&&&&&1&2&3&1&2&2\\
\cline{7-12}
&&&&&&1&3&2&1&3&3\\
\cline{7-12}
&&&&&&3&2&1&3&2&2\\
\cline{7-12}
&&&&&&3&2&3&3&2&2\\
\cline{7-12}
&&&&&&3&3&2&3&3&3\\
\cline{7-12}
&&&&&&3&1&2&3&3&1\\
\cline{7-12}
&&&&&&3&3&2&3&1&3\\
\hline
2&2&1&2&2&1&1&1&2&1&1&2\\
\cline{7-12}
&&&&&&1&2&1&1&2&1\\
\cline{7-12}
&&&&&&1&2&3&1&2&3\\
\cline{7-12}
&&&&&&1&3&2&1&3&2\\
\cline{7-12}
&&&&&&3&2&1&3&2&1\\
\cline{7-12}
&&&&&&3&2&3&3&2&3\\
\cline{7-12}
&&&&&&3&3&2&3&3&2\\
\cline{7-12}
&&&&&&3&1&2&3&3&2\\
\cline{7-12}
&&&&&&3&3&2&3&1&2\\
\hline
\end{tabular}}
\caption{The possible  $SU(3)_{C}$ and $SU(2)_{L}$ numbers of the fields in T-L-5.}
\end{sidewaystable}

%%%%%%%%%%%%%%%%%%%%%%%%%%%%%%%%%%%%%%%%%%%%%%%%%%%%%%%%%%%%%%%%%%%%%%%%
\begin{sidewaystable}[p]\scriptsize
\adjustbox{valign=t}{
\renewcommand\arraystretch{0.76}
\begin{tabular}{||c|c|c|c|c|c|c|c|c|c|c|c||}
\hline
\multicolumn{6}{||c|}{Particles}&\multicolumn{6}{c||}{$U(1)_{Y}$}\\
\hline
$q_{1}$&$q_{2}$&$q_{3}$&$q_{4}$&$q_{5}$&$q_{6}$&$I_{1}$&$I_{2}$&$I_{3}$&$\psi_{1}$&$\psi_{2}$&$\psi_{3}$\\
\hline
$d_{R}$&$d_{R}$&$d_{R}$&$d_{R}$&$u_{R}$&$u_{R}$&$-\frac{2}{3}$&$\alpha$&$-\alpha+\frac{2}{3}$&$0$&$-\alpha$&$\alpha-\frac{1}{3}$\\
\hline
$d_{R}$&$d_{R}$&$d_{R}$&$u_{R}$&$d_{R}$&$u_{R}$&$-\frac{2}{3}$&$\alpha$&$-\alpha-\frac{1}{3}$&$0$&$-\alpha$&$\alpha+\frac{2}{3}$\\
\hline
$d_{R}$&$d_{R}$&$u_{R}$&$d_{R}$&$d_{R}$&$u_{R}$&$-\frac{2}{3}$&$\alpha$&$-\alpha-\frac{1}{3}$&$0$&$-\alpha$&$\alpha-\frac{1}{3}$\\
\hline
$d_{R}$&$d_{R}$&$d_{R}$&$u_{R}$&$u_{R}$&$d_{R}$&$-\frac{2}{3}$&$\alpha$&$-\alpha-\frac{1}{3}$&$-1$&$-\alpha-1$&$\alpha+\frac{2}{3}$\\
\hline
$d_{R}$&$d_{R}$&$u_{R}$&$d_{R}$&$u_{R}$&$d_{R}$&$-\frac{2}{3}$&$\alpha$&$-\alpha-\frac{1}{3}$&$-1$&$-\alpha-1$&$\alpha-\frac{1}{3}$\\
\hline
$d_{R}$&$d_{R}$&$u_{R}$&$u_{R}$&$d_{R}$&$d_{R}$&$-\frac{2}{3}$&$\alpha$&$-\alpha-\frac{4}{3}$&$-1$&$-\alpha-1$&$\alpha+\frac{2}{3}$\\
\hline
\end{tabular}}
\adjustbox{valign=t}{
\renewcommand\arraystretch{0.6}
\begin{tabular}{||c|c|c|c|c|c|c|c|c|c|c|c||}
\hline
\multicolumn{6}{||c|}{Particles}&\multicolumn{6}{c||}{$U(1)_{Y}$}\\
\hline
$q_{1}$&$q_{2}$&$q_{3}$&$q_{4}$&$q_{5}$&$q_{6}$&$I_{1}$&$I_{2}$&$I_{3}$&$I_{4}$&$\psi_{1}$&$\psi_{2}$\\
\hline
$d_{R}$&$d_{R}$&$d_{L}$&$d_{L}$&$u_{L}$&$u_{L}$&\multirow{6}*{$-\frac{2}{3}$}&\multirow{6}*{$\alpha$}&\multirow{6}*{$-\alpha-\frac{1}{3}$}&\multirow{6}*{$-\frac{1}{2}$}&\multirow{6}*{$-\alpha-\frac{1}{2}$}&\multirow{6}*{$\alpha+\frac{1}{6}$}\\
\cline{1-6}
$d_{R}$&$d_{R}$&$d_{L}$&$u_{L}$&$d_{L}$&$u_{L}$&&&&&&\\
\cline{1-6}
$d_{R}$&$d_{R}$&$u_{L}$&$d_{L}$&$d_{L}$&$u_{L}$&&&&&&\\
\cline{1-6}
$d_{R}$&$d_{R}$&$u_{L}$&$u_{L}$&$d_{L}$&$d_{L}$&&&&&&\\
\cline{1-6}
$d_{R}$&$d_{R}$&$u_{L}$&$d_{L}$&$u_{L}$&$d_{L}$&&&&&&\\
\cline{1-6}
$d_{R}$&$d_{R}$&$d_{L}$&$u_{L}$&$u_{L}$&$d_{L}$&&&&&&\\
\hline
\end{tabular}}
\caption{The possible $U(1)_{Y}$ numbers of the fields in Topology-L-5.}
\end{sidewaystable}

\begin{sidewaystable}[p]\scriptsize
\adjustbox{valign=t}{
\renewcommand\arraystretch{1.36}
\begin{tabular}{||c|c|c|c|c|c|}
\hline
\multicolumn{6}{||c|}{$SU(3)_{C}$}\\
\hline
$I_{1}$&$I_{2}$&$I_{3}$&$\psi_{1}$&$\psi_{2}$&$\psi_{3}$\\
\hline
$I_{1}$&$\psi_{1}$&$\psi_{2}$&$I_{2}$&$I_{3}$&$I_{4}$\\
\hline
$\overline{3}$&$\overline{3}$&$\overline{3}$&3&3&1\\
\hline
$\overline{3}$&$\overline{3}$&$\overline{3}$&3&3&8\\
\hline
$\overline{3}$&$\overline{3}$&6&3&3&8\\
\hline
$\overline{3}$&6&$\overline{3}$&3&3&8\\
\hline
$\overline{3}$&6&6&3&3&8\\
\hline
$\overline{3}$&$\overline{3}$&$\overline{3}$&3&$\overline{6}$&1\\
\hline
$\overline{3}$&$\overline{3}$&$\overline{3}$&3&$\overline{6}$&8\\
\hline
$\overline{3}$&6&$\overline{3}$&3&$\overline{6}$&8\\
\hline 
$\overline{3}$&$\overline{3}$&$\overline{3}$&$\overline{6}$&3&1\\
\hline
$\overline{3}$&$\overline{3}$&$\overline{3}$&$\overline{6}$&3&8\\
\hline
$\overline{3}$&$\overline{3}$&6&$\overline{6}$&3&8\\
\hline
$\overline{3}$&1&3&$\overline{3}$&8&3\\
\hline
$\overline{3}$&1&$\overline{6}$&$\overline{3}$&8&3\\
\hline
$\overline{3}$&8&3&$\overline{3}$&8&3\\
\hline
$\overline{3}$&8&3&$\overline{3}$&8&$\overline{6}$\\
\hline
$\overline{3}$&8&$\overline{6}$&$\overline{3}$&8&3\\
\hline
$\overline{3}$&8&3&6&8&3\\
\hline
$\overline{3}$&8&3&6&8&$\overline{6}$\\
\hline
$\overline{3}$&8&$\overline{6}$&6&8&3\\
\hline
$\overline{3}$&3&1&8&$\overline{3}$&$\overline{3}$\\
\hline
$\overline{3}$&$\overline{6}$&1&8&$\overline{3}$&$\overline{3}$\\
\hline
$\overline{3}$&3&8&8&$\overline{3}$&$\overline{3}$\\
\hline
$\overline{3}$&3&8&8&$\overline{3}$&6\\
\hline
$\overline{3}$&$\overline{6}$&8&8&$\overline{3}$&$\overline{3}$\\
\hline
$\overline{3}$&3&8&8&6&6\\
\hline
$\overline{3}$&3&1&1&$\overline{3}$&$\overline{3}$\\
\hline
$\overline{3}$&3&8&1&$\overline{3}$&$\overline{3}$\\
\hline
$\overline{3}$&3&8&1&$\overline{3}$&6\\
\hline
\end{tabular}}\!\!
\adjustbox{valign=t}{
\renewcommand\arraystretch{1.405}
\begin{tabular}{|c|c|c|c|c|c||}
\hline
\multicolumn{6}{|c||}{$SU(3)_{C}$}\\
\hline
$I_{1}$&$I_{2}$&$I_{3}$&$\psi_{1}$&$\psi_{2}$&$\psi_{3}$\\
\hline
$I_{1}$&$\psi_{1}$&$\psi_{2}$&$I_{2}$&$I_{3}$&$I_{4}$\\
\hline
$\overline{3}$&3&8&8&6&$\overline{3}$\\
\hline
$\overline{3}$&$\overline{6}$&8&8&6&$\overline{3}$\\
\hline
%%%%%%%%%%%%%%%%%%%%%%%%%%%%%%%%%%%%
6&$\overline{3}$&$\overline{3}$&3&3&1\\
\hline
6&$\overline{3}$&$\overline{3}$&3&3&8\\
\hline
6&$\overline{3}$&6&3&3&8\\
\hline
6&6&$\overline{3}$&3&3&8\\
\hline
6&6&6&3&3&8\\
\hline
6&1&3&$\overline{3}$&8&3\\
\hline
6&1&$\overline{6}$&$\overline{3}$&8&3\\
\hline
6&8&3&$\overline{3}$&8&3\\
\hline
6&8&3&$\overline{3}$&8&$\overline{6}$\\
\hline
6&8&$\overline{6}$&$\overline{3}$&8&3\\
\hline
6&8&3&6&8&3\\
\hline
6&8&3&6&8&$\overline{6}$\\
\hline
6&8&$\overline{6}$&6&8&3\\
\hline
6&3&1&8&$\overline{3}$&$\overline{3}$\\
\hline
6&$\overline{6}$&1&8&$\overline{3}$&$\overline{3}$\\
\hline
6&3&8&8&$\overline{3}$&$\overline{3}$\\
\hline
6&3&8&8&$\overline{3}$&6\\
\hline
6&$\overline{6}$&8&8&$\overline{3}$&$\overline{3}$\\
\hline
6&3&8&8&6&6\\
\hline
6&3&8&8&6&$\overline{3}$\\
\hline
6&$\overline{6}$&8&8&6&$\overline{3}$\\
\hline
6&$\overline{3}$&$\overline{3}$&$\overline{6}$&$\overline{6}$&1\\
\hline
6&$\overline{3}$&$\overline{3}$&$\overline{6}$&$\overline{6}$&8\\
\hline
6&3&8&1&6&3\\
\hline
6&3&8&1&6&$\overline{6}$\\
\hline
\end{tabular}}\quad
\adjustbox{valign=t}{
\renewcommand\arraystretch{0.3}
\begin{tabular}{||c|c|c|c|c|c|c|c|c|c|c|c|}
\hline
\multicolumn{12}{||c|}{$SU(2)_{L}$}\\
\hline
$q_{1}$&$q_{2}$&$q_{3}$&$q_{4}$&$q_{5}$&$q_{6}$&$I_{1}$&$I_{2}$&$I_{3}$&$\psi_{1}$&$\psi_{2}$&$\psi_{3}$\\
\hline
$q_{1}$&$q_{2}$&$q_{3}$&$q_{4}$&$q_{5}$&$q_{6}$&$I_{1}$&$\psi_{1}$&$\psi_{2}$&$I_{2}$&$I_{3}$&$I_{4}$\\
\hline
1&1&1&1&1&1&1&1&1&1&1&1\\
\cline{7-12}
&&&&&&1&2&2&2&2&2\\
\cline{7-12}
&&&&&&1&3&3&3&3&3\\
\hline
1&1&1&1&2&2&1&2&2&1&1&2\\
\cline{7-12}
&&&&&&1&1&1&2&2&1\\
\cline{7-12}
&&&&&&1&3&3&2&2&3\\
\cline{7-12}
&&&&&&1&2&2&3&3&2\\
\hline
2&2&1&1&1&1&1&1&1&1&1&1\\
\cline{7-12}
&&&&&&1&2&2&2&2&2\\
\cline{7-12}
&&&&&&1&3&3&3&3&3\\
\cline{7-12}
&&&&&&3&2&2&2&2&2\\ 
\cline{7-12}
&&&&&&3&3&3&3&3&3\\
\hline
1&1&2&2&1&1&1&1&1&1&1&2\\
\cline{7-12}
&&&&&&1&2&2&2&2&1\\
\cline{7-12}
&&&&&&1&2&2&2&2&3\\
\cline{7-12}
&&&&&&1&3&3&3&3&2\\
\hline
1&1&1&2&1&2&1&1&2&1&1&1\\
\cline{7-12}
&&&&&&1&2&1&2&2&2\\
\cline{7-12}
&&&&&&1&2&3&2&2&2\\
\cline{7-12}
&&&&&&1&3&2&3&3&3\\
\hline
1&1&2&1&1&2&1&1&2&1&1&2\\
\cline{7-12}
&&&&&&1&2&1&2&2&1\\
\cline{7-12}
&&&&&&1&2&3&2&2&3\\
\cline{7-12}
&&&&&&1&3&2&3&3&2\\
\hline
1&1&2&1&2&1&1&2&1&1&1&1\\
\cline{7-12}
&&&&&&1&1&2&2&2&2\\
\cline{7-12}
&&&&&&1&2&3&3&3&3\\
\cline{7-12}
&&&&&&1&3&2&2&2&2\\
\hline
1&1&1&2&2&1&1&2&1&1&1&2\\
\cline{7-12}
&&&&&&1&1&2&2&2&2\\
\cline{7-12}
&&&&&&1&2&3&3&3&2\\
\cline{7-12}
&&&&&&1&3&2&2&2&3\\
\hline
%%%%%%%%%%%%%%%%%%%%%%%%%%%%%
1&1&2&2&2&2&1&2&2&1&1&1\\
\cline{7-12}
&&&&&&1&2&2&1&1&3\\
\cline{7-12}
&&&&&&1&1&1&2&2&2\\
\cline{7-12}
&&&&&&1&3&3&2&2&2\\
\cline{7-12}
&&&&&&1&1&3&2&2&2\\
\cline{7-12}
&&&&&&1&3&1&2&2&2\\
\cline{7-12}
&&&&&&1&2&2&3&3&1\\
\cline{7-12}
&&&&&&1&2&2&3&3&3\\
\hline
2&2&1&1&2&2&1&2&2&1&1&2\\
\cline{7-12}
&&&&&&1&1&1&2&2&1\\
\cline{7-12}
&&&&&&1&3&3&2&2&3\\
\cline{7-12}
&&&&&&1&2&2&3&3&2\\
\cline{7-12}
&&&&&&3&1&1&2&2&1\\
\cline{7-12}
&&&&&&3&3&3&2&2&3\\
\cline{7-12}
&&&&&&3&2&2&1&3&2\\
\hline
\end{tabular}}\!\!
\adjustbox{valign=t}{
\renewcommand\arraystretch{0.3}
\begin{tabular}{|c|c|c|c|c|c|c|c|c|c|c|c||}
\hline
\multicolumn{12}{|c||}{$SU(2)_{L}$}\\
\hline
$q_{1}$&$q_{2}$&$q_{3}$&$q_{4}$&$q_{5}$&$q_{6}$&$I_{1}$&$I_{2}$&$I_{3}$&$\psi_{1}$&$\psi_{2}$&$\psi_{3}$\\
\hline
$q_{1}$&$q_{2}$&$q_{3}$&$q_{4}$&$q_{5}$&$q_{6}$&$I_{1}$&$\psi_{1}$&$\psi_{2}$&$I_{2}$&$I_{3}$&$I_{4}$\\
\hline
2&2&1&1&2&2&3&2&2&3&1&2\\
\cline{7-12}
&&&&&&3&2&2&3&3&2\\
\hline
2&2&2&2&1&1&1&1&1&1&1&2\\
\cline{7-12}
&&&&&&1&2&2&2&2&1\\
\cline{7-12}
&&&&&&1&2&2&2&2&3\\
\cline{7-12}
&&&&&&1&3&3&3&3&2\\
\cline{7-12}
&&&&&&3&2&2&2&2&1\\
\cline{7-12}
&&&&&&3&2&2&2&2&3\\
\cline{7-12}
&&&&&&3&3&3&3&3&2\\
\cline{7-12}
&&&&&&3&1&3&1&3&2\\
\cline{7-12}
&&&&&&3&3&1&3&1&2\\ 
\hline
2&2&1&2&1&2&1&1&2&1&1&1\\
\cline{7-12}
&&&&&&1&2&1&2&2&2\\
\cline{7-12}
&&&&&&1&2&3&2&2&2\\
\cline{7-12}
&&&&&&1&3&2&3&3&3\\
\cline{7-12}
&&&&&&3&1&2&1&3&1\\
\cline{7-12}
&&&&&&3&2&1&2&2&2\\
\cline{7-12}
&&&&&&3&2&3&2&2&2\\
\cline{7-12}
&&&&&&3&3&2&3&3&3\\
\cline{7-12}
&&&&&&3&3&2&3&1&3\\
\hline    
2&2&2&1&1&2&1&1&2&1&1&2\\
\cline{7-12}
&&&&&&1&2&1&2&2&1\\
\cline{7-12}
&&&&&&1&2&3&2&2&3\\
\cline{7-12}
&&&&&&1&3&2&3&3&2\\
\cline{7-12}
&&&&&&3&1&2&1&3&2\\
\cline{7-12}
&&&&&&3&2&1&2&2&1\\
\cline{7-12}
&&&&&&3&2&3&2&2&3\\
\cline{7-12}
&&&&&&3&3&2&3&3&2\\
\cline{7-12}
&&&&&&3&3&2&3&1&2\\
\hline
2&2&2&1&2&1&1&1&2&2&2&2\\
\cline{7-12}
&&&&&&1&2&1&1&1&1\\
\cline{7-12}
&&&&&&1&2&3&3&3&3\\
\cline{7-12}
&&&&&&1&3&2&2&2&2\\
\cline{7-12}
&&&&&&3&2&1&3&1&1\\
\cline{7-12}
&&&&&&3&2&3&1&3&3\\
\cline{7-12}
&&&&&&3&2&3&3&3&3\\
\cline{7-12}
&&&&&&3&3&2&2&2&2\\
\cline{7-12}
&&&&&&3&1&2&2&2&2\\
\hline
2&2&1&2&2&1&1&1&2&2&2&1\\
\cline{7-12}
&&&&&&1&2&1&1&1&2\\
\cline{7-12}
&&&&&&1&2&3&3&3&2\\
\cline{7-12}
&&&&&&1&3&2&2&2&3\\
\cline{7-12}
&&&&&&3&2&1&3&1&2\\
\cline{7-12}
&&&&&&3&2&3&1&3&2\\
\cline{7-12}
&&&&&&3&2&3&3&3&2\\
\cline{7-12}
&&&&&&3&3&2&2&2&3\\
\cline{7-12}
&&&&&&3&1&2&2&2&1\\
\hline
\end{tabular}}
\caption{The possible  $SU(3)_{C}$ and $SU(2)_{L}$ numbers of the fields in T-L-6/7.}
\end{sidewaystable}

\begin{sidewaystable}[p]\scriptsize
\adjustbox{valign=t}{
\renewcommand\arraystretch{0.85}
\begin{tabular}{||c|c|c|c|c|c|c|c|c|c|c|c|}
\hline
\multicolumn{6}{||c|}{Particles}&\multicolumn{6}{c|}{$U(1)_{Y}$}\\
\hline
$q_{1}$&$q_{2}$&$q_{3}$&$q_{4}$&$q_{5}$&$q_{6}$&$I_{1}$&$I_{2}$&$I_{3}$&$\psi_{1}$&$\psi_{2}$&$\psi_{3}$\\
\hline
$q_{1}$&$q_{2}$&$q_{3}$&$q_{4}$&$q_{5}$&$q_{6}$&$I_{1}$&$\psi_{1}$&$\psi_{2}$&$I_{2}$&$I_{3}$&$I_{4}$\\
\hline
$d_{R}$&$d_{R}$&$d_{R}$&$d_{R}$&$u_{R}$&$u_{R}$&$-\frac{2}{3}$&$\alpha+\frac{2}{3}$&$-\alpha$&$\alpha$&$-\alpha-\frac{2}{3}$&$\alpha+\frac{1}{3}$\\
\hline
$d_{R}$&$d_{R}$&$d_{R}$&$u_{R}$&$d_{R}$&$u_{R}$&$-\frac{2}{3}$&$\alpha-\frac{1}{3}$&$-\alpha$&$\alpha$&$-\alpha-\frac{2}{3}$&$\alpha-\frac{2}{3}$\\
\hline
$d_{R}$&$d_{R}$&$u_{R}$&$d_{R}$&$d_{R}$&$u_{R}$&$-\frac{2}{3}$&$\alpha-\frac{1}{3}$&$-\alpha$&$\alpha$&$-\alpha-\frac{2}{3}$&$\alpha+\frac{1}{3}$\\
\hline
$d_{R}$&$d_{R}$&$d_{R}$&$u_{R}$&$u_{R}$&$d_{R}$&$-\frac{2}{3}$&$\alpha+\frac{2}{3}$&$-\alpha-1$&$\alpha$&$-\alpha-\frac{2}{3}$&$\alpha+\frac{1}{3}$\\
\hline
$d_{R}$&$d_{R}$&$u_{R}$&$d_{R}$&$u_{R}$&$d_{R}$&$-\frac{2}{3}$&$\alpha+\frac{2}{3}$&$-\alpha-1$&$\alpha$&$-\alpha-\frac{2}{3}$&$\alpha+\frac{4}{3}$\\
\hline
\end{tabular}}\!\!
\adjustbox{valign=t}{
\renewcommand\arraystretch{0.672}
\begin{tabular}{|c|c|c|c|c|c|c|c|c|c|c|c||}
\hline
\multicolumn{6}{|c|}{Particles}&\multicolumn{6}{c||}{$U(1)_{Y}$}\\
\hline
$q_{1}$&$q_{2}$&$q_{3}$&$q_{4}$&$q_{5}$&$q_{6}$&$I_{1}$&$I_{2}$&$I_{3}$&$I_{4}$&$\psi_{1}$&$\psi_{2}$\\
\hline
$q_{1}$&$q_{2}$&$q_{3}$&$q_{4}$&$q_{5}$&$q_{6}$&$I_{1}$&$\psi_{1}$&$\psi_{2}$&$I_{2}$&$I_{3}$&$I_{4}$\\
\hline
$d_{R}$&$d_{R}$&$d_{L}$&$d_{L}$&$u_{L}$&$u_{L}$&\multirow{6}*{$-\frac{2}{3}$}&\multirow{6}*{$\alpha+\frac{1}{6}$}&\multirow{6}*{$-\alpha-\frac{1}{2}$}&\multirow{6}*{$\alpha$}&\multirow{6}*{$-\alpha-\frac{2}{3}$}&\multirow{6}*{$\alpha+\frac{1}{3}$}\\
\cline{1-6}
$d_{R}$&$d_{R}$&$d_{L}$&$u_{L}$&$d_{L}$&$u_{L}$&&&&&&\\
\cline{1-6}
$d_{R}$&$d_{R}$&$u_{L}$&$d_{L}$&$d_{L}$&$u_{L}$&&&&&&\\
\cline{1-6}
$d_{R}$&$d_{R}$&$u_{L}$&$u_{L}$&$d_{L}$&$d_{L}$&&&&&&\\
\cline{1-6}
$d_{R}$&$d_{R}$&$u_{L}$&$d_{L}$&$u_{L}$&$d_{L}$&&&&&&\\
\hline
\end{tabular}}
\caption{The possible $U(1)_{Y}$ numbers of the fields in Topology-L-6/7.}
\end{sidewaystable}

\begin{sidewaystable}[p]\scriptsize
\adjustbox{valign=t}{
\renewcommand\arraystretch{1.197}
\begin{tabular}{||c|c|c|c|c|c|}
\hline
\multicolumn{6}{||c|}{$SU(3)_{C}$}\\
\hline
$\psi_{1}$&$I_{1}$&$I_{2}$&$\psi_{2}$&$\psi_{3}$&$I_{3}$\\
\hline
3&1&$\overline{3}$&3&1&$\overline{3}$\\
\cline{3-6}
&&$\overline{3}$&3&8&$\overline{3}$\\
\cline{3-6}
&&$\overline{3}$&3&8&6\\
\cline{3-6}
&&6&3&8&$\overline{3}$\\
\cline{3-6}
&&6&3&8&6\\
\hline
3&8&$\overline{3}$&3&1&$\overline{3}$\\
\cline{3-6}
&&$\overline{3}$&3&8&$\overline{3}$\\
\cline{3-6}
&&$\overline{3}$&3&8&6\\
\cline{3-6}
&&$\overline{3}$&$\overline{6}$&1&$\overline{3}$\\
\cline{3-6}
&&$\overline{3}$&$\overline{6}$&8&$\overline{3}$\\
\cline{3-6}
&&6&3&8&$\overline{3}$\\
\cline{3-6}
&&6&3&8&6\\
\cline{3-6}
&&6&$\overline{6}$&8&$\overline{3}$\\
\hline
8&3&3&$\overline{3}$&$\overline{3}$&1\\
\cline{3-6}
&&3&$\overline{3}$&$\overline{3}$&8\\
\cline{3-6}
&&3&$\overline{3}$&6&8\\
\cline{3-6}
&&$\overline{6}$&$\overline{3}$&$\overline{3}$&1\\
\cline{3-6}
&&$\overline{6}$&$\overline{3}$&$\overline{3}$&8\\
\cline{3-6}
&&3&6&$\overline{3}$&8\\
\cline{3-6}
&&3&6&6&8\\
\cline{3-6}
&&$\overline{6}$&6&$\overline{3}$&8\\
\hline
$\overline{3}$&6&1&8&3&3\\
\cline{3-6}
&&1&8&3&$\overline{6}$\\
\cline{3-6}
&&8&8&3&3\\
\cline{3-6}
&&8&8&3&$\overline{6}$\\
\cline{3-6}
&&8&8&$\overline{6}$&3\\
\hline
6&$\overline{3}$&8&1&3&3\\
\hline
\end{tabular}}\!\!
\adjustbox{valign=t}{
\renewcommand\arraystretch{1.197}
\begin{tabular}{|c|c|c|c|c|c||}
\hline
\multicolumn{6}{|c||}{$SU(3)_{C}$}\\
\hline
$\psi_{1}$&$I_{1}$&$I_{2}$&$\psi_{2}$&$\psi_{3}$&$I_{3}$\\
\hline
6&$\overline{3}$&8&1&$\overline{6}$&3\\
\cline{3-6}
&&8&8&3&3\\
\cline{3-6}
&&8&8&$\overline{6}$&3\\
\cline{3-6}
&&8&8&3&$\overline{6}$\\
\hline
$\overline{6}$&8&$\overline{3}$&3&1&$\overline{3}$\\
\cline{3-6}
&&$\overline{3}$&3&8&$\overline{3}$\\
\cline{3-6}
&&$\overline{3}$&3&8&6\\
\cline{3-6}
&&$\overline{3}$&$\overline{6}$&1&$\overline{3}$\\
\cline{3-6}
&&$\overline{3}$&$\overline{6}$&8&$\overline{3}$\\
\hline
8&$\overline{6}$&3&$\overline{3}$&$\overline{3}$&1\\
\cline{3-6}
&&3&$\overline{3}$&$\overline{3}$&8\\
\cline{3-6}
&&3&$\overline{3}$&6&8\\
\cline{3-6}
&&$\overline{6}$&$\overline{3}$&$\overline{3}$&1\\
\cline{3-6}
&&$\overline{6}$&$\overline{3}$&$\overline{3}$&8\\
\hline
$\overline{3}$&$\overline{3}$&1&1&3&3\\
\cline{3-6}
&&1&8&3&3\\
\cline{3-6}
&&1&8&3&$\overline{6}$\\
\cline{3-6}
&&8&1&3&3\\
\cline{3-6}
&&8&1&$\overline{6}$&3\\
\cline{3-6}
&&8&8&3&3\\
\cline{3-6}
&&8&8&3&$\overline{6}$\\
\cline{3-6}
&&8&8&$\overline{6}$&3\\
\hline
1&3&3&$\overline{3}$&$\overline{3}$&1\\
\cline{3-6}
&&3&$\overline{3}$&$\overline{3}$&8\\
\cline{3-6}
&&3&$\overline{3}$&6&8\\
\cline{3-6}
&&3&6&$\overline{3}$&8\\
\cline{3-6}
&&3&6&6&8\\
\hline
\end{tabular}}\quad
\adjustbox{valign=t}{
\renewcommand\arraystretch{0.5}
\begin{tabular}{||c|c|c|c|c|c|c|c|c|c|c|c|}
\hline
\multicolumn{12}{||c|}{$SU(2)_{L}$}\\
\hline
$q_{1}$&$q_{2}$&$q_{3}$&$q_{4}$&$q_{5}$&$q_{6}$&$\psi_{1}$&$I_{1}$&$I_{2}$&$\psi_{2}$&$\psi_{3}$&$I_{3}$\\
\hline
1&1&1&1&1&1&1&1&1&1&1&1\\
\cline{7-12}
&&&&&&2&2&2&2&2&2\\
\cline{7-12}
&&&&&&3&3&3&3&3&3\\
\hline
1&1&2&2&2&2&1&1&1&2&2&1\\
\cline{7-12}
&&&&&&1&1&1&2&2&3\\
\cline{7-12}
&&&&&&2&2&2&1&1&2\\
\cline{7-12}
&&&&&&2&2&2&1&3&2\\
\cline{7-12}
&&&&&&2&2&2&3&1&2\\
\cline{7-12}
&&&&&&2&2&2&3&3&2\\
\cline{7-12}
&&&&&&3&3&3&2&2&1\\
\cline{7-12}
&&&&&&3&3&3&2&2&3\\
\hline
1&2&1&2&2&2&1&1&2&1&1&2\\
\cline{7-12}
&&&&&&1&1&2&1&3&2\\
\cline{7-12}
&&&&&&2&2&1&2&2&1\\
\cline{7-12}
&&&&&&2&2&1&2&2&3\\
\cline{7-12}
&&&&&&2&2&3&2&2&1\\
\cline{7-12}
&&&&&&2&2&3&2&2&3\\
\cline{7-12}
&&&&&&3&3&2&3&1&2\\
\cline{7-12}
&&&&&&3&3&2&3&3&2\\
\cline{7-12}
1&2&2&1&2&2&1&1&2&2&2&1\\
\cline{7-12}
&&&&&&1&1&2&2&2&3\\
\cline{7-12}
&&&&&&2&2&1&1&1&2\\
\cline{7-12}
&&&&&&2&2&1&3&1&2\\
\cline{7-12}
&&&&&&2&2&3&1&3&2\\
\cline{7-12}
&&&&&&2&2&3&3&3&2\\
\cline{7-12}
&&&&&&3&3&2&2&2&1\\
\cline{7-12}
&&&&&&3&3&2&2&2&3\\
\hline
1&2&2&2&1&2&1&1&2&2&1&2\\
\cline{7-12}
&&&&&&1&1&2&2&3&2\\
\cline{7-12}
&&&&&&2&2&1&1&2&1\\
\cline{7-12}
&&&&&&2&2&3&3&2&3\\
\cline{7-12}
&&&&&&2&2&1&3&2&3\\
\cline{7-12}
&&&&&&2&2&3&1&2&1\\
\cline{7-12}
&&&&&&3&3&2&2&1&2\\
\cline{7-12}
&&&&&&3&3&2&2&3&2\\
\hline
1&2&2&2&2&1&1&1&2&2&1&1\\
\hline
\end{tabular}}\!\!
\adjustbox{valign=t}{
\renewcommand\arraystretch{0.5}
\begin{tabular}{|c|c|c|c|c|c|c|c|c|c|c|c||}
\hline
\multicolumn{12}{|c||}{$SU(2)_{L}$}\\
\hline
$q_{1}$&$q_{2}$&$q_{3}$&$q_{4}$&$q_{5}$&$q_{6}$&$\psi_{1}$&$I_{1}$&$I_{2}$&$\psi_{2}$&$\psi_{3}$&$I_{3}$\\
\hline
1&2&2&2&2&1&1&1&2&2&3&3\\
\cline{7-12}
&&&&&&2&2&1&1&2&2\\
\cline{7-12}
&&&&&&2&2&3&3&2&2\\
\cline{7-12}
&&&&&&2&2&1&3&2&2\\
\cline{7-12}
&&&&&&2&2&3&1&2&2\\
\cline{7-12}
&&&&&&3&3&2&2&1&1\\
\cline{7-12}
&&&&&&3&3&2&2&3&3\\
\hline
2&1&1&2&2&2&1&2&1&2&2&1\\
\cline{7-12}
&&&&&&1&2&1&2&2&3\\
\cline{7-12}
&&&&&&2&1&2&1&1&2\\
\cline{7-12}
&&&&&&2&1&2&1&3&2\\
\cline{7-12}
&&&&&&2&3&2&3&1&2\\
\cline{7-12}
&&&&&&2&3&2&3&3&2\\
\cline{7-12}
&&&&&&3&2&3&2&2&1\\
\cline{7-12}
&&&&&&3&2&3&2&2&3\\
\hline
2&1&2&1&2&2&1&2&1&1&1&2\\
\cline{7-12}
&&&&&&1&2&1&3&1&2\\
\cline{7-12}
&&&&&&2&1&2&2&2&1\\
\cline{7-12}
&&&&&&2&1&2&2&2&3\\
\cline{7-12}
&&&&&&2&3&2&2&2&1\\
\cline{7-12}
&&&&&&2&3&2&2&2&3\\
\cline{7-12}
&&&&&&3&2&3&1&3&2\\
\cline{7-12}
&&&&&&3&2&3&3&3&2\\
\hline
2&1&2&2&1&2&1&2&1&1&2&1\\
\cline{7-12}
&&&&&&1&2&1&3&2&3\\
\cline{7-12}
&&&&&&2&1&2&2&1&2\\
\cline{7-12}
&&&&&&2&1&2&2&3&2\\
\cline{7-12}
&&&&&&2&3&2&2&1&2\\
\cline{7-12}
&&&&&&2&3&2&2&3&2\\
\cline{7-12}
&&&&&&3&2&3&1&2&1\\
\cline{7-12}
&&&&&&3&2&3&3&2&3\\
\hline
2&1&2&2&2&1&1&2&1&1&2&2\\
\cline{7-12}
&&&&&&1&2&1&3&2&2\\
\cline{7-12}
&&&&&&2&1&2&2&1&1\\
\cline{7-12}
&&&&&&2&1&2&2&3&3\\
\cline{7-12}
&&&&&&2&3&2&2&1&1\\
\hline
\end{tabular}}
\caption{The possible $SU(3)_{C}$ and $SU(2)_{L}$ numbers of the fields in Topology-L-8.}\label{T-L-8_SU32}
\end{sidewaystable}

\begin{sidewaystable}[p]\scriptsize
\adjustbox{valign=t}{
\renewcommand\arraystretch{0.51}
\begin{tabular}{||c|c|c|c|c|c|c|c|c|c|c|c|}
\hline
\multicolumn{12}{||c|}{$SU(2)_{L}$}\\
\hline
$q_{1}$&$q_{2}$&$q_{3}$&$q_{4}$&$q_{5}$&$q_{6}$&$\psi_{1}$&$I_{1}$&$I_{2}$&$\psi_{2}$&$\psi_{3}$&$I_{3}$\\
\hline
2&1&2&2&2&1&2&3&2&2&3&3\\
\cline{7-12}
&&&&&&3&2&3&1&2&2\\
\cline{7-12}
&&&&&&3&2&3&3&2&2\\
\hline
2&2&1&1&2&2&1&2&2&2&2&1\\
\cline{7-12}
&&&&&&1&2&2&2&2&3\\
\cline{7-12}
&&&&&&2&1&1&1&1&2\\
\cline{7-12}
&&&&&&2&1&3&1&3&2\\
\cline{7-12}
&&&&&&2&3&1&3&1&2\\
\cline{7-12}
&&&&&&2&3&3&3&3&2\\
\cline{7-12}
&&&&&&3&2&2&2&2&1\\
\cline{7-12}
&&&&&&3&2&2&2&2&3\\
\hline
2&2&1&2&1&2&1&2&2&2&1&2\\
\cline{7-12}
&&&&&&1&2&2&2&3&2\\
\cline{7-12}
&&&&&&2&1&1&1&2&1\\
\cline{7-12}
&&&&&&2&1&3&1&2&1\\
\cline{7-12}
&&&&&&2&3&1&3&2&3\\
\cline{7-12}
&&&&&&2&3&3&3&2&3\\
\cline{7-12}
&&&&&&3&2&2&2&1&2\\
\cline{7-12}
&&&&&&3&2&2&2&3&2\\
\hline
2&2&1&2&2&1&1&2&2&2&1&1\\
\cline{7-12}
&&&&&&1&2&2&2&3&3\\
\cline{7-12}
&&&&&&2&1&1&1&2&2\\
\cline{7-12}
&&&&&&2&1&3&1&2&2\\
\cline{7-12}
&&&&&&2&3&1&3&2&2\\
\cline{7-12}
&&&&&&2&3&3&3&2&2\\
\cline{7-12}
&&&&&&3&2&2&2&1&1\\
\cline{7-12}
&&&&&&3&2&2&2&3&3\\
\hline
2&2&2&1&1&2&1&2&2&1&2&1\\
\cline{7-12}
&&&&&&1&2&2&3&2&3\\
\cline{7-12}
&&&&&&2&1&1&2&1&2\\
\cline{7-12}
&&&&&&2&1&3&2&3&2\\
\cline{7-12}
&&&&&&2&3&1&2&1&2\\
\cline{7-12}
&&&&&&2&3&3&2&3&2\\
\cline{7-12}
&&&&&&3&2&2&1&2&1\\
\cline{7-12}
&&&&&&3&2&2&3&2&3\\
\hline
2&2&2&1&2&1&1&2&2&1&2&2\\
\cline{7-12}
&&&&&&1&2&2&3&2&2\\
\hline
\end{tabular}}\!\!
\adjustbox{valign=t}{
\renewcommand\arraystretch{0.505}
\begin{tabular}{|c|c|c|c|c|c|c|c|c|c|c|c|}
\hline
\multicolumn{12}{|c|}{$SU(2)_{L}$}\\
\hline
$q_{1}$&$q_{2}$&$q_{3}$&$q_{4}$&$q_{5}$&$q_{6}$&$\psi_{1}$&$I_{1}$&$I_{2}$&$\psi_{2}$&$\psi_{3}$&$I_{3}$\\
\hline
2&2&2&1&2&1&2&1&1&2&1&1\\
\cline{7-12}
&&&&&&2&1&3&2&3&3\\
\cline{7-12}
&&&&&&2&3&1&2&1&1\\
\cline{7-12}
&&&&&&2&3&3&2&3&3\\
\cline{7-12}
&&&&&&3&2&2&1&2&2\\
\cline{7-12}
&&&&&&3&2&2&3&2&2\\
\hline
2&2&2&2&1&1&1&2&2&1&1&1\\
\cline{7-12}
&&&&&&1&2&2&3&3&3\\
\cline{7-12}
&&&&&&2&1&1&2&2&2\\
\cline{7-12}
&&&&&&2&1&3&2&2&2\\
\cline{7-12}
&&&&&&2&3&1&2&2&2\\
\cline{7-12}
&&&&&&2&3&3&2&2&2\\
\cline{7-12}
&&&&&&3&2&2&1&1&1\\
\cline{7-12}
&&&&&&3&2&2&3&3&3\\
\hline
2&2&1&1&1&1&1&2&2&2&2&2\\
\cline{7-12}
&&&&&&2&1&1&1&1&1\\
\cline{7-12}
&&&&&&2&3&3&3&3&3\\
\cline{7-12}
&&&&&&3&2&2&2&2&2\\
\hline
2&1&2&1&1&1&1&2&1&1&1&1\\
\cline{7-12}
&&&&&&2&1&2&2&2&2\\
\cline{7-12}
&&&&&&2&3&2&2&2&2\\
\cline{7-12}
&&&&&&3&2&3&3&3&3\\
\hline
2&1&1&2&1&1&1&2&1&2&2&2\\
\cline{7-12}
&&&&&&2&1&2&1&1&1\\
\cline{7-12}
&&&&&&2&3&2&3&3&3\\
\cline{7-12}
&&&&&&3&2&3&2&2&2\\
\hline
2&1&1&1&2&1&1&2&1&2&1&1\\
\cline{7-12}
&&&&&&2&1&2&1&2&2\\
\cline{7-12}
&&&&&&2&3&2&3&2&2\\
\cline{7-12}
&&&&&&3&2&3&2&3&3\\
\hline
2&1&1&1&1&2&1&2&1&2&1&2\\
\cline{7-12}
&&&&&&2&1&2&1&2&1\\
\cline{7-12}
&&&&&&2&3&2&3&2&3\\
\cline{7-12}
&&&&&&3&2&3&2&3&2\\
\hline
1&2&2&1&1&1&1&1&2&2&2&2\\
\cline{7-12}
&&&&&&2&2&1&1&1&1\\
\cline{7-12}
&&&&&&2&2&3&3&3&3\\
\hline
\end{tabular}}\!\!
\adjustbox{valign=t}{
\renewcommand\arraystretch{0.5}
\begin{tabular}{|c|c|c|c|c|c|c|c|c|c|c|c||}
\hline
\multicolumn{12}{|c||}{$SU(2)_{L}$}\\
\hline
$q_{1}$&$q_{2}$&$q_{3}$&$q_{4}$&$q_{5}$&$q_{6}$&$\psi_{1}$&$I_{1}$&$I_{2}$&$\psi_{2}$&$\psi_{3}$&$I_{3}$\\
\hline
1&2&2&1&1&1&3&3&2&2&2&2\\
\hline
1&2&1&2&1&1&1&1&2&1&1&1\\
\cline{7-12}
&&&&&&2&2&1&2&2&2\\
\cline{7-12}
&&&&&&2&2&3&2&2&2\\
\cline{7-12}
&&&&&&3&3&2&3&3&3\\
\hline
1&2&1&1&2&1&1&1&2&1&2&2\\
\cline{7-12}
&&&&&&2&2&1&2&1&1\\
\cline{7-12}
&&&&&&2&2&3&2&3&3\\
\cline{7-12}
&&&&&&3&3&2&3&2&2\\
\hline
1&2&1&1&1&2&1&1&2&1&2&1\\
\cline{7-12}
&&&&&&2&2&1&2&1&2\\
\cline{7-12}
&&&&&&2&2&3&2&3&2\\
\cline{7-12}
&&&&&&3&3&2&3&2&3\\
\hline
1&1&2&2&1&1&1&1&1&2&2&2\\
\cline{7-12}
&&&&&&2&2&2&1&1&1\\
\cline{7-12}
&&&&&&2&2&2&3&3&3\\
\cline{7-12}
&&&&&&3&3&3&2&2&2\\
\hline
1&1&2&1&2&1&1&1&1&2&1&1\\
\cline{7-12}
&&&&&&2&2&2&1&2&2\\
\cline{7-12}
&&&&&&2&2&2&3&2&2\\
\cline{7-12}
&&&&&&3&3&3&2&3&3\\
\hline
1&1&2&1&1&2&1&1&1&2&1&2\\
\cline{7-12}
&&&&&&2&2&2&1&2&1\\
\cline{7-12}
&&&&&&2&2&2&3&2&3\\
\cline{7-12}
&&&&&&3&3&3&2&3&2\\
\hline
1&1&1&2&2&1&1&1&1&1&2&2\\
\cline{7-12}
&&&&&&\textcolor{blue}{2}&\textcolor{blue}{2}&\textcolor{blue}{2}&\textcolor{blue}{2}&\textcolor{blue}{1}&\textcolor{blue}{1}\\
\cline{7-12}
&&&&&&2&2&2&2&3&3\\
\cline{7-12}
&&&&&&3&3&3&3&2&2\\
\hline
1&1&1&2&1&2&1&1&1&1&2&1\\
\cline{7-12}
&&&&&&2&2&2&2&1&2\\
\cline{7-12}
&&&&&&2&2&2&2&3&2\\
\cline{7-12}
&&&&&&3&3&3&3&2&3\\
\hline
1&1&1&1&2&2&1&1&1&1&1&2\\
\cline{7-12}
&&&&&&2&2&2&2&2&1\\
\cline{7-12}
&&&&&&2&2&2&2&2&3\\
\cline{7-12}
&&&&&&3&3&3&3&2&2\\
\hline
\end{tabular}}
\caption{The possible $SU(2)_{L}$ numbers of the fields in Topology-L-8. The blue row is the one that we select in the example model.}\label{T-L-8_SU2}
\end{sidewaystable}

\begin{sidewaystable}[p]\scriptsize
\adjustbox{valign=t}{
\renewcommand\arraystretch{1.2}
\begin{tabular}{||c|c|c|c|c|c|c|c|c|c|c|c|}
\hline
\multicolumn{6}{||c|}{Particles}&\multicolumn{6}{|c|}{$U(1)_{Y}$}\\
\hline
$q_{1}$&$q_{2}$&$q_{3}$&$q_{4}$&$q_{5}$&$q_{6}$&$\psi_{1}$&$I_{1}$&$I_{2}$&$\psi_{2}$&$\psi_{3}$&$I_{3}$\\
\hline
$u_{R}$&$u_{R}$&$d_{R}$&$d_{R}$&$d_{R}$&$d_{R}$&$\alpha$&$-\alpha+\frac{2}{3}$&$\alpha+\frac{2}{3}$&$-\alpha+\frac{1}{3}$&$\alpha+\frac{1}{3}$&$-\alpha$\\
\hline
$u_{R}$&$d_{R}$&$d_{R}$&$u_{R}$&$d_{R}$&$d_{R}$&$\alpha$&$-\alpha+\frac{2}{3}$&$\alpha-\frac{1}{3}$&$-\alpha+\frac{1}{3}$&$\alpha+\frac{1}{3}$&$-\alpha$\\
\hline
$u_{R}$&$d_{R}$&$d_{R}$&$d_{R}$&$d_{R}$&$u_{R}$&$\alpha$&$-\alpha+\frac{2}{3}$&$\alpha-\frac{1}{3}$&$-\alpha+\frac{1}{3}$&$\alpha-\frac{2}{3}$&$-\alpha$\\
\hline
$u_{R}$&$d_{R}$&$u_{R}$&$d_{R}$&$d_{R}$&$d_{R}$&$\alpha$&$-\alpha+\frac{2}{3}$&$\alpha-\frac{1}{3}$&$-\alpha+\frac{4}{3}$&$\alpha-\frac{2}{3}$&$-\alpha+1$\\
\hline
$u_{R}$&$d_{R}$&$d_{R}$&$d_{R}$&$u_{R}$&$d_{R}$&$\alpha$&$-\alpha+\frac{2}{3}$&$\alpha-\frac{1}{3}$&$-\alpha+\frac{1}{3}$&$\alpha-\frac{2}{3}$&$-\alpha+1$\\
\hline
$d_{R}$&$d_{R}$&$u_{R}$&$d_{R}$&$u_{R}$&$d_{R}$&$\alpha$&$-\alpha-\frac{1}{3}$&$\alpha-\frac{1}{3}$&$-\alpha+\frac{1}{3}$&$\alpha-\frac{2}{3}$&$-\alpha+1$\\
\hline
\end{tabular}}\!\!
\adjustbox{valign=t}{
\renewcommand\arraystretch{1.2}
\begin{tabular}{|c|c|c|c|c|c|c|c|c|c|c|c||}
\hline
\multicolumn{6}{|c|}{Particles}&\multicolumn{6}{|c||}{$U(1)_{Y}$}\\
\hline
$q_{1}$&$q_{2}$&$q_{3}$&$q_{4}$&$q_{5}$&$q_{6}$&$\psi_{1}$&$I_{1}$&$I_{2}$&$\psi_{2}$&$\psi_{3}$&$I_{3}$\\
\hline
$q_{1,L}$&$q_{2,L}$&$d_{R}$&$d_{R}$&$q_{5,L}$&$q_{6,L}$&$\alpha$&$-\alpha+\frac{1}{6}$&$\alpha+\frac{1}{6}$&$-\alpha-\frac{1}{6}$&$\alpha-\frac{1}{6}$&$-\alpha$\\
\hline
$q_{1,L}$&$q_{2,L}$&$d_{R}$&$q_{4,L}$&$q_{5,L}$&$d_{R}$&$\alpha$&$-\alpha+\frac{1}{6}$&$\alpha+\frac{1}{6}$&$-\alpha-\frac{1}{6}$&$\alpha+\frac{1}{3}$&$-\alpha$\\
\hline
$q_{1,L}$&$q_{2,L}$&$q_{3,L}$&$d_{R}$&$d_{R}$&$q_{6,L}$&$\alpha$&$-\alpha+\frac{1}{6}$&$\alpha+\frac{1}{6}$&$-\alpha+\frac{1}{3}$&$\alpha-\frac{1}{6}$&$-\alpha$\\
\hline
$q_{1,L}$&$q_{2,L}$&$q_{3,L}$&$q_{4,L}$&$d_{R}$&$d_{R}$&$\alpha$&$-\alpha+\frac{1}{6}$&$\alpha+\frac{1}{6}$&$-\alpha+\frac{1}{3}$&$\alpha+\frac{1}{3}$&$-\alpha$\\
\hline
$q_{1,L}$&$d_{R}$&$q_{3,L}$&$d_{R}$&$q_{5,L}$&$q_{6,L}$&$\alpha$&$-\alpha+\frac{1}{6}$&$\alpha-\frac{1}{3}$&$-\alpha+\frac{1}{3}$&$\alpha-\frac{2}{3}$&$-\alpha+\frac{1}{2}$\\
\hline
$q_{1,L}$&$d_{R}$&$q_{3,L}$&$q_{4,L}$&$q_{5,L}$&$d_{R}$&$\alpha$&$-\alpha+\frac{1}{6}$&$\alpha-\frac{1}{3}$&$-\alpha+\frac{1}{3}$&$\alpha-\frac{1}{6}$&$-\alpha+\frac{1}{2}$\\
\hline
\end{tabular}}
\caption{The possible $U(1)_{Y}$ numbers of the fields in Topology-L-8.}\label{T-L-8_U1-1}
\end{sidewaystable}

\begin{sidewaystable}[p]\scriptsize
\adjustbox{valign=t}{
\renewcommand\arraystretch{1.2}
\begin{tabular}{||c|c|c|c|c|c|c|c|c|c|c|c|}
\hline
\multicolumn{6}{||c|}{Particles}&\multicolumn{6}{|c|}{$U(1)_{Y}$}\\
\hline
$q_{1}$&$q_{2}$&$q_{3}$&$q_{4}$&$q_{5}$&$q_{6}$&$\psi_{1}$&$I_{1}$&$I_{2}$&$\psi_{2}$&$\psi_{3}$&$I_{3}$\\
\hline
$q_{1,L}$&$u_{R}$&$q_{3,L}$&$d_{R}$&$d_{R}$&$d_{R}$&$\alpha$&$-\alpha+\frac{1}{6}$&$\alpha+\frac{2}{3}$&$-\alpha+\frac{1}{3}$&$\alpha+\frac{1}{3}$&$-\alpha$\\
\hline
$q_{1,L}$&$u_{R}$&$d_{R}$&$d_{R}$&$q_{5,L}$&$d_{R}$&$\alpha$&$-\alpha+\frac{1}{6}$&$\alpha+\frac{2}{3}$&$-\alpha-\frac{1}{6}$&$\alpha+\frac{1}{3}$&$-\alpha$\\
\hline
$d_{R}$&$q_{2,L}$&$q_{3,L}$&$d_{R}$&$d_{R}$&$u_{R}$&$\alpha$&$-\alpha-\frac{1}{3}$&$\alpha+\frac{1}{6}$&$-\alpha-\frac{1}{6}$&$\alpha-\frac{1}{6}$&$-\alpha-\frac{1}{2}$\\
\hline
$d_{R}$&$q_{2,L}$&$q_{3,L}$&$u_{R}$&$d_{R}$&$d_{R}$&$\alpha$&$-\alpha-\frac{1}{3}$&$\alpha+\frac{1}{6}$&$-\alpha-\frac{1}{6}$&$\alpha+\frac{5}{6}$&$-\alpha-\frac{1}{2}$\\
\hline
$d_{R}$&$q_{2,L}$&$d_{R}$&$d_{R}$&$q_{5,L}$&$u_{R}$&$\alpha$&$-\alpha-\frac{1}{3}$&$\alpha+\frac{1}{6}$&$-\alpha-\frac{2}{3}$&$\alpha-\frac{1}{6}$&$-\alpha-\frac{1}{2}$\\
\hline
$d_{R}$&$q_{2,L}$&$d_{R}$&$u_{R}$&$q_{5,L}$&$d_{R}$&$\alpha$&$-\alpha-\frac{1}{3}$&$\alpha+\frac{1}{6}$&$-\alpha-\frac{2}{3}$&$\alpha+\frac{5}{6}$&$-\alpha-\frac{1}{2}$\\
\hline
\end{tabular}}\!\!
\adjustbox{valign=t}{
\renewcommand\arraystretch{1.2}
\begin{tabular}{|c|c|c|c|c|c|c|c|c|c|c|c||}
\hline
\multicolumn{6}{|c|}{Particles}&\multicolumn{6}{|c||}{$U(1)_{Y}$}\\
\hline
$q_{1}$&$q_{2}$&$q_{3}$&$q_{4}$&$q_{5}$&$q_{6}$&$\psi_{1}$&$I_{1}$&$I_{2}$&$\psi_{2}$&$\psi_{3}$&$I_{3}$\\
\hline
$d_{R}$&$d_{R}$&$u_{R}$&$q_{4,L}$&$q_{5,L}$&$d_{R}$&$\alpha$&$-\alpha-\frac{1}{3}$&$\alpha-\frac{1}{3}$&$-\alpha+\frac{1}{3}$&$\alpha-\frac{1}{6}$&$-\alpha+\frac{1}{2}$\\
\hline
$u_{R}$&$d_{R}$&$d_{R}$&$q_{4,L}$&$q_{5,L}$&$d_{R}$&$\alpha$&$-\alpha+\frac{2}{3}$&$\alpha-\frac{1}{3}$&$-\alpha+\frac{1}{3}$&$\alpha-\frac{1}{6}$&$-\alpha+\frac{1}{2}$\\
\hline
$d_{R}$&$d_{R}$&$u_{R}$&$d_{R}$&$q_{5,L}$&$q_{6,L}$&$\alpha$&$-\alpha-\frac{1}{3}$&$\alpha-\frac{1}{3}$&$-\alpha+\frac{1}{3}$&$\alpha-\frac{2}{3}$&$-\alpha+\frac{1}{2}$\\
\hline
$u_{R}$&$d_{R}$&$d_{R}$&$d_{R}$&$q_{5,L}$&$q_{6,L}$&$\alpha$&$-\alpha+\frac{2}{3}$&$\alpha-\frac{1}{3}$&$-\alpha+\frac{1}{3}$&$\alpha-\frac{2}{3}$&$-\alpha+\frac{1}{2}$\\
\hline
$d_{R}$&$q_{2,L}$&$d_{R}$&$q_{4,L}$&$u_{R}$&$d_{R}$&$\alpha$&$-\alpha-\frac{1}{3}$&$\alpha+\frac{1}{6}$&$-\alpha-\frac{2}{3}$&$\alpha+\frac{1}{3}$&$-\alpha$\\
\hline
$d_{R}$&$q_{2,L}$&$u_{R}$&$q_{4,L}$&$d_{R}$&$d_{R}$&$\alpha$&$-\alpha-\frac{1}{3}$&$\alpha+\frac{1}{6}$&$-\alpha+\frac{1}{3}$&$\alpha+\frac{1}{3}$&$-\alpha$\\
\hline
\end{tabular}}
\caption{The possible $U(1)_{Y}$ numbers of the fields in Topology-L-8.}\label{T-L-8_U1-2}
\end{sidewaystable}

%%%%%%%%%%%%%%%%%%%%%%%%%%%%%%%%%%%%%%%%%%%%%%%%%%%%%%%%%%%%%%%%%%%%%%%%%%%%%%%%%%%%%%%%%%%%%%%%%%%%%%%%%%%%%%%%5
\begin{sidewaystable}[p]\scriptsize
\adjustbox{valign=t}{
\renewcommand\arraystretch{0.75}
\begin{tabular}{||c|c|c|c|c|c|}
\hline
\multicolumn{6}{||c|}{$SU(3)_{C}$}\\
\hline
$I_{1}$&$I_{2}$&$I_{3}$&$\psi_{1}$&$\psi_{2}$&$\psi_{3}$\\
\hline
$\overline{3}$&$\overline{3}$&$\overline{3}$&3&3&1\\
\cline{4-6}
&&&3&3&8\\
\cline{4-6}
&&&1&$\overline{3}$&3\\
\cline{4-6}
&&&$\overline{3}$&1&$\overline{3}$\\
\cline{4-6}
&&&3&$\overline{6}$&8\\
\cline{4-6}
&&&$\overline{6}$&3&8\\
\cline{4-6}
&&&$\overline{3}$&8&$\overline{3}$\\
\cline{4-6}
&&&$\overline{3}$&8&6\\
\cline{4-6}
&&&8&$\overline{3}$&3\\
\cline{4-6}
&&&8&$\overline{3}$&$\overline{6}$\\
\cline{4-6}
&&&6&8&$\overline{3}$\\
\cline{4-6}
&&&8&6&3\\
\hline
$\overline{3}$&$\overline{3}$&6&3&3&8\\
\cline{4-6}
&&&3&$\overline{6}$&1\\
\cline{4-6}
&&&3&$\overline{6}$&8\\
\cline{4-6}
&&&1&$\overline{3}$&3\\
\cline{4-6}
&&&$\overline{3}$&1&6\\
\cline{4-6}
&&&$\overline{6}$&3&8\\
\cline{4-6}
&&&$\overline{3}$&8&$\overline{3}$\\
\cline{4-6}
&&&$\overline{3}$&8&6\\
\cline{4-6}
&&&8&$\overline{3}$&3\\
\cline{4-6}
&&&6&8&$\overline{3}$\\
\cline{4-6}
&&&8&6&3\\
\hline
$\overline{3}$&6&$\overline{3}$&3&3&8\\
\cline{4-6}
&&&3&$\overline{6}$&1\\
\cline{4-6}
&&&3&$\overline{6}$&8\\
\cline{4-6}
&&&$\overline{6}$&3&8\\
\cline{4-6}
&&&$\overline{3}$&8&$\overline{3}$\\
\cline{4-6}
&&&8&$\overline{3}$&3\\
\cline{4-6}
&&&8&$\overline{3}$&6\\
\cline{4-6}
&&&6&8&6\\
\cline{4-6}
&&&8&6&3\\
\hline
$\overline{3}$&6&6&3&3&8\\
\cline{4-6}
&&&3&$\overline{6}$&8\\
\cline{4-6}
&&&$\overline{6}$&3&8\\
\cline{4-6}
&&&$\overline{3}$&8&$\overline{3}$\\
\hline
\end{tabular}}\!\!
\adjustbox{valign=t}{
\renewcommand\arraystretch{0.665}
\begin{tabular}{|c|c|c|c|c|c||}
\hline
\multicolumn{6}{|c||}{$SU(3)_{C}$}\\
\hline
$I_{1}$&$I_{2}$&$I_{3}$&$\psi_{1}$&$\psi_{2}$&$\psi_{3}$\\
\hline
$\overline{3}$&6&6&8&$\overline{3}$&3\\
\cline{4-6}
&&&6&8&6\\
\cline{4-6} 
&&&8&6&$\overline{6}$\\
\hline
6&$\overline{3}$&$\overline{3}$&3&3&1\\
\cline{4-6}
&&&3&3&8\\
\cline{4-6}
&&&1&6&3\\
\cline{4-6}
&&&6&1&$\overline{3}$\\
\cline{4-6}
&&&$\overline{3}$&8&$\overline{3}$\\
\cline{4-6}
&&&$\overline{3}$&8&6\\
\cline{4-6}
&&&8&$\overline{3}$&3\\
\cline{4-6}
&&&8&$\overline{3}$&$\overline{6}$\\
\cline{4-6}
&&&6&8&$\overline{3}$\\
\cline{4-6}
&&&8&6&3\\
\cline{4-6}
&&&$\overline{6}$&$\overline{6}$&8\\
\hline
6&$\overline{3}$&6&3&3&8\\
\cline{4-6}
&&&$\overline{3}$&8&$\overline{3}$\\
\cline{4-6}
&&&$\overline{3}$&8&6\\
\cline{4-6}
&&&8&$\overline{3}$&3\\
\cline{4-6}
&&&6&8&$\overline{3}$\\
\cline{4-6}
&&&8&6&$\overline{6}$\\
\cline{4-6}
&&&$\overline{6}$&$\overline{6}$&8\\
\hline
6&6&$\overline{3}$&3&3&8\\
\cline{4-6}
&&&$\overline{3}$&8&$\overline{3}$\\
\cline{4-6}
&&&8&$\overline{3}$&3\\
\cline{4-6}
&&&6&8&6\\
\cline{4-6}
&&&8&6&$\overline{6}$\\
\cline{4-6}
&&&$\overline{6}$&$\overline{6}$&1\\
\cline{4-6}
&&&$\overline{6}$&$\overline{6}$&8\\
\hline
6&6&6&3&3&8\\
\cline{4-6}
&&&$\overline{3}$&8&$\overline{3}$\\
\cline{4-6}
&&&1&6&$\overline{6}$\\
\cline{4-6}
&&&6&1&6\\
\cline{4-6}
&&&8&$\overline{3}$&3\\
\cline{4-6}
&&&8&$\overline{3}$&$\overline{6}$\\
\cline{4-6}
&&&6&8&6\\
\cline{4-6}
&&&8&6&3\\
\cline{4-6}
&&&$\overline{6}$&$\overline{6}$&8\\
\hline
\end{tabular}}\quad
\adjustbox{valign=t}{
\renewcommand\arraystretch{1.205}
\begin{tabular}{||c|c|c|c|c|c|c|c|c|c|c|c||}
\hline
\multicolumn{12}{||c||}{$SU(2)_{L}$}\\
\hline
$q_{1}$&$q_{2}$&$q_{3}$&$q_{4}$&$q_{5}$&$q_{6}$&$I_{1}$&$I_{2}$&$I_{3}$&$\psi_{1}$&$\psi_{2}$&$\psi_{3}$\\
\hline
$q_{1}$&$q_{2}$&$q_{3}$&$q_{4}$&$q_{5}$&$q_{6}$&$I_{1}$&$I_{2}$&$I_{3}$&$I_{4}$&$I_{5}$&$I_{6}$\\
\hline
1&1&1&1&1&1&1&1&1&1&1&1\\
\cline{7-12}
&&&&&&1&1&1&2&2&2\\
\cline{7-12}
&&&&&&1&1&1&3&3&3\\
\hline
1&1&1&1&2&2&1&1&1&1&1&1\\
\cline{7-12}
&&&&&&1&1&1&2&2&2\\
\hline
\end{tabular}}
\caption{The possible quantum numbers of the fields in Topology-L-9 and 10.}
\end{sidewaystable}

\begin{sidewaystable}[p]\scriptsize
\adjustbox{valign=t}{
\renewcommand\arraystretch{1.1}
\begin{tabular}{||c|c|c|c|c|c|}
\hline
\multicolumn{6}{||c|}{$SU(3)_{C}$}\\
\hline
$I_{1}$&$I_{2}$&$I_{3}$&$\psi_{1}$&$\psi_{2}$&$\psi_{3}$\\
\hline
$I_{1}$&$I_{2}$&$\psi_{2}$&$\psi_{1}$&$I_{3}$&$I_{4}$\\
\hline
$\overline{3}$&$\overline{3}$&$\overline{3}$&1&3&3\\
\cline{3-6}
&&$\overline{3}$&8&3&3\\
\cline{3-6}
&&6&8&3&3\\
\cline{3-6}
&&6&1&3&$\overline{6}$\\
\cline{3-6}
&&6&8&3&$\overline{6}$\\
\cline{3-6}
&&$\overline{3}$&8&3&$\overline{6}$\\
\cline{3-6}
&&$\overline{3}$&1&$\overline{6}$&3\\
\cline{3-6}
&&$\overline{3}$&8&$\overline{6}$&3\\
\cline{3-6}
&&8&1&$\overline{3}$&8\\
\cline{3-6}
&&8&8&$\overline{3}$&8\\
\cline{3-6}
&&1&8&$\overline{3}$&8\\
\cline{3-6}
&&3&1&1&$\overline{3}$\\
\cline{3-6}
&&3&8&1&$\overline{3}$\\
\cline{3-6}
&&1&1&$\overline{3}$&1\\
\cline{3-6}
&&8&8&$\overline{3}$&1\\
\cline{3-6}
&&3&1&8&$\overline{3}$\\
\cline{3-6}
&&3&8&8&$\overline{3}$\\
\cline{3-6}
&&$\overline{6}$&8&8&$\overline{3}$\\
\cline{3-6}
&&8&1&6&8\\
\cline{3-6}
&&8&8&6&8\\
\cline{3-6}
&&$\overline{6}$&1&8&6\\
\cline{3-6}
&&3&8&8&6\\
\cline{3-6}
&&$\overline{6}$&8&8&6\\
\hline
$\overline{3}$&6&$\overline{3}$&8&3&3\\
\cline{3-6}
&&6&8&3&3\\
\cline{3-6}
&&$\overline{3}$&8&3&$\overline{6}$\\
\cline{3-6}
&&6&8&3&$\overline{6}$\\
\cline{3-6}
&&$\overline{3}$&8&$\overline{6}$&3\\
\cline{3-6}
&&1&8&$\overline{3}$&8\\
\cline{3-6}
&&8&8&$\overline{3}$&8\\
\cline{3-6}
&&3&8&8&$\overline{3}$\\
\cline{3-6}
&&8&8&$\overline{3}$&1\\
\cline{3-6}
&&3&8&1&$\overline{3}$\\
\hline
\end{tabular}%
}
\caption{The possible quantum numbers of the fields in Topology-L-11 and 12.}
\end{sidewaystable}

\begin{sidewaystable}[p]\scriptsize
\adjustbox{valign=t}{
\renewcommand\arraystretch{0.85}
%
}
\caption{The possible quantum numbers of the fields in Topology-L-13 and 14.}
\end{sidewaystable}

\begin{sidewaystable}[p]\scriptsize
\adjustbox{valign=t}{
\renewcommand\arraystretch{1.345}
%
}\caption{The possible quantum numbers of the fields in Topology-L-15 and 16.}
\end{sidewaystable}

\begin{sidewaystable}[p]\scriptsize
\adjustbox{valign=t}{
\renewcommand\arraystretch{0.845}
%
}\caption{The possible quantum numbers of the fields in Topology-L-15 and 16.}
\end{sidewaystable}
\end{rotatepage}
\end{appendices}

%%%references%%%
\newpage
\bibliographystyle{utphys}
\bibliography{references}

\end{document}